\documentclass[twocolumn]{aastex631}

\newcommand{\msun}{$M_{\odot}$}
\newcommand{\lsun}{$L_{\odot}$}
\newcommand{\h}{$h^{-1}$}
\newcommand{\kms}{km~s$^{-1}$}

\newcommand{\hi}{H{\sc\,i}}
\newcommand{\gapp}{$_>\atop{^\sim}$}
\newcommand{\lapp}{$_<\atop{^\sim}$}

\shorttitle{Searching for Dwarf Galaxies with ALFALFA and WIYN}
\shortauthors{Rhode et al.}

\begin{document}

\title{A Search for Gas-Rich Dwarf Galaxies in the Local Universe with ALFALFA \\and the WIYN One Degree Imager}

\correspondingauthor{Katherine Rhode}
\author[0000-0001-8283-4591]{Katherine L. Rhode}
\affiliation{Department of Astronomy, Indiana University, 727 East Third Street, 
Bloomington, IN 47405, USA}
\email{krhode@indiana.edu}

\author[0000-0002-3222-2949]{Nicholas J. Smith}
\affiliation{Department of Astronomy, Indiana University, 727 East
 Third Street, Bloomington, IN 47405, USA}

\author[0000-0003-4364-0799]{William F. Janesh}
\affiliation{Department of Astronomy, Case Western Reserve University, 10900 Euclid Avenue, Cleveland, OH 44106, USA}
 
\author[0000-0001-8483-603X]{John J. Salzer}
\affiliation{Department of Astronomy, Indiana University, 727 East
 Third Street, Bloomington, IN 47405, USA}
 
\author[0000-0002-9798-5111]{Elizabeth A. K. Adams}
\affiliation{ASTRON, Netherlands Institute for Radio Astronomy, Oude Hoogeveensedijk 4, 7991 PD Dwingeloo, The Netherlands}
\affiliation{Kapteyn Astronomical Institute, University of Groningen, Landleven 12, 9747 AD, Groningen, The Netherlands}

\author[0000-0001-5334-5166]{Martha P. Haynes}
\affiliation{Cornell Center for Astrophysics and Planetary Science, 
Space Sciences Building, Cornell University, Ithaca, NY 14853, USA}

\author[0000-0001-9165-8905]{Steven Janowiecki}
\affiliation{University of Texas, Hobby-Eberly Telescope, McDonald Observatory, TX 79734, USA}

\author[0000-0002-1821-7019]{John M. Cannon}
\affiliation{Department of Physics \& Astronomy, Macalester College, 1600 Grand Avenue, Saint Paul, MN 55105}



\begin{abstract}
We present results from an optical search for Local Group dwarf galaxy
candidates associated with the Ultra-Compact High Velocity Clouds (UCHVCs)
discovered by the ALFALFA neutral hydrogen survey.  The ALFALFA UCHVCs
are isolated, compact HI clouds with projected sizes, velocities, and
estimated HI masses that suggest they may be nearby dwarf galaxies,
but that have no clear counterpart in existing optical survey data.
We observed 26 UCHVCs with the WIYN 3.5-m telescope and One Degree
Imager (ODI) in two broadband filters and searched the images
for resolved stars with properties that match those of stars in typical dwarf galaxies at distances $<$2.5 Mpc.  We identify one promising dwarf galaxy candidate
at a distance of $\sim$570~kpc associated with the UCHVC AGC~268071,
and five other candidates that may deserve additional follow-up.  We
carry out a detailed analysis of ODI imaging of a UCHVC that is close
in both projected distance and radial velocity to the outer-halo Milky
Way globular cluster Pal~3. We also use our improved detection methods
to reanalyze images of five UCHVCs that were found to have possible
optical counterparts during the first phase of the project,
and confirm the detection of a possible stellar counterpart to the UCHVC AGC~249525
at an estimated distance of $\sim$2 Mpc. We compare
the optical and HI properties of the dwarf galaxy candidates to the
results from recent theoretical simulations that model 
satellite galaxy populations in group environments,
as well as to the observed properties of galaxies in and around the Local Group. 

\end{abstract}



\section{Introduction}
\label{sec:introduction}

The Arecibo Legacy Fast ALFA survey (ALFALFA; \citealt{giovanelli05a,
haynes18a}) was a survey for \hi\ emission carried out from
2005$-$2012 with the Arecibo L-band Feed Array (ALFA) instrument on
the 305-m Arecibo radio telescope.  The survey covered $\sim$7000
sq. degrees of sky in drift-scan mode, 
detecting sources of 21-cm line emission without regard to the
locations of optical sources.  The final catalog produced by the ALFALFA 
survey includes $\sim$31,500 \hi\ sources out to a redshift of $z\sim 0.06$, 
a large fraction of which had never before been observed via their 21-cm
emission \citep{haynes18a}.

Because of its excellent sensitivity, ALFALFA was able to detect
objects with only modest amounts of \hi\ ($\sim$$10^4 - 10^5$ \msun) out
to distances beyond the edge of the Local Group. A new class of potentially nearby, low-mass \hi\ sources was
identified in the ALFALFA survey data and dubbed Ultra-Compact High
Velocity Clouds (UCHVCs; \citet{giovanelli10a}). Roughly 100 of these
objects were found in ALFALFA; \citet{adams13a} laid out a 
set of criteria for sources that qualify as UCHVCs and presented an initial catalog.  
These objects are compact (angular diameter $<$30\arcmin), 
low-mass (\hi\ masses of $10^5 - 10^6$ \msun\ at
1~Mpc), isolated from other sources in terms of both sky position and velocity,
%
and have
velocities that help distinguish them
from Galactic High Velocity Clouds
while making them likely to be located within the Local Volume. 
Perhaps most importantly, ALFALFA sources are classified as
UCHVCs only if they lack an obvious optical counterpart when
matched with existing optical survey data from the Sloan Digital Sky
Survey (SDSS; \citealt{york00a,eisenstein11a}) or the Digitized Sky Survey (DSS).

The properties of the UCHVCs imply that these objects could represent a
population of low-mass, gas-bearing dark matter halos located in and around
the Local Group; this is the ``minihalo hypothesis'' described in
\citet{giovanelli10a} and \citet{adams13a}. 
Accordingly, our group has been observing a selection of the UCHVCs at optical wavelengths in
order to
investigate their nature and
the possibility that the UCHVCs are 
nearby gas-rich dwarf galaxies that have previously gone undetected.
We image the UCHVC locations at optical wavelengths
with the WIYN 3.5-m Observatory\footnote{The WIYN Observatory is a
  joint facility of the NSF’s National Optical-Infrared Astronomy
  Research Laboratory, Indiana University, the University of
  Wisconsin-Madison, Pennsylvania State University,
and Purdue University.} and
systematically search the area around the location of the
\hi\ detection to identify any stellar populations that may be
associated with the neutral gas.
The first UCHVC we observed with WIYN led to the discovery of the 
nearby gas-rich dwarf galaxy Leo~P \citep{giovanelli13a,rhode13a} which
is located just outside the Local Group at $\sim$1.6 Mpc, is extremely
metal-deficient,
and is the lowest-mass galaxy known that is actively forming stars
\citep{skillman13a,mcquinn15b}.

After the discovery of Leo~P, we initiated a systematic optical observing campaign that targeted
UCHVCs from the catalog presented in \citet{adams13a} as well as additional sources drawn from the low-velocity ALFALFA grids.  We used wide-field imaging in two broadband filters, resolved stellar photometry, and a color magnitude diagram (CMD) filtering technique to search for the presence of possible stellar populations
associated with the HI clouds over a distance range of 250 kpc to 2.5 Mpc.  
The first part of the campaign was carried out with the WIYN One Degree Imager with a partially-filled focal plane -- an instrument referred to as pODI -- and 
focused mainly on objects with higher \hi\ column density values.  In addition, we used the Very Large Array (VLA) and the Westerbork Synthesis Radio Telescope (WSRT) to acquire and analyze \hi\ synthesis data for several of the targets in the pODI sample, in order to obtain more detailed information about the objects' \hi\ distributions and kinematics \citep[e.g.,][]{adams15a,adams16a,paine20a, bralts20a}. Optical imaging studies were also carried out by other groups to search for counterparts to the UCHVCs and to other potentially nearby gas-rich sources found in surveys like GALFA-HI \citep{bellazzini15a, bellazzini15b, sand15a}; the results from these 
studies will be discussed further in later sections of this paper.

A total of 23 UCHVCs that were observed with WIYN pODI were analyzed in the first phase of the project.  In \citet{janesh15a}, we described the WIYN survey and methods
and presented some initial results for AGC~198606, 
a UCHVC that is
close in both projected spatial location and velocity to the Local Group dwarf galaxy Leo~T \citep{irwin07a,ryan-weber08a}.
Results for another UCHVC, AGC~249525, were initially presented in \citet{janesh17a} and the complete set of results for the first 23 UCHVCs we analyzed was published in
\citet{janesh19a} (hereafter J19). J19 identified a total of five UCHVCs 
that had possible faint stellar counterparts 
in the pODI imaging data. 
These five counterparts had estimated distances ranging from 
$\sim$350~kpc to 1.6~Mpc and estimated total optical magnitudes between $M_V$ $=$ $-$1.4 and $-$7.1 mag.

The second phase of our UCHVC follow-up imaging campaign was carried out with the upgraded One Degree Imager (ODI) camera on WIYN,
which was commissioned in 2015 and which provides a field-of-view (FOV) more than three times larger than that of pODI. 
In a series of observing runs between 2015 and 2021,
we acquired deep WIYN ODI images of more than 30 additional UCHVC targets in two broadband filters (SDSS $g$ and $i$).  During this phase of the project, we made improvements to our detection and analysis methods and then applied the updated methods to the ODI sample of objects.  We also used the improved methods to reanalyze the imaging data of the five UCHVCs in the pODI sample for which J19 identified potential stellar counterparts.

In this paper, we present the results from the second and final phase of our campaign to obtain follow-up optical imaging of ALFALFA UCHVCs.  In Section~\ref{sec:sample}, we describe the selection criteria and properties of the UCHVCs we observed and analyzed.  Section~\ref{sec:data reduction} summarizes the procedures we used to acquire imaging observations and produce source catalogs and calibrated broadband optical photometry for each UCHVC field. In Section~\ref{sec:search method}, we describe the steps we carried out to search for stellar counterparts to the HI sources and to assess the significance and validity of the dwarf galaxy candidates 
that we have identified.  We also discuss the improvements that were made to our analysis methods in the second phase of the project. 
Section~\ref{sec:results} presents the full results of this analysis.  In the last section of the paper, we examine our results in the context of recent relevant theoretical work and other observational studies aimed at detecting and studying gas-rich dwarf galaxies in and around the Local Group. 

\section{ The Sample of UCHVCs Analyzed for This Work}
\label{sec:sample}

The UCHVCs we observed were drawn from the ALFALFA survey data \citep{haynes18a} using criteria laid out in detail in \citet{adams13a}. The primary criteria that were used to select sources are as follows:

\begin{itemize}
    \item The source must have an \hi\ major axis, $a$, less than 30$\arcmin$ in order to have a size consistent with expectations for the 
baryonic component of low-mass dark matter halos. This corresponds to a diameter of $\sim$2 kpc at a
distance of 250 kpc, $\sim$9 kpc at 1 Mpc, and $\sim$20 kpc at 2.5 Mpc.

\item The source must have an ALFALFA signal-to-noise ratio greater than 8, as defined in Equation 2 in the ALFALFA 40\% catalog \citep{haynes11a}. 

\item The source must have no more than three neighboring HI structures within 3 degrees on the
sky or 15~\kms\ in velocity space in order to ensure that it is sufficiently isolated. It must
also have at least a 15-degree separation from any previously known high-velocity cloud (HVC) complexes.

\item The source must have a heliocentric velocity between $-$500 and 1000 \kms, so that
it is likely to reside in the Local Volume.

\item The source must have an absolute velocity with respect to the Local Standard of Rest (LSR) greater than 120 \kms, to avoid contamination with Galactic HVCs.  This criterion was in some cases relaxed when all other criteria were met, since some nearby dwarf galaxies 
have velocities lower than this value 
\citep[e.g., Leo~T, with $v_{\rm LSR}$ $\sim$60~\kms;][]{irwin07a,ryan-weber08a}.

\item The source must have no clear optical counterparts in either SDSS or DSS.

\end{itemize}

Approximately 100 ALFALFA sources were identified as UCHVCs, and we selected a sample of 
roughly 60
sources from that original list that were best suited for optical follow-up (e.g., strong HI detections, relatively high column densities, and/or potentially interesting for other reasons, such as sky position or velocity); see J19 for more details about how the optical follow-up sample was selected.  As mentioned in the Introduction,
23 of these objects were analyzed in the first phase of the project and presented in J19.  We observed 35 more objects after the pODI instrument 
was replaced with ODI.
The substantially larger FOV of ODI (see Section~\ref{sec:data reduction}) necessitated modifications to our processing and detection steps, so it made sense to split the project into two phases in this way. 

The final sample of objects being presented here includes 25 objects that were observed with ODI and for which we were able to obtain a complete set of observations, i.e., high-quality, deep imaging in both filters ($g$ and $i$). 
In addition, we include one more UCHVC (AGC~227977) that
already had existing pODI imaging data, but that we had planned to re-observe with ODI under better sky conditions.  We were not able to acquire a complete set of high-quality ODI images for this object, but in the end we deemed the existing pODI images to be of sufficient quality 
that we could include the object in the current sample.
Nine more UCHVCs were observed, 
but the imaging data 
could not be fully analyzed for a range of reasons (e.g., observed under poor sky conditions, good-quality images obtained in only one filter);
the objects are mentioned here for completeness.

The ALFALFA-derived properties of the 26 UCHVCs we analyzed for the second phase of the project are presented in Table~\ref{table: uchvcs sample}.  The table includes the source designation, sky position, \hi\ line flux ($S21$), recessional velocity ($cz$), \hi\ line FWHM ($W_{50}$), mean angular diameter of the \hi\ detection ($a$), mean \hi\ column density (log~${\rm N_{HI}}$), \hi\ mass at an assumed distance of 1 Mpc, the dates of the observations, and the detection status from our analysis.

Information about the nine other UCHVCs that could not be analyzed as part of this phase of the project
are presented in Table~\ref{table: insufficient}.
The columns are the same as those shown in Table~\ref{table: uchvcs sample} except that 
no detection status is given since a full analysis could not be done.  We include information about the observations we were able to acquire when applicable.  The objects that fall into this category typically had poor-quality data and/or imaging in only one filter.  

Table~\ref{table: uchvcs sample} shows that the objects we analyzed possess a fairly broad range of \hi\ properties. 
The objects in our sample have \hi\ flux values that 
range from 0.49 to 7.95 Jy~\kms, with most sources between 0.6 to 2.0 Jy~\kms. 
In general, we prioritized \hi\ sources with higher column densities 
during the data acquisition phase of the project, 
under the assumption that these targets would
be more likely to host a dwarf galaxy;  
one (albeit indirect) consequence of this is that the sources we were unable to analyze because of insufficient data (Table~\ref{table: insufficient})
tend to have smaller total HI fluxes than the objects in the sample shown in Table~\ref{table: uchvcs sample}. The \hi\ major axes of our objects range from 5.3$\arcmin$ to 17.2$\arcmin$ (approximately
1.5 to 5.0 kpc at a distance of 1 Mpc) and they span a recessional velocity range from $-$452 to
$+$320 \kms. 

\section{Observations \& Initial Data Reduction}
\label{sec:data reduction}

We obtained optical imaging of 
the UCHVC sources with the WIYN 3.5-m telescope at Kitt Peak National Observatory over the course of several observing seasons; the year and semester of the observations of each source are given in Table~\ref{table: uchvcs sample}. Observations carried out prior to 2015 were taken with pODI, which had a central imaging area 
that consisted of a 3 x 3 array of orthogonal transfer arrays (OTAs).  Each individual OTA is comprised of an 8 x 8 grid of CCD detectors.  The pODI configuration provided a 24\arcmin\ by 24\arcmin\ FOV and a pixel scale of 0.11\arcsec~pixel$^{-1}$. The upgraded ODI camera 
includes
a 5 x 6 arrangement of OTAs, yielding a 40\arcmin\ x 48\arcmin\ FOV and the same pixel scale as pODI. 

Each of the UCHVCs was imaged with a series of nine 300-second exposures in both the SDSS \textit{g} and \textit{i} filters, and the telescope was dithered slightly between exposures in order to fill
in the gaps between the CCD detectors and the OTAs; this yielded a
total integration time of 45 minutes in each filter. The images were transferred to the One Degree Imager Pipeline,
Portal, and Archive (ODI-PPA) system\footnote{The ODI Portal,
  Pipeline, and Archive (ODI-PPA) system is a joint development
  project of the WIYN Consortium, Inc., in partnership with Indiana
  University's Pervasive Technology Institute (PTI) and NSF’s
  NOIRLab.} \citep{gopu14a,young14a} for storage and processing. The QuickReduce
data reduction pipeline \citep{kotulla14a} was used to carry out pixel
masking, crosstalk and persistence correction, overscan subtraction,
bias and dark subtraction, flat-field correction, pupil ghost
correction, and cosmic ray removal on the images. 

After the
QuickReduce processing, the images were further reduced and stacked using the odi-tools software routines, a suite of Python routines developed by 
William Janesh and Owen Boberg to help facilitate the analysis of the J19 UCHVCs sample and other 
pODI and ODI data \citep{janesh18a}. The images were illumination-corrected and reprojected to the same pixel scale and coordinate reference frame. Sets of images taken in the same
filter were then 
scaled to a common flux level and 
combined into a single science-ready image.
The combined images in the two
filters were aligned and then trimmed to 20000 pixels by 22000 pixels ($\sim$37\arcmin\ x 40\arcmin) to cut out the regions without full coverage in the dither pattern. 
The mean full-width at half-maximum of the point-spread function (FWHM PSF)
in the final combined images was typically \lapp
1.0\arcsec\
in both filters.  In cases where we observed an object on multiple nights with varying sky conditions, we used the images with the best seeing values to construct the final combined 
image in a given filter.  

All of the UCHVC targets fall within the footprint of SDSS, so we 
measured the magnitudes and colors of 
several hundred
SDSS stars that appeared within each of the ODI frames in order to derive photometric calibration
coefficients (zero points and color terms) that we could then use to
calculate calibrated magnitudes and colors for the other sources in the final combined images.
Typical uncertainties on the photometric zero points in the calibrated $g$
and $i$ magnitude equations are $\sim$0.02 mag. 

\section{Searching for Stellar Populations Associated with the HI~Clouds}
\label{sec:search method}

\subsection{Applying a Color-Magnitude Diagram Filter to the Stellar Catalogs}
\label{sec:cmd filter}

When we analyze our optical images of the UCHVCs, 
the goal is to search for
collections of stars -- which we refer to as ``stellar overdensities''
-- with properties like those of the
stellar populations one would expect to be associated with a Local Group dwarf
galaxy.  We have developed a largely automated procedure to carry out this
search as well as to characterize the statistical significance of any stellar
overdensities we find. The overall method we use was laid out in
\citet{janesh15a}; as mentioned earlier, we have since made some improvements to the method, which are described in this section.  Additional details can be found in \citet{smith22a}.

We use a processing pipeline that calls Python routines and IRAF tasks and includes both automated steps and steps that require input from the user.  The main steps in the process are as follows. 
First, we carry out
detection and aperture photometry of the sources in the WIYN images that are
above a certain count threshold -- typically, 3.5 to 4 times the
standard deviation of the sky background level in the images.  The
source lists generated from the two images are then matched, and
extended objects are eliminated 
based on the difference in magnitudes measured in two different apertures. 
We measure the instrumental magnitude of each source using a small aperture with a radius equal to the average FWHM of the image (magnitude $m1$) and a larger aperture with a radius equal to twice that value (magnitude $m2$). 
Bright point sources fall along a clearly-defined sequence when plotted in the $m2$ vs. $m1-m2$ plane; we use that sequence to define the center of our selection region and then select sources within 2.5$\sigma$ of that center.  The selection region is thus wider at fainter magnitudes where the photometric uncertainties  are larger; see \citet{janesh18a} and \citet{smith22a} for more details.
Aperture photometry of the 
point sources that survive the extended source cut is then carried out in both
images, using an aperture radius equal to the average FWHM PSF of the
image.  An aperture correction is computed from photometric measurements of
selected bright stars 
across the image
and then applied to the instrumental magnitudes of all of the point
sources. Final calibrated magnitudes and colors are calculated for each point
source in the field by applying the appropriate photometric calibration coefficients
along with Galactic extinction corrections calculated from the \citet{schlegel98a} dust maps and the relations in \citet{schlafly11a}. Typical 5-$\sigma$ detection limits in the images are $g$$\sim$25.4 mag and $i$$\sim$24.4 mag. 

\begin{figure}[h!]
    \plotone{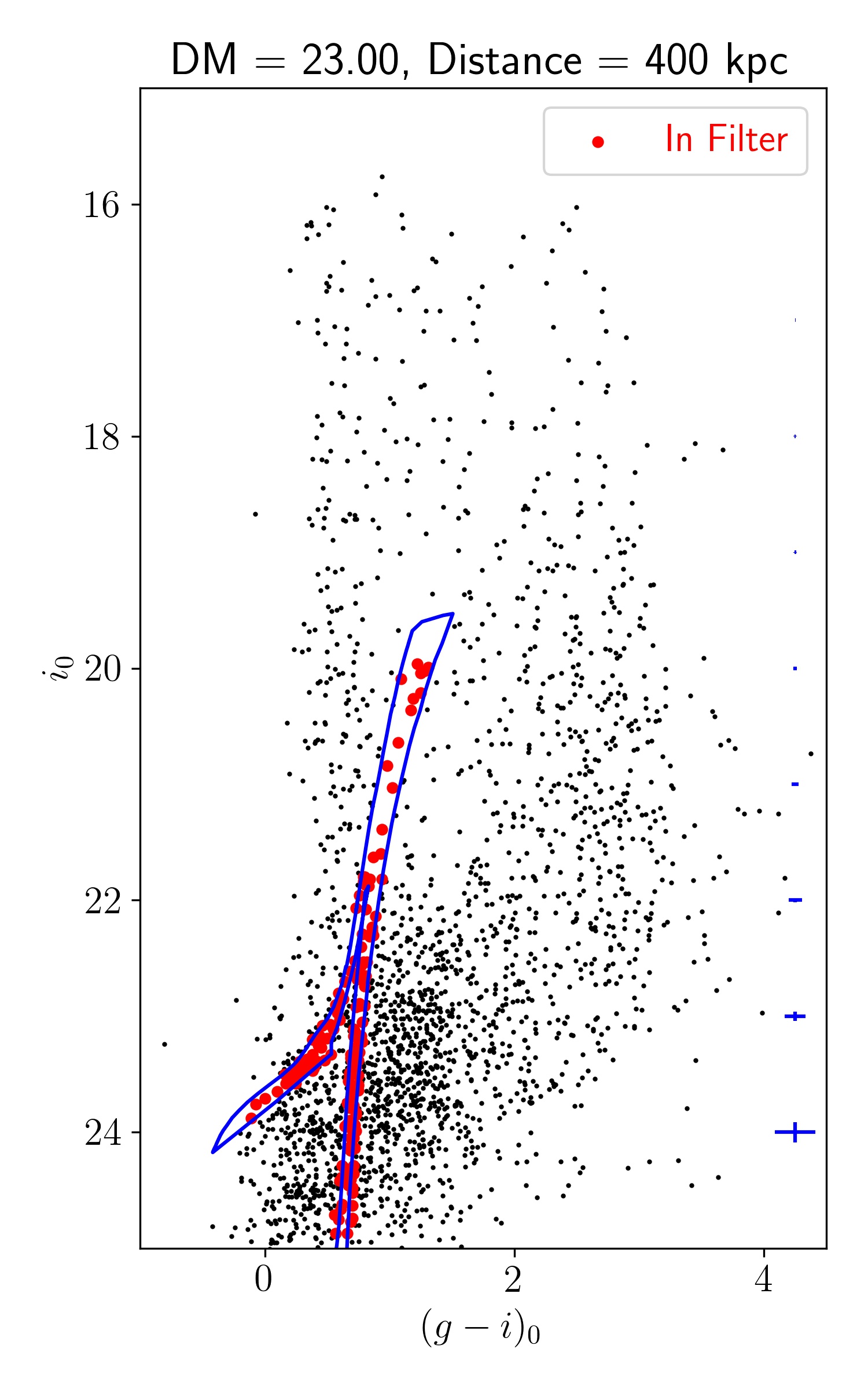}
    \caption{An example of the CMD filter that is applied to the photometric catalog derived from the WIYN ODI imaging of each UCHVC field in order to select stars with properties expected for a dwarf galaxy at a given distance (see Sec.~\ref{sec:cmd filter}). Black points are point sources located within the images and red points are those that coincide with the location of the filter given their $g$ and $i$ magnitudes and associated 1-$\sigma$ uncertainties. Error bars showing the typical  photometric uncertainties at various magnitudes are plotted in blue on the right-hand side. The filter shown in this example
    corresponds to a distance of 400~kpc.} 
    \label{fig:cmd filter}
\end{figure}

Next we construct a color-magnitude diagram (CMD) filter that can be applied to the final photometry catalog for a given field.
The construction of the filter was described in detail in \citet{janesh15a} and
followed the basic method outlined in \citet{walsh09a} using stellar isochrones from \citet{girardi04a}. Because UCHVCs are expected to host old, metal-poor stars, the CMD filter is designed to select stars with ages between 8 and 14 Gyr and metallicities between Z = 0.0001 and 0.0004.  
In \citet{janesh15a}, we used our WIYN observations of Leo~P to test the CMD filtering method and demonstrated that the filter and our detection methods are effective at finding a stellar population that consists of only a modest population of red giant branch (RGB) stars.
The position and extent of the filter on the CMD is defined by the locations of the isochrones for the chosen age and metallicity ranges, as well as by the distance to the putative stellar population.  We shift the filter in the vertical (magnitude) direction on the CMD over a range of magnitudes in increments of 0.01~mag, to cover a corresponding distance range of 250~kpc to 2.5~Mpc. Applying the CMD filter at each distance step to our photometric catalog yields a subset of stars in a given UCHVC field that, given their $g$ and $i$ photometric measurements and associated uncertainties, lie in a region of the CMD that is consistent with the location and boundaries of the filter.  
Figure~\ref{fig:cmd filter} shows an example of a CMD filter being applied to our data; the black dots show the CMD positions of the point sources that appear in
one of the UCHVC fields, the blue solid line marks the location of the CMD filter corresponding to a distance of 400 kpc and the age and metallicity ranges mentioned above, and the red dots are the sources 
that are selected by the CMD filter in this case. 

\begin{figure}[h!]
    \plotone{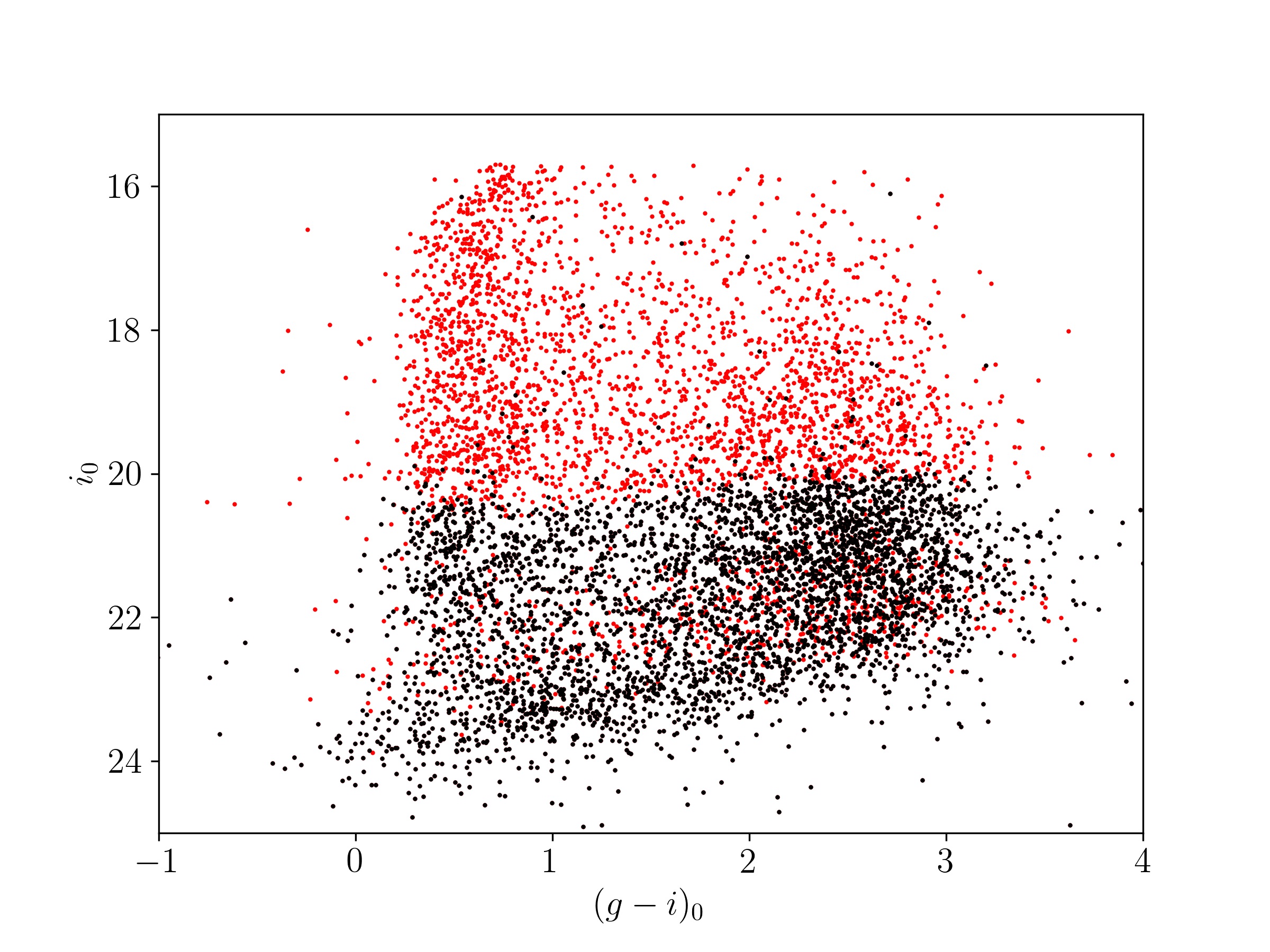}
    \caption{An example of the effect of the cross-matching step that is carried out on source catalogs of fields in which potential stellar overdensities have been identified.  Sources in the field are cross-matched with Gaia and SDSS catalogs in order to identify and remove contaminating foreground objects (Galactic stars) and background objects (galaxies and AGN) from our star lists.  Sources that are retained 
    are shown with black points and sources that are removed 
    are shown with red points.   In this example field, approximately one-third
    of the objects in our original source list were removed as likely Galactic stars and 
    $\sim$10\%
    were removed as likely background galaxies or AGN; these percentages are fairly typical, although the exact proportions of foreground and background objects and remaining point sources depend on quantities like the Galactic latitude and exact detection limits of each set of images.}
    \label{fig:cross-match}
\end{figure}

\subsection{Searching for Overdensities in the Spatial Distribution of CMD-Filtered Stars}
\label{sec:smoothing}

Once we have identified the set 
of stars 
selected by a given CMD filter,
we search for clustering among the stars to identify any stellar populations that might be associated with the \hi\ source. We begin by smoothing the spatial distribution of the stars with a Gaussian kernel.  We use smoothing kernels with a diameter
of both 2$\arcmin$ and 3$\arcmin$ in order to mimic the range of typical angular sizes that a dwarf galaxy might have if it were located within the distance range being probed.  
We then estimate the density of stars at each location in the smoothed image as a function of the mean density of the CMD-filtered stars across the image.  Any region of the image where the stellar density exceeds the mean value represents a potential stellar overdensity associated with the \hi\ source.  

In order to determine the significance of a given overdensity (i.e., how likely it is to be a genuine overdensity rather than a result of random fluctuations in the point source spatial distribution), we carry out a series of 10,000 Monte Carlo experiments corresponding to each CMD filter position.  We distribute the same number of CMD-filtered stars at random locations across the field and repeat the smoothing steps.  Any overdensity identified in the real data can then be compared to those found in the randomly-generated data.  The significance of the overdensity in the real data is quantified by calculating the number of overdensities in the random realizations that are less dense than the real overdensity. 
After carrying out these steps for every distance and for both smoothing radius sizes, 
we can compare the overall results across the entire parameter space that has been searched 
and look for the most significant detections that appear within a given UCHVC field.  

\subsection{Assessing the Detected Stellar Overdensities and Cross-Matching with the Gaia and SDSS Catalogs}
\label{sec:assessing}

The next stage of the analysis includes a variety of steps that are applied 
as appropriate, depending on the results of the CMD filtering and smoothing process. 
The first step in all cases is to examine the location and significance of the overdensities revealed by the CMD filtering process at every step of the distance range being sampled (250~kpc to 2.5~Mpc).
We focus on detections that have high significance 
(approximately 90\% or higher) 
and that are close in projection to the \hi\ centroid position -- usually within 8$\arcmin$, which translates to a separation of 2.3~kpc at a distance of 1~Mpc. We sometimes consider
overdensities that have significance in the 80$-$90\% range, or are up to 10$\arcmin$ away from the \hi\ centroid, 
if their other characteristics suggest they could be a genuine dwarf galaxy candidate. 
If no overdensities are found that meet these criteria, we classify the UCHVC target as a non-detection and do not carry out any further analysis on the imaging data for that UCHVC. If multiple unique detections (i.e., at different sky positions and/or distances) are found, each is assessed with the next steps. We note also that the steps described here are carried out on the results produced using both of the smoothing kernels (i.e., smoothing kernels with diameters of 2$\arcmin$ and 3$\arcmin$), and we take the results derived from both kernel sizes into account when selecting and classifying the various overdensities that are identified. 

Once we have identified possible stellar overdensities associated with the \hi\ sources, we examine the properties of the individual stars that 
are within 3$\arcmin$ of the center of the detected overdensity.
We look at the CMD locations of the stars that compose
the detection to judge whether the stars match the expected morphology of the RGB
of a genuine dwarf galaxy at the specific distance to which the filter location corresponds.  We generally look for a detected overdensity that has stars that populate a fair fraction of the 
RGB region of the CMD filter, as well as stars that lie in the Horizontal Branch (HB) region when applicable.  We also compare the numbers and positions of the stars in the detected overdensity to those of the stars in a reference circle of similar size 
that is placed at a random location near the edges of the image, far from the \hi\ centroid position (to ensure that it does not overlap with any detection circles).  If the CMD of the stars in the detection circle is similar in appearance and number of stars to the CMD of the reference circle, then the detection is deemed less convincing.  

We also inspect how closely the stars within the detection CMD are clustered on the image.  In some cases, the stars that comprise a given detected overdensity may not actually be very close to each other spatially (e.g., the detection might consist of 
several stars that are scattered around the edges of a particular smoothed region that makes up the detected overdensity), which makes the detection less convincing as a genuine dwarf galaxy candidate. 
In addition, we examine the radial profiles of the individual stars within the detection, and the objects in the region around the detection area, because in some cases (for example) clusters of faint, distant unresolved background galaxies can masquerade as collections of nearby stars with colors like RGB stars. To help with this step, we also perform a quick cross-reference of the stars within the over-density with source catalogs from the SDSS survey, to see whether the sources in and around the detection are actually galaxies. 
Finding a detected overdensity that is close (in projection) to an obvious background galaxy group or cluster suggests that the detection may actually be made up of unresolved background galaxies rather than being a genuine stellar association in or near the Local Group. 

After these steps are completed, the UCHVC targets with detected overdensities are assigned into separate categories based on our initial assessment 
of the likelihood 
that we may have detected the presence of a genuine dwarf galaxy.
UCHVCs with 
overdensities that no longer seem convincing after these checks are designated as "non-detections" (see Table~\ref{table: uchvcs sample}) and are not analyzed further.  

For the UCHVC targets 
that remain, we cross-match the point source catalogs with the Gaia\footnote{This work has made use of data from the European Space Agency (ESA) mission Gaia (https://www.cosmos.esa.int/gaia), processed by the Gaia Data Processing and Analysis Consortium (DPAC, https://www.cosmos.esa.int/web/gaia/dpac/consortium). Funding for the DPAC has
been provided by national institutions, in particular the institutions participating in the Gaia Multilateral Agreement.} EDR3 catalog \citep{gaia16,gaia21}  and remove foreground stars (by rejecting any source with a proper motion greater than three times the uncertainty on that quantity) as well as known AGN \citep[by removing sources that match those in the list of AGN used to calibrate the Gaia EDR3 celestial reference frame;][]{gaia21}.
We also make use of the star/galaxy classification 
in the SDSS catalog
to remove any sources that are likely to be galaxies.
Before implementing the latter step, we examined how reliable the SDSS star/galaxy classification was by comparing it to the classification used by the
Cosmic Assembly Near-infrared Deep Extralactic Legacy Survey\footnote{This work is based on observations taken by the CANDELS Multi-Cycle Treasury Program
with the NASA/ESA HST, which is operated by the Association of Universities for Research in Astronomy, Inc., under NASA contract NAS5-26555.} \citep[CANDELS;][]{grogin11a,koekemoer11a,stefanon17a}.
CANDELS was carried out with the Hubble Space Telescope (HST) and therefore genuine point sources can be distinguished from background galaxies more effectively in CANDELS data than in data from a ground-based survey.  
We found that for sources in common between SDSS and CANDELS, the CANDELS star/galaxy classification matched the SDSS classification $>$94\% of the time.  
Since mis-identification occurred for only a small fraction of sources, and the number of objects in our source lists that are identified as galaxies in SDSS is also relatively small, we decided that applying the SDSS classification to our source lists was warranted because it would help us weed out 
spurious dwarf galaxy candidate detections.
Figure~\ref{fig:cross-match} shows that the majority of sources that are eliminated when we cross-match our point source catalogs with Gaia EDR3 and SDSS are bright (with $i$ $<$ about 20 mag) and are excluded because they are Galactic foreground stars.

After removing the contaminants found in the catalog cross-matching, 
we repeat the CMD filtering and smoothing steps, 
again assess each over-density that is detected in a given UCHVC field, and update the initial classification that was assigned.  
The final results of the search process are sorted into three categories: "non-detections" (UCHVCs that have no convincing evidence for an optical counterpart);  "possible detections" (UCHVCs that may have an optical counterpart that warrants further follow-up), and "best detections" (UCHVCs that have a convincing or likely optical counterpart).
%

For objects that fall within the latter two categories, we estimate the \hi\ and optical properties for the UCHVC and its counterpart at the relevant distance.
We take the distance of the object to be the CMD filter distance that corresponds to the most significant detection that occurs for a given overdensity.  We also tabulate the range of distances over which a given detection remains above some threshold for significance -- 90\% for overdensities that have a peak significance $\geq$90\%, and 80\% for overdensities with a peak significance between 80$-$90\% -- and use this to help define an uncertainty on the estimated distance (see Sections~\ref{sec:results} and \ref{sec:discussion}).


\subsection{
Quantifying the Detection Completeness}
\label{sec:completeness}

As part of the analysis, we carry out 
artificial star
tests 
to quantify the photometric depth and detection limits of each set of UCHVC images. 
Before running the full set of tests,
we confirmed that completeness testing performed on a 
4000x4000-pixel subsection of the images yields the same results as 
tests performed on the full-size ODI images. Since source crowding is not an issue in our UCHVC target fields (they are empty of strongly clustered sources, with the exception of a field that includes the globular cluster Pal~3; see Sec.~\ref{sec: pal3} for more discussion),
we found that running artificial star tests on portions of the images is a valid approach.
The rest of the steps were carried out on these smaller-sized images. 

We characterized the point spread function (PSF) within each of the images
by fitting the light distributions of several dozen
bright but unsaturated stars within the frames.  We then generated artificial stars and injected them into the images, 400 stars at a time in steps of 0.2~mag, until we had spanned the range of stellar magnitudes present in the real data.
We followed our detection procedures to recover and measure the magnitudes of the artificial stars and used that information to create a completeness curve (which quantifies the proportion of stars recovered as a function of magnitude) for that image and filter.  We created 50\% completeness curves for the CMD of a given UCHVC field by combining the completeness information in each of the filters.  
Specifically, we constructed a grid of $i$ and $g-i$ values and then computed the completeness at a given position in the grid by multiplying the $i$ magnitude completeness by the appropriate completeness in the $g$ filter based on the color at that grid position; for example, for the point at $i$ $=$ 25 mag and $g-i$ $=$ 1, we multiplied the detection completeness in $i$ at 25 mag by the completeness in $g$ at 26 mag. We carried out this calculation over the entire range of source magnitudes and colors that appear in our CMDs (i.e., $i$ $\geq$ 26 mag, $g-i$ from $-$1.5 to $+$4).  We then interpolated between the grid values to define the 50\% completeness curve for a given UCHVC field.
These 50\% completeness curves are included in the results presented in Section~\ref{sec:results}. 

\subsection{Changes to the Original Detection Methods}
\label{sec:pipeline improvements}

Before analyzing the WIYN ODI data set,
we examined each of the detection steps carried out on the J19 
sample to explore whether modifying the steps might increase our chances of detecting dwarf galaxy candidates associated with the UCHVCs. 
We made two substantial modifications to our process.

The first was a change in the 
CMD filtering step.
As described in Sec.~\ref{sec:cmd filter}, the boundaries of the CMD filter are defined by the
set of isochrones that characterizes 
the expected stellar population of a UCHVC.  When we apply the filter, we select sources that have $i$ magnitudes and $g-i$ colors that would fall within the filter given the relevant photometric uncertainties.
When we examined this step in the detection process, we 
discovered 
a numerical error in the original code that was applied to the J19 WIYN pODI data 
which made the CMD filter less restrictive than intended.  The effect of the error was more pronounced at faint magnitudes, where the photometric errors are large, but 
%
%
the impact was
mitigated because the 
code also included a photometric error cut that removed sources with 
$i$ magnitude errors $\geq$0.2 mag or $g-i$ color errors $\geq$0.28 mag.
Nevertheless, the end result of this coding error
was that the pODI data set yielded a relatively large number of detections 
that were later rejected as false when they were examined individually. 

After correcting this 
error and then 
testing various options for how to apply the CMD filter to the ODI imaging data, we decided to 
implement a relatively restrictive source selection, allowing only those objects
that are consistent with the boundaries of the CMD filter given the 1-$\sigma$ photometric errors 
of the object. We also implemented a change so that the code uses the uncertainties in the $g$ magnitude and $i$ magnitude (instead of the uncertainties in the $i$ magnitude and $g-i$ color, as was the case in the original code) to determine whether a source should be accepted. More specifically, we imposed the condition that sources were selected only if the $g$ magnitude and $i$ magnitude errors were both $\leq$0.2~mag.  This prevented sources with, for example, small $i$ uncertainties but larger $g$ uncertainties from passing the filtering step. 
The final result is 
a more restrictive CMD filter than the filter originally applied to the J19 pODI data set; the revised filtering process only 
selects stars with reliable 
%
photometry in both filters that match closely with the metallicity and age range of the expected stellar population, which reduces the number of spurious detections appearing in each of the UCHVC fields.  

The second major change made to the detection process was to add the cross-matching between our sources and objects in the Gaia and SDSS catalogs that was described in Section~\ref{sec:assessing}. This added step serves to much more effectively remove both foreground and background contaminants from each of the UCHVC fields (see Fig.~\ref{fig:cross-match}), which was 
especially helpful for processing 
the ODI data set, given the large field of view and the large numbers of sources detected in each field. 
 
We used the improved series of steps to process all of the objects in the ODI sample listed in Table~\ref{table: uchvcs sample}.  We also decided to reanalyze the WIYN pODI imaging data for the five UCHVCs in the J19 sample that yielded
possible dwarf galaxy candidates.
Results from the current sample of UCHVCs, as well as the re-analysis of the five pODI objects from J19, are presented in the next section. 

\section{Results}
\label{sec:results}

\subsection{Results for the Sample of UCHVCs Analyzed for This Work} 
\label{sec:results odi}

We carried out the above-described processing and analysis steps for the 26 objects in the sample listed in Table~\ref{table: uchvcs sample}. Twenty of the UCHVCs in this sample had no stellar overdensity in the images that satisfied our criteria for a dwarf galaxy candidate, and are thus classified as "Non-Detections" and labeled "ND" in the table.  The first 19 non-detections are discussed in Sec.~\ref{sec: non-detections}, and one object in this category warrants a separate discussion in Sec.~\ref{sec: pal3}. Five objects classified as "Possible Detections" are described in Sec.~\ref{sec: possible detections}, and the one object classified as a "Best Detection" is discussed in Sec.~\ref{sec: best detection}.

\subsubsection{Objects Classified as "Non-Detections"}
\label{sec: non-detections}

For most of the objects in this category, the classification was straightforward because no stellar overdensities were found that had significance in the $\sim$80$-$90\% range or higher and were near
the \hi\ centroid. 
In a few other cases, significant overdensities were identified initially, but then not found to be believable upon close inspection of the individual sources located within the overdensity and in the surrounding region.  
One object 
(AGC~335755)
made it through the first round of screening 
to the final stages of analysis -- i.e., the cross-matching with the Gaia and SDSS catalogs -- but 
was no longer deemed convincing after the contaminating foreground and background objects were removed from the source list.  
In any case, these 19 UCHVCs show no evidence that they host a dwarf galaxy candidate within the distance range that we tested (250 kpc to 2.5 Mpc), given the detection limits of our images.  
We note that other searches for optical counterparts to compact \hi\ clouds have detected dwarf galaxies at larger distances \citep[$\sim$3$-$20~Mpc;][]{bellazzini15a, bellazzini15b, sand15a, tollerud15a}, so 
there is still the possibility that these UCHVCs are associated with a stellar counterpart 
that is farther away
than 2.5 Mpc and therefore beyond the scope of our survey. 
On the other hand, our images of the UCHVCs in this category contain no faint optical counterparts (either resolved-star counterparts or unresolved faint optical emission) like the 
ones detected in those studies. 

\subsubsection{An \hi\ Cloud Near in a Globular Cluster?}
\label{sec: pal3}

One more UCHVC
in the 
"Non Detection" category
requires a more detailed discussion.
AGC~501816 was included in the ODI sample because it lacked an obvious optical counterpart and its other properties satisfy the criteria for UCHVCs, except that its $|v_{LSR}|$ value, 99~\kms, is below the nominal 120~\kms\ minimum requirement. In addition to being of interest as a possible compact \hi\ cloud, 
AGC~501816 is notable because of its proximity to the Milky Way globular cluster (GC) Pal~3. The centroid of the \hi\ source 
lies 7.3\arcmin\ away in projection from Pal~3 and the GC appears within our ODI images. This projected angular separation translates to only 195~pc at the distance of Pal~3 \citep[91.9 kpc;][]{hilker06a}. Moreover, the GC and the UCHVC have similar heliocentric velocities: Pal~3 has a $v_{\rm helio}$ of 94.0$\pm$0.8~\kms, only 13~\kms\ below that of AGC~501816 (Table~\ref{table: uchvcs sample}).

\begin{figure*}[h!]
\centering
    \includegraphics[width=0.6\textwidth]{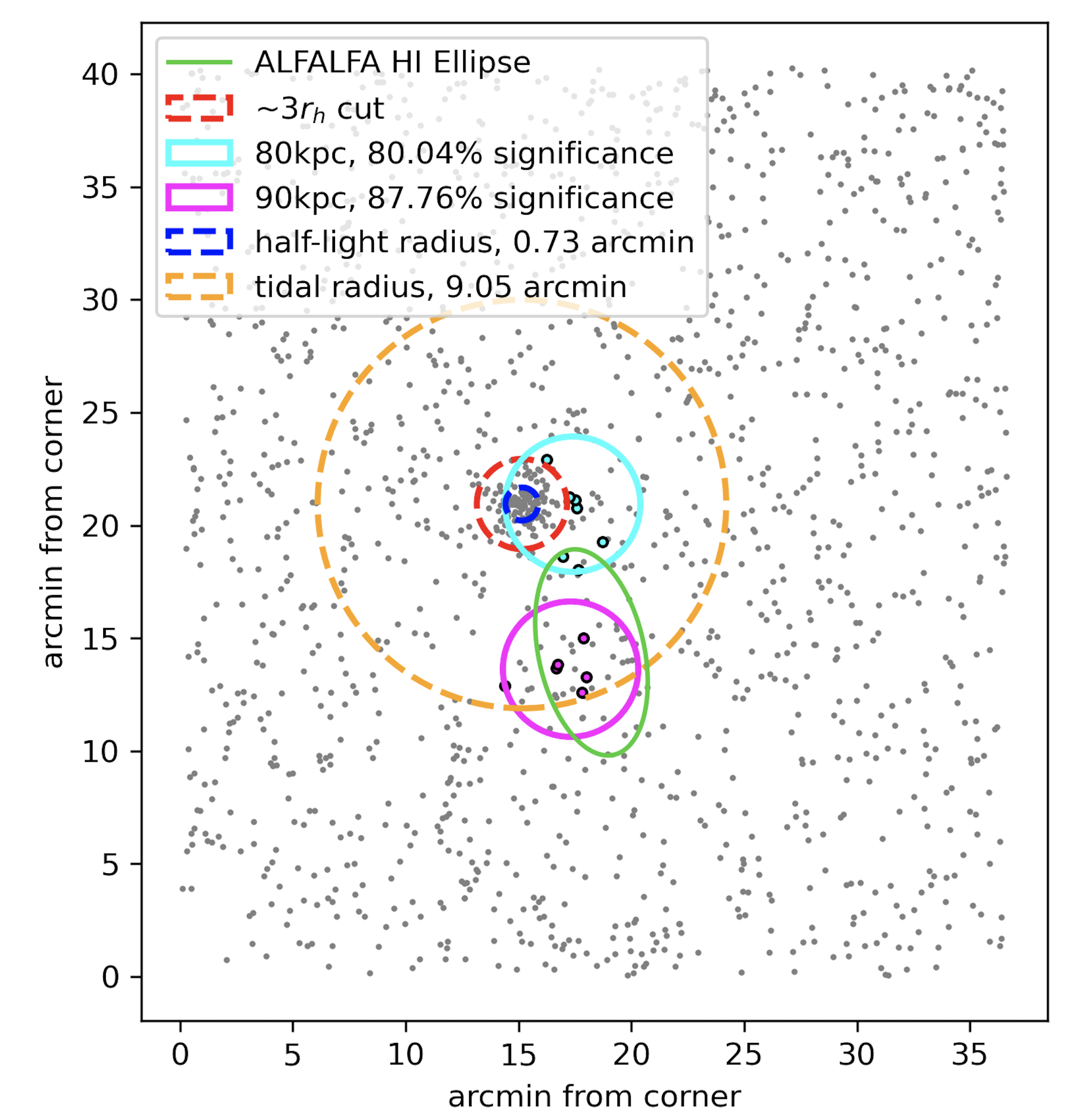}
    \includegraphics[width=0.5\textwidth]{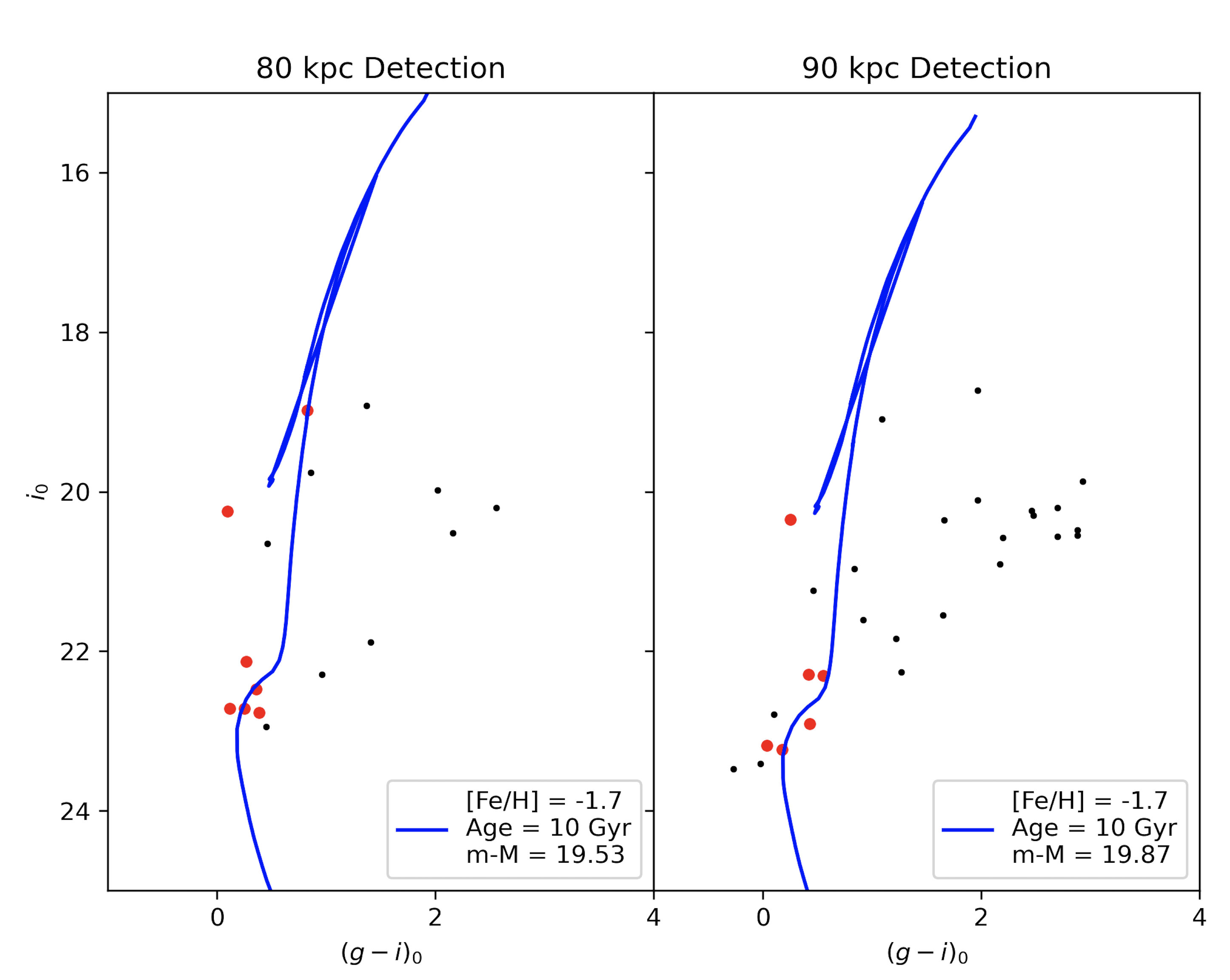}
    \caption{
    Top: Diagram showing the positions of sources across the field-of-view of our ODI images of 
    AGC~501816,  which includes the 
    Galactic GC Pal~3.  Gray points mark the locations of objects
    in the final stellar catalog (after cross-matching with the Gaia EDR3 and SDSS catalogs and removing contaminants).  
    The ALFALFA \hi\ source is shown with a solid green ellipse.  Pal~3 
    is the 
    cluster of point sources $\sim$7 arcminutes away from the \hi\ centroid.  The half-light radius and tidal radius of the GC from \citet{baumgardt18a} are marked, along with
    the 3~$r_h$ circle used to remove Pal~3 stars before we searched for stellar overdensities associated with the \hi.  
    Two detected overdensities are marked with cyan and purple circles.
    The stellar overdensities and the \hi\ source are either inside or overlapping with the tidal radius of Pal~3.
    Bottom: The CMDs for the two 
   stellar overdensities detected in the AGC~501816 field after all stars within 3~$r_h$ of Pal~3's center were removed.  Gray points are all sources inside a 3\arcmin\ radius around the peak of the overdensity and red points are sources within that radius that fall within the CMD filter at the given distance. 
   For reference, we have over-plotted an isochrone from \citet{girardi04a} (solid blue line) with the age and metallicity of Pal~3 (10~Gyr, [Fe/H] $=$ $-$1.7 or Z = 0.0004; \citet{hilker06a}), shifted to match the distance at which each overdensity is detected (80~kpc or 90~kpc).   }
\label{fig:Pal3 field}
\end{figure*}

\begin{figure*}[h!]
    \plotone{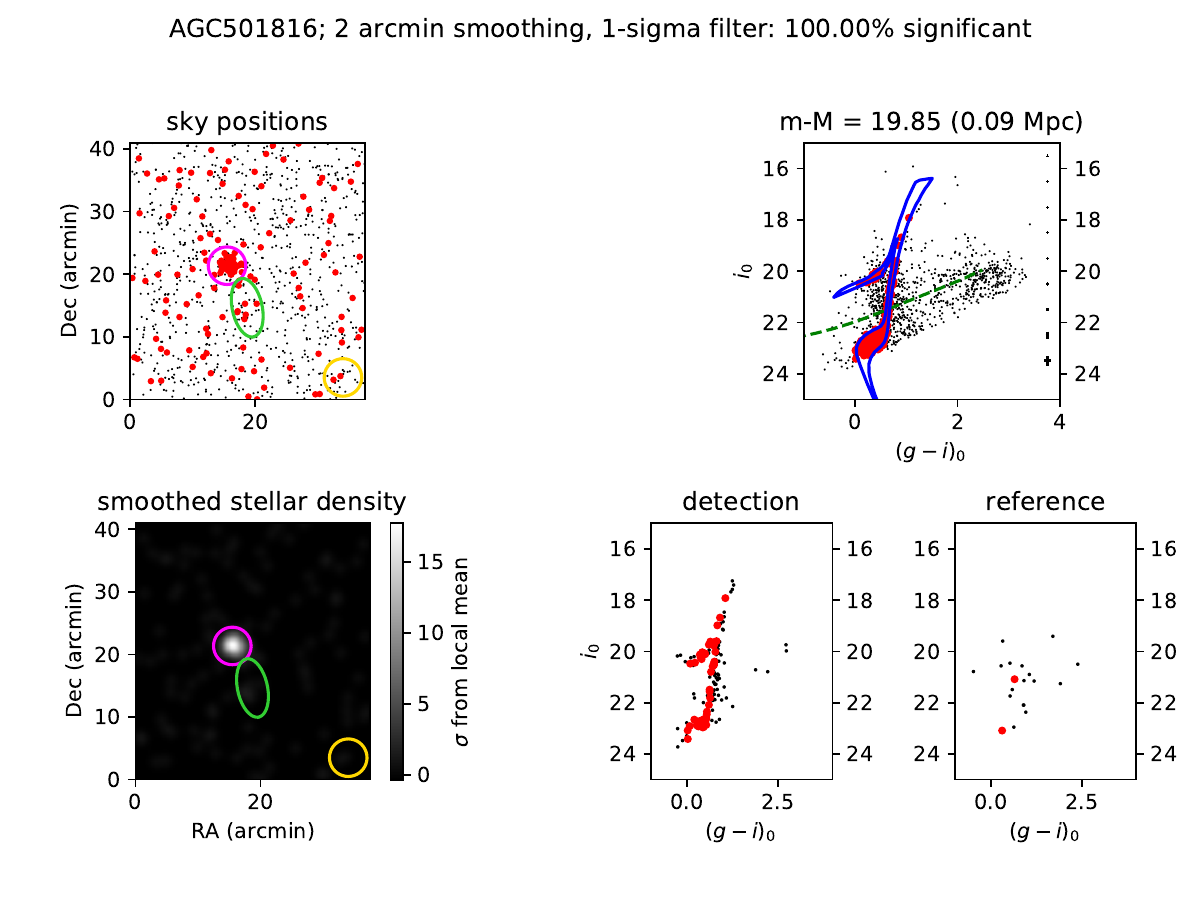}
    \caption{
The UCHVC AGC~501816 is located on the sky roughly 7$\arcmin$ from the outer Galactic halo globular cluster Pal~3.  We ran the detection pipeline on the images of the AGC~501816 field with Pal~3 present and with Pal~3 removed; this set of plots shows some of the results from the searches carried out with the GC present in the images.  The upper left plot shows the locations of the stars detected in the field (gray points), the stars selected by the CMD filter (red points), and the \hi\ ellipse for the UCHVC from ALFALFA (green ellipse).  The region of the detected overdensity is marked with a magenta circle of radius 3\arcmin.  A yellow circle of the same size is placed at a random location in the outskirts of the field and used to generate a comparison CMD; the sources within the detection and refernce circle are compared as part of the assessment process described in Sec.~\ref{sec:assessing}.  The plot on the lower left shows the smoothed surface density map of the CMD-selected stars, with the same magenta and yellow circles and the green ellipse as are shown in the upper left plot. 
The CMD in the upper right shows all the sources in the field (black points), the CMD filter (blue solid line) and the CMD-selected stars (red points).  The CMDs for objects located within the the detection circle and comparison (reference) circle appear on the lower right. The detection pipeline found Pal~3 at 100\% significance at a distance range of 86$-$106~kpc, which brackets the actual distance of the cluster.  It was also found at a distance of $\sim$160~kpc because of a combination of factors, including the shape of the CMD filter; see Sec.~\ref{sec: pal3} for a full discussion.}
\label{fig:four plot pal3}
\end{figure*}

Pal~3 is located in the outer stellar halo of the Milky Way
and is one of only five Galactic GCs
with galactocentric distances of $\sim$90~kpc or greater
(\citealp{harris96a}, 2010 edition).
It is fairly faint for a GC, with $M_V$ $=$
$-$5.7 (\citealp{harris96a}, 2010 edition), has a modest mass (1.9 x 10$^4$ \msun; \citet{baumgardt18a}), 
and is markedly extended in size, with a half-light radius ($r_h$) of
$\sim$0.7\arcmin \citep{baumgardt18a}, or $\sim$19~pc at the 91.9~kpc distance.  This is several times
larger than the median half-light radii of the GCs in both the Milky
Way and Andromeda, which is $\sim$2$-$3~pc \citep{vandenbergh10a}.
CMD fitting 
indicates that the cluster belongs to
the metal-poor subpopulation of the Milky Way GC system, with [Fe/H]
$=$ $-$1.7, and has an estimated age of $\sim$10~Gyr \citep{hilker06a}.
The location and orbital properties of Pal~3 have prompted studies of its
origin, with some concluding
that it may have been accreted into the outer Galactic halo along with its
parent dwarf satellite galaxy \citep{palma02a}, and others suggesting
it could have been ejected from the Phoenix dwarf irregular
galaxy and captured by the Milky Way \citep{sharina18a}.  
However, stellar abundance studies
seem to suggest instead that Pal~3 co-evolved with the rest of the Galactic GC system \citep{koch09a}.

It has long been understood that GCs like Pal~3 and the other members of the Galactic GC system may contain gas and dust that originated in the outer atmospheres of evolved 
stars and was subsequently lost to the interstellar medium
within the cluster (e.g., \citealp{roberts59a,roberts60a}). Much of
the gas is expected to be removed from the GCs through a variety of
mechanisms, including winds from 
stars and stellar remnants, and dynamical pressure
as the cluster orbits the Galaxy (e.g., \citealp{frank76a}, 
\citealp{spergel91a}).
However some gas may remain and be detectable
as, for example, diffuse atomic hydrogen
that emits 21-cm
radiation, or perhaps hotter gas that produces 
UV or X-ray emission.

Many searches have been carried out over the past several decades using a range of approaches to look for hydrogen gas in and around 
Milky Way GCs; in general these studies have had limited success, 
often 
resulting in upper limits or ambiguous detections 
(e.g., \citealp{heiles66a,kerr72a,knapp73a,erkes75a,birkinshaw83a,faulkner91a, freire01a,vanloon06a}, and many others).
One study of particular relevance is 
\citet{vanloon09a}, which was carried out as part of the Galactic Arecibo L-band Feed Array (GALFA) survey. \citet{vanloon09a} observed
four GCs and set 3-$\sigma$ limits of 6$-$51 \msun\ on the amount of gas
in the four clusters, and also identified a compact high-velocity
\hi\ cloud roughly 1 degree away from the outer halo GC Pal~4. They
conclude that the cloud may be physically associated with Pal~4 or may simply be a chance superposition, since the two objects differ by $\sim$40~\kms\ in velocity and the cloud has a relatively large separation ($\sim$2~kpc) and a high
\hi\ mass (3 x 10$^6$ \msun) if it is actually at the same distance as
Pal~4 (109~kpc).

In Figure~\ref{fig:Pal3 field}, we show the location of Pal~3 and the UCHVC AGC~501816 in our ODI images.  The ALFALFA \hi\ ellipse is marked, as are the tidal and half-light radii of the globular cluster.
We analyzed the ODI imaging data of this field in two ways:  with the stars that make up Pal~3 included in the source lists, and with the stars in Pal~3 removed.  
The first approach 
provides a useful check of our methods for finding  metal-poor, old stellar associations in our images, since it allows us to see how effectively the pipeline detects the Pal~3 stars, and whether it finds them at the appropriate distance. The second approach 
allows us to carry out the usual 
search for stellar populations in a wide area
around 
AGC~501816, as we had done with all of the other UCHVCs in the sample.  

We first carried out the detection steps described in Sec.~\ref{sec:search method} with the Pal~3 stars in place.  We used the same CMD filter as before, but we changed the lower limit of the distance range to 25~kpc (down from 250~kpc) in order to include Pal~3's much closer distance in our search process. 
Searching the field with the Pal~3 stars in place yielded multiple detections at the location of Pal~3 with high statistical significance (100\%).  The distances corresponding to this local maximum in significance ranged from 
86$-$106~kpc,
a distance range that encompasses the 
actual measured distance of Pal~3 (91.9~kpc). 
For the detections within this distance range, the stars selected by the CMD filter populated 
the RGB, HB, Main Sequence Turnoff (MSTO) and the upper Main Sequence.  The stellar densities for these detections were more than 15 times the mean density across the field.
  Figure~\ref{fig:four plot pal3} shows a typical example of the detection results with the Pal~3 stars present in the images; in this example, the stellar overdensity is detected at a distance of $\sim$93~kpc.

The pipeline also detected the Pal~3 stars at high significance (100\%) at a distance of $\sim$160~kpc.  
%
A close examination of the results shows that this is caused by a combination of factors:  the distance range being sampled, 
the shape of the lower portion of the CMD filter, and the detection limits of our data.  When we use the CMD filter to search 
for stellar populations at close distances -- i.e., between $\sim$25~kpc and $\sim$150~kpc -- 
the main sequence turnoff (MSTO) portion of the filter is included in the filtering process.
This portion of the filter is wide compared to the other CMD features and allows stars with a 
broader range of magnitudes and colors to be selected.  At nearby distances, the position of the MSTO feature also coincides with regions of the CMD where our photometric errors are larger (i.e., at $i$ magnitudes between $\sim$22$-$24).  The end result is likely that a higher level of contamination exists in the sample of filter-selected stars. The strong detection at 160~kpc occurred because that is the point at which the broad MSTO feature in the filter falls 
just {\it below} the typical detection limits of our data, and therefore only stars that coincide with the narrower RGB and HB portions of the filter are selected.  This lowers the mean surface density of stars selected across the field and allows the bright stars in Pal~3 that intersect other parts of the filter to yield a strong detection with a high surface density relative to the mean density in the field. This type of false detection 
should not occur in any of our other searches, 
because we normally only sample the distance range 250~kpc to 2.5~Mpc, which means that the MSTO feature is well below our detection limits. The bottom line is that the results of the search carried out with the Pal~3 stars in place confirms that our detection method can easily find a genuine old, metal-poor stellar population present in the images and that it yields a strong detection at the correct distance. 
%

We next removed Pal~3 by excluding stars within 3~$r_h$ of the cluster center
in order to search for stars that might be associated with the \hi\ source a few arc~minutes away.  
%
A few overdensities with high significance 
(\gapp90\%) were identified, so we carried out the final set of detection steps -- i.e., cross-matching the CMD-filtered source list with Gaia and SDSS to remove likely contaminants and then repeating the detection process on the more restricted catalog. 
The process yielded two possible stellar overdensities, both of which are marked in the map of the field in Figure~\ref{fig:Pal3 field}.  One detection, with a final significance of 87.76\%,
closely coincides with the \hi\ centroid and is found at a distance of $\sim$90~kpc.
The detection is made up of only six stars
that lie in the MS turnoff portion of the CMD filter, which is the broadest part of the filter and also (at the relevant distance) falls well below the 50\% detection limit of our images.
A second detection has a statistical significance of only 80.04\% and is made up of only seven stars, but is worth mentioning because it is 
located very close to Pal~3, and is well inside the tidal radius of the cluster. Five of the seven stars lie in the MSTO region of the CMD filter and the others coincide with the HB and lower RGB. The distance yielded by the CMD filter for this object is $\sim$80~kpc.

The locations of the two detected overdensities are marked in the diagram of the field presented in Figure~\ref{fig:Pal3 field}.  The CMDs for the sources that make up the overdensities are shown in the lower panels of that figure. For reference, a single isochrone with the age and metallicity of Pal~3 is plotted on each CMD, shifted to the distance at which the stellar overdensity is detected.


The question that arises at this point is what these two modestly significant detections might represent.  One clear possibility is that the detected stellar overdensities are not genuine stellar populations associated with Pal~3 or AGC~501816 but are instead simply chance superpositions of stars that happen to have magnitudes and colors that fall 
within the CMD filter at the given distances.
Another option is that we have detected stars that are somehow associated with Pal~3, with the \hi\ source, or with both objects.  The latter might be possible if, for example, the \hi\ and stars originated in Pal~3
and then were subsequently removed somehow from the cluster, or the gas was stripped first and then stars formed within it later.

%
Evidence that argues against the idea that the gas could have been stripped from Pal~3
is that if the \hi\ were at the same distance as Pal~3 and the detected overdensities ($\sim$80$-$90~kpc), the mass of AGC~501816 \citep[based on combining the ALFALFA flux $S_{21}$ with the distance using Equation~7 in][]{adams13a} 
would be $\sim$2$-$3 x 10$^3$~\msun. 
This is a substantial amount of \hi\ gas to have originated in a globular cluster, especially since it is found at a projected separation of $\sim$7\arcmin, or $\sim$180 pc at a distance of 90~kpc. Most other studies have estimated gas masses on the order of tens of \msun\ or less for globular clusters. Another issue is that the gas appears at a velocity that makes it difficult to distinguish from Galactic neutral hydrogen.  As mentioned, the \hi\ source was included in the UCHVC sample slated for optical follow-up because of its other properties and its proximity to Pal~3, and despite the fact that its $|v_{LSR}|$ value, 99~\kms, is lower than the 120~\kms\ minimum threshold.  
The ALFALFA detection grid in the vicinity of AGC~501816 and Pal~3
shows strong, widespread Galactic \hi\ emission primarily at substantially 
lower velocities, but 
also shows a modest amount of weak, diffuse Galactic emission 
in the velocity range $\sim$100$-$125~\kms; thus we cannot be certain that 
AGC~501816 is not gas associated with the Milky Way.
We 
conclude that additional observations of both the \hi\ and the optical sources in the field around Pal~3 and AGC~501816 
are needed in order to definitively determine whether or not there is any 
connection
between these two objects.

\begin{figure*}[h!]
\plotone{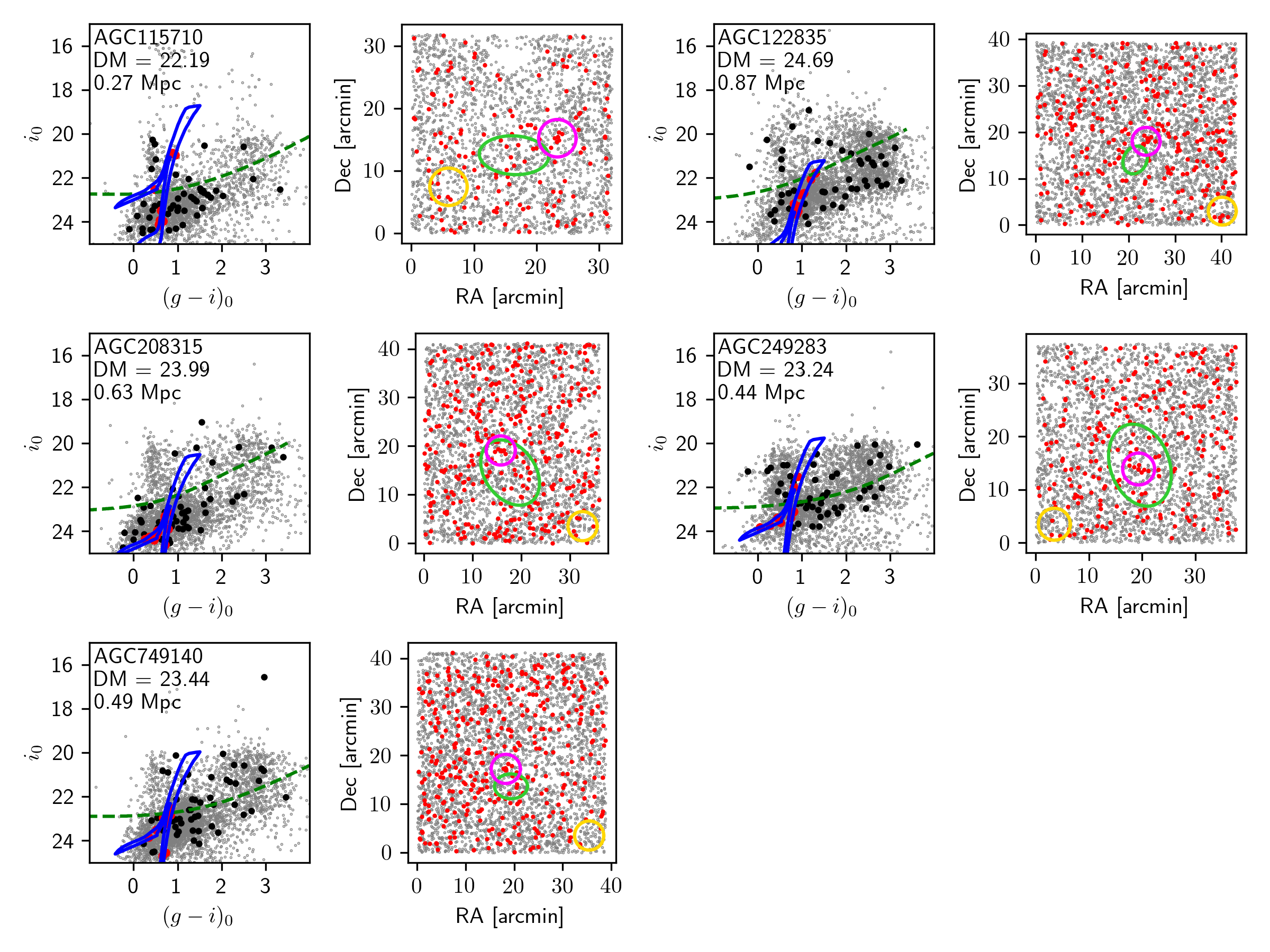}
    \caption{The CMDs and spatial positions of the point sources detected in the images of five UCHVCs classified as 
"Possible Detections".  The UCHVC name appears in the upper left corner of the CMD. Each CMD shows the location of the filter (blue solid line), the distance modulus and distance of the detected stellar overdensity, and the 50\% completeness level (green dashed line).  In the CMDs, the gray points are the point sources that survive cross-matching with Gaia and SDSS, sources within 3\arcmin\ of the overdensity peak are plotted in black, and sources that also fall within the CMD filter are over-plotted in red. 
In the spatial position plots, the gray points are point sources that survive cross-matching with Gaia and SDSS and red points are sources that are selected by the CMD filter.  Also marked are the overdensity peak (magenta circle of 3\arcmin\ radius), the \hi\ ellipse (green), and the reference circle (yellow). 
}
\label{fig:cmd five PD}
\end{figure*}

\subsubsection{Five Objects Classified as "Possible Detections"}
\label{sec: possible detections}

Five of the UCHVCs we analyzed in the sample of 26 yielded stellar overdensity detections that we categorize as 
"Possible Detections" because they are only modestly convincing, but may warrant future follow-up observations.
These five detections survived 
all of the analysis steps, including the
cross-matching step with the Gaia and SDSS catalogs, and still met our criteria for identifying possible dwarf galaxy candidates.  However, they were only significant in one of the two smoothing kernels (in all cases, the 2\arcmin\ kernel) and some of their other properties
suggest they may not be genuine stellar counterparts to the UCHVCs.
Figure~\ref{fig:cmd five PD} shows the CMDs and spatial locations of the sources in the fields of each of the five UCHVCs and Table~\ref{table: detections} provides information about the detected stellar overdensity, including the significance, estimated distance, the sky coordinates, 
the log of the \hi\ gas mass (if the UCHVC were at the same distance as the detected overdensity),  the estimated $M_V$ and stellar mass of the stellar counterpart (the calculation of these quantities is described in Sec.~\ref{sec:optical properties}), and the estimated ratio of the atomic gas mass to the stellar mass ($M_{HI}/M_*$). 
Details about each of these five objects are provided below.

{\it AGC~115710:---} This field yielded the stellar overdensity that had both the highest significance (99.55\%) and the closest CMD filter distance (270~kpc).  The CMD filter includes several bright objects that might be suitable targets for follow-up spectroscopy. On the other hand, the overdensity is relatively far from the \hi\ centroid (7.4\arcmin\ or $\sim$580~pc in projection), and a few background galaxies appear in the area around the overdensity.

{\it AGC~122835.---} This UCHVC field yielded an overdensity with one of the highest levels of significance, 95.30\%. 
The same overdensity was detected both before and after the catalog cross-matching step, and even increased in significance between the two iterations. The distance to the stellar overdensity (based on the location of the CMD filter) is estimated at 870~kpc and the projected separation from the \hi\ centroid is 4.5\arcmin, which corresponds to a physical separation of 1.1~kpc at the 870~kpc distance.  The stars in the detected overdensity do a fair job of filling in the RGB region of the CMD filter, from the top of the RGB down to the faint limits of the images.  On the other hand, 
the CMD of the sources in the reference circle 
looks similar, both in terms of the number of stars and their locations in the color-magnitude plane. 

{\it AGC~208315.---} The stellar overdensity detected near this UCHVC is only modestly significant (86.87\%) and has an estimated distance of 630~kpc.  It is located approximately 4.9\arcmin\ ($\sim$900~pc) in projection from the \hi\ centroid but nevertheless lies almost entirely within the \hi\ ellipse (Fig.~\ref{fig:cmd five PD}).
A small collection of stars is readily visible in the spatial plot and the region within the detection circle is more populated than the region within the reference circle.  However, the stars that make up the overdensity do not fill in the RGB region of the CMD filter and all of them fall below the 50\% completeness line in the CMD.  


{\it AGC~249283.---} 
The stellar overdensity associated with this UCHVC is of interest mainly because it lies directly on top of the \hi\ cloud, with a projected separation from the \hi\ centroid of 0.8\arcmin. This translates to $\sim$100~pc at the estimated distance given by the CMD filter, which is 440 kpc.
The stars that make up the overdensity lie within both the RGB and HB regions of the CMD filter, and the detection circle is clearly more populated than the reference circle.  The major drawback of this detection is its lower significance, which decreased from $\sim$88\% to 83.13\% after the catalog-matching step.


{\it AGC~749140.---} The overdensity detected in this field has a significance of 93.04\% and a CMD filter distance of 490~kpc, and became the prevalent detection after the cross-matching  step.  
It is about 3.7\arcmin\ (527~pc) from the \hi\ centroid.  The stars within the overdensity do appear more clustered than the stars in the surrounding image; however, none of the stars in the CMD filter are bright, and the detection is due to the presence of objects in the lower part of the RGB only, where the photometric uncertainties begin to increase. 

\subsubsection{Object Classified as the "Best Detection"}
\label{sec: best detection}

The images of the UCHVC AGC268071 yielded the most convincing dwarf galaxy candidate identified in the ODI imaging data; this object falls in the 
"Best Detection" category.
The initial run of the detection pipeline on the ODI images of AGC~268071 yielded possible detections of an optical counterpart at a range of distances with statistical significance above 90\%, so we carried out the catalog cross-matching steps and then 
searched with the more restricted catalog.  
The end result was a detection of a stellar association with high statistical significance that is located 9.2\arcmin\ from the \hi\ centroid.  The same detection appears at 95.7\% significance when we use a 2\arcmin\ smoothing kernel and at 97.3\% significance when we use a 3\arcmin\ kernel. The estimated distance for this putative 
dwarf galaxy counterpart
is 570~kpc, with a possible range between 490 and 590 kpc.

\begin{figure*}[h!]
    \plotone{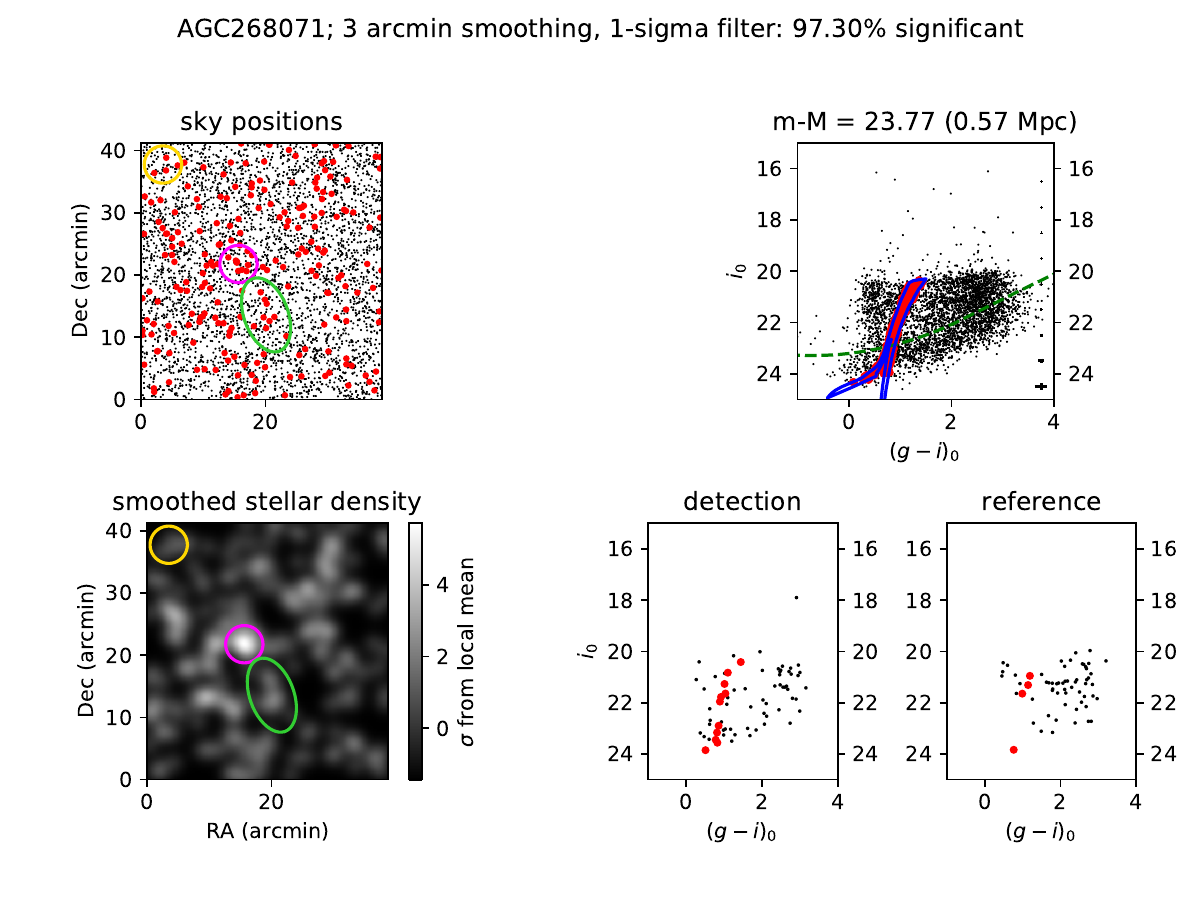}
    \caption{
CMDs and spatial positions of the point sources detected in the images of the UCHVC AGC~268071. A stellar overdensity with a statistical significance of 97.30\% is detected $\sim$9 arcmin from the centroid of the \hi\ source. The upper left plot shows the locations of the stars detected in the field (gray points), the stars selected by the CMD filter (red points), the \hi\ ellipse from ALFALFA (green ellipse), the 3\arcmin-radius detection circle (magenta) and a reference circle of the same size placed at a random location in the outskirts of the field.  The smoothed surface density map of the CMD-filtered stars is shown on the lower left, with the detection and reference circles and \hi\ ellipse marked as in the upper left plot. The CMD in the upper right shows all the sources in the field (black points), the CMD filter (blue solid line) and the CMD-selected stars (red points).  The CMDs for objects located with the the detection circle and comparison (reference) circle appear on the lower right. 
}
\label{fig:four plot agc268071}
\end{figure*}

The diagram showing the results (including the CMDs for the field, the spatial locations of the stars relative to the \hi, and the smoothed stellar distribution) for the 3\arcmin\ smoothing kernel is shown in Figure~\ref{fig:four plot agc268071}.
The stellar overdensity detected with the 3\arcmin\ smoothing kernel includes 11 stars that are noticeably clustered in the spatial distribution on the sky and that populate the RGB portion of the CMD filter reasonably well.  Furthermore, there are appreciably more stars in the detection circle (11 stars) compared to a reference circle (four stars) of the same size placed at a random position within the field (lower right of Fig.~\ref{fig:four plot agc268071}).   On the other hand, there are a few background galaxies that appear in the ODI image in the general area of the overdensity, so it is certainly possible that the detection is at least partially due to the presence of a collection of unresolved background galaxies that happen to have magnitudes and colors that fall within the CMD filter.  Also concerning is the fact that the stellar overdensity is located relatively far away from the \hi\ centroid; at a distance of 570~kpc, the 9.2\arcmin\ angular separation between the stellar overdensity and the \hi\ translates to a projected physical separation of 1.5~kpc.  It may be worth noting that AGC~268071 has the largest W$_{50}$ value of any of the UCHVCs included in our sample (Table~\ref{table: uchvcs sample}) and in the UCHVCs catalog published in \citet{adams13a}. Follow-up spectroscopy of some of the stars in the detected overdensity would be useful for determining whether the sources are genuinely associated with the \hi\ source found by ALFALFA,
or are simply the result of a random clustering of foreground and background objects.  This would be challenging because the detection CMD in Figure~\ref{fig:four plot agc268071} shows that the stars in the detected overdensity are faint, with the brightest having an $i$ magnitude of 20.4 mag.

\subsection{Estimating the Optical Properties of the Candidate Stellar Populations}
\label{sec:optical properties}

We can utilize the information provided by the ODI imaging data to calculate rough estimates of the optical brightness of the potential stellar populations we have detected.  For completeness, we carry out this calculation for all six of the sources in the current sample that we classify as either "possible detections" or the "best detection".  We calculate two estimates of the photometric properties of the detected stellar overdensities and combine those with the estimated distance to produce a faint and bright limit for the magnitude of the dwarf galaxy candidate that may be associated with the \hi.

To calculate the faint estimate, we simply combine the flux from each of the CMD-selected stars that make up the overdensity.  This is a conservative estimate of the brightness and assumes that these are the only stars associated with the \hi\ (i.e., that there is no underlying fainter stellar population present). 

To arrive at a bright limit for the magnitude of the possible stellar population, we carry out aperture photometry of the images at the location of the stellar overdensity.  First we mask out bright objects that are not part of the detected overdensity (e.g., extended background galaxies and stars that are much brighter than the CMD-selected stars) as well as background pixels that deviate appreciably from the median sky background level in that part of the image. We replace the masked pixels with the median sky 
value and then measure the flux in the region of the overdensity.  We set our aperture size to be based
on the apparent angular size of the nearby dwarf galaxy Leo~T \citep{irwin07a}
if it were located at the physical distance of our detection. Therefore, the aperture ranges in size from roughly 1.4 to 4 arc~minutes in diameter, given the range of estimated distances of the detected stellar overdensities (Table~\ref{table: detections}). 

We also calculate a $g-i$ color that corresponds to the faint and bright estimates, by subtracting the faint $i$ magnitude from the faint $g$ magnitude, and doing the same thing for the bright magnitudes. 
We use the computed $g,i$ magnitudes and colors, the estimated distance, and the photometric conversion relations in \citet{jester05a} to calculate the total absolute magnitude in the Johnson $V$ filter, $M_V$.  

We can combine these magnitude and color estimates to calculate at least an approximate stellar mass range for the candidate stellar populations we have identified.  We combine the relations in \citet{bell03a} with our magnitudes and colors to estimate the possible range of mass-to-light ratios for each object and then apply those values to the faint and bright magnitude limits to calculate the range of stellar masses.  We also calculate the ratio of the \hi\ mass to the stellar mass for the faint and bright cases.  

The optical magnitude and color estimates, and the associated \hi\ mass to stellar light ratios, are included in Table~\ref{table: detections}. 
The faint and bright estimates for a given object can vary from each other by a few magnitudes, which means the optical brightness of the putative stellar counterparts are highly uncertain.  Nevertheless, having at least a rough estimate allows us to compare the properties of these objects to the properties of simulated and observed dwarf galaxies in the Local Group (see Sections~\ref{sec: model compare} and \ref{sec:LG dwarfs}). 

\subsection{Revisiting the Stellar Overdensities Identified in the UCHVC Sample from J19}
\label{sec:revisit podi}

Because of the improvements we had made to the stellar overdensity search process,
%
it seemed appropriate
to 
reprocess and reassess 
the five potential stellar overdensities that were detected in the 
J19 sample.
We carried out a full analysis 
of the imaging data 
for those five objects, beginning with running source detection and photometry on the combined $g$ and $i$ image pairs
and proceeding all the way through the Gaia and SDSS catalog cross-matching steps and the final assessment of the results.
Information about the five UCHVCs that were reanalyzed in this way is listed in 
Table~\ref{table: podi uchvcs sample}; the table columns are the same as those in the corresponding tables for the primary sample of UCHVCs (Tables~\ref{table: uchvcs sample} and \ref{table: insufficient}). 

After processing and analyzing these objects with the updated procedures, we found that four out of the five possible overdensities identified by J19 no longer had convincing detections. 
Although there were a few cases where a stellar overdensity with a significance of $\sim$80$-$90\% or above was found somewhere in the images, further scrutiny 
showed that 
in all cases the overdensity was 
too far away from the \hi\ source, and/or the CMD of the stars within the detection circle looked too much like the CMD of the stars in the reference circle.  Accordingly, we now categorize these objects as non-detections and are mark them as "ND" in Table~\ref{table: podi uchvcs sample}.
Our overall conclusion is that the changes to the detection pipeline described in Sec.~\ref{sec:pipeline improvements} -- especially the reduction in foreground and background objects from our star lists made possible by the catalog cross-matching -- resulted in a more stringent process, which in turn showed that four of the five most significant overdensities presented in J19 are not likely to be genuine dwarf galaxies. 

%
%
%

One UCHVC field, however, yielded a highly statistically significant dwarf galaxy candidate that meets our criteria for a "best detection".  The UCHVC is AGC~249525, and the corresponding optical counterpart was first highlighted in \citet{janesh17a} 
and was one of the most convincing dwarf galaxy candidates identified in the full sample of objects presented in J19. Using our updated detection methods, we find a highly significant overdensity that matches the characteristics of the original detection.

AGC~249525 has an \hi\ mass of $\sim$10$^7$ \msun\ (see Table~\ref{table: podi uchvcs sample}), a column density log($N_{HI}$) of 19.60 atoms~cm$^{-2}$ and an apparent HI major axis of $\sim$9~arcmin.  This source is one of the UCVHCs that was observed with the 
WSRT and the VLA and has therefore been mapped at higher resolution \citep{adams16a,paine20a,bralts20a}.  \citet{bralts20a} analyzed the VLA observations of AGC~249525 and concluded that they show a velocity gradient that is indicative of rotation support, but that more work is necessary to draw firmer conclusions about the \hi\ kinematics of this object.

The stellar overdensity detected with the earlier version of the pipeline and presented in \citet{janesh17a} and \citet{janesh19a} had a statistical significance of $\sim$98\%. It was located well within the ALFALFA \hi\ ellipse and was directly coincident with the highest-density contour in the \hi\ synthesis maps from the WSRT presented in \citet{adams16a}.  \citet{janesh17a} reported a distance of 1.60~Mpc, with an error (which they estimated by determining the range of distances over which the significance stayed above 95\%, for this object) of $\pm$0.45~Mpc.  Their estimated stellar mass for the optical counterpart was between $\sim$2 x 10$^4$ and 4 x 10$^5$ \msun.  

Our updated analysis yields
a possible stellar counterpart at the same location as the one found by \citet{janesh17a}, but with an increased statistical significance of 99.26\%.  The range of distances over which the detected overdensity has a significance above 90\% is 1.93$-$2.12~Mpc, and the highest significance corresponds to a distance of 2.09~Mpc.  
This distance is slightly larger than the J19 distance but the two distance ranges overlap.  

\begin{figure*}[h!]
    \plotone{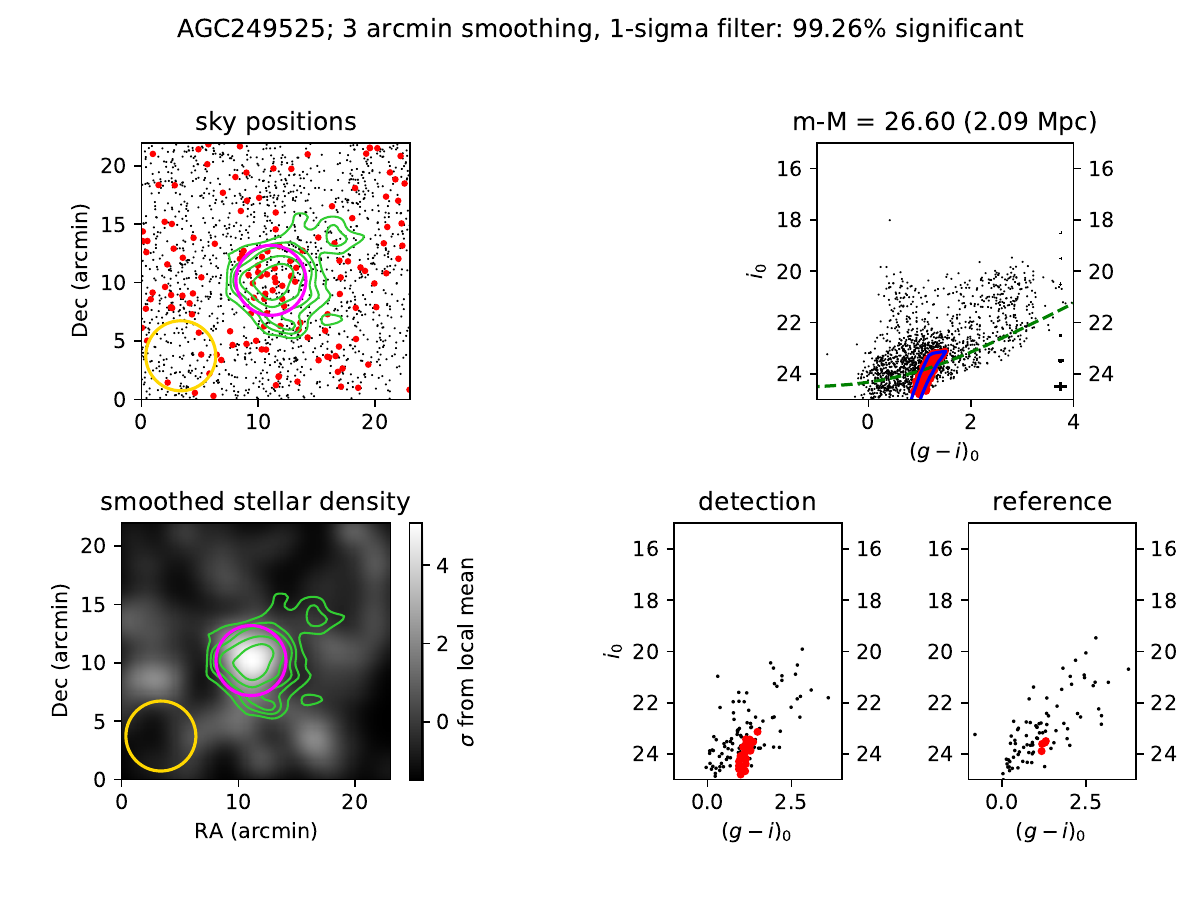}
    \caption{
CMDs and spatial positions of the point sources detected in the images of the UCHVC AGC~249525, which was included in the sample presented in \citet{janesh19a} and which we have reprocessed and analyzed with our updated detection pipeline. A stellar overdensity with a statistical significance of 99.26\% is detected that coincides with the location of the \hi\ source (green contours). The upper left plot shows the locations of the stars detected in the field (gray points), the stars selected by the CMD filter (red points),
the 3\arcmin-radius detection circle (magenta) and a reference circle of the same size placed at a random location along the edges of the field.  The smoothed surface density map of the CMD-filtered stars appears on the lower left, with the detection and reference circles 
marked in the same way as in the upper left plot. The CMD in the upper right shows all the sources in the field (black points), the CMD filter (blue solid line) and the CMD-selected stars (red points).  The CMDs for objects located with the the detection circle and comparison (reference) circle appear on the lower right.  
The \hi\ contours from the \citet{adams16a} WSRT study are shown with 
with green solid lines on the upper left and lower left spatial plots; the 
contour levels are [9, 15, 20, 30, and 40] x 10$^{18}$ atoms cm$^{-2}$. 
}
\label{fig:four plot agc249525}
\end{figure*}

The overdensity found by the new analysis is made up of 25 point sources that are selected by the CMD filter and located within a region 3\arcmin\ in radius.  These 25 sources all fall within the upper portion of the RGB in the CMD filter
(see Figure~\ref{fig:four plot agc249525}).  The number of sources in this detected overdensity is more than twice the number of sources that make up the other overdensities in the 
sample of 
UCHVCs presented in the current paper (see Table \ref{table: detections}). The \hi\ contours from the WSRT observations presented in \citet{adams16a} are over-plotted on the spatial plots of the field that are shown in Figure~\ref{fig:four plot agc249525}; the position of the 
stellar overdensity found with the updated pipeline remains coincident with the \hi\ contour with the highest $N_{\rm HI}$ value. 

A distance of 2~Mpc is near the upper end of the distance range over which we search for stellar counterparts, and at these large distances the upper portion of the RGB coincides with the region of the CMD where the photometric errors become larger, our detection completeness is lower, and it becomes more difficult to distinguish true point sources from extended background objects in our ground-based images.  Therefore follow-up observations -- deeper, higher-resolution imaging and/or spectroscopy of the sources that make up the detection -- 
are needed in order to explore whether this overdensity is a genuine stellar counterpart to the UCHVC or is simply a collection of foreground and/or background objects that happen to coincide with the \hi\ and lie in the relevant portion of the CMD.  

We carried out the steps described in Section~\ref{sec:optical properties} to estimate the optical properties and the distance-dependent \hi\ properties of the possible stellar counterpart to AGC~249525; these are listed in Table~\ref{table: podi detection}.
The $V$-band absolute magnitude for the stellar counterpart is estimated to be in the range $-$5.99 to $-$6.87 mag.
As one might expect (since the analysis was carried out on the same images, and we find the overdensity at a similar distance),
these brightness estimates overlap the original brightness range estimated in J19 ($M_V$ $=$ $-$4.5 to $-$7.1 mag).

\section{Summary and Discussion}
\label{sec:discussion}

\subsection{Summary of Results}
\label{sec:summary}

The UCHVCs investigated in this study were selected from the ALFALFA survey because they had
relatively compact angular sizes, were isolated from other \hi\ sources in both velocity space and location on the sky, had measured velocities that suggested they were within the Local Volume and were not likely to be Galactic \hi, and did not have a clear optical counterpart in existing catalogs and survey data.
Our goal was to determine whether any of the UCHVCs might actually host an as-yet undiscovered dwarf galaxy, perhaps even one like the 
dwarf galaxy Leo~P, which was first identified as an ALFALFA UCHVC and then imaged in the optical with WIYN 
\citep{giovanelli13a, rhode13a}.  If the UCHVCs do indeed host a dwarf galaxy, they would be some of the faintest, lowest-mass neutral-gas-bearing dwarf galaxies known in the nearby Universe, 
with absolute V-band magnitudes in the range $\sim$$-$2 to $-$8, stellar masses in the range $\sim$10$^3$ to 10$^6$ \msun, but with a gas content in the range $\sim$10$^5$ to 10$^6$ \msun\ (for an assumed distance of 1 Mpc).

In the current paper, we have presented results from the second and final phase of our campaign to image a large sample of ALFALFA UCHVCs with WIYN.  For this phase of the campaign, we made several improvements to our procedures for finding
possible stellar counterparts to the \hi\ clouds and we analyzed high-quality $g-$ and $i-$band imaging 
of 26 UCHVCs
to look for dwarf galaxy candidates.  

From the sample of 26 objects, we identified six UCHVCs with at least one significant detection.  We also carried out a detailed analysis of ODI imaging of a UCHVC located a few arc~minutes away from 
the Galactic outer halo GC Pal~3.
Our best dwarf galaxy candidate is associated with the UCHVC AGC~268071; the stellar overdensity we identify has a statistical significance of 97.30\%, a distance of $\sim$570~kpc, and is located roughly 9\arcmin\ ($\sim$1.5~kpc) from the centroid of the \hi\ source. 

We 
used our improved methods to reprocess and analyze the original WIYN imaging of the 
five UCHVCs from the first phase of the project that had possible stellar counterparts. 
From this set of five UCHVCs, we find one stellar counterpart that qualifies as a convincing detection 
based on our new procedures and criteria.
The detected stellar overdensity appears 
at the same location and a similar distance ($\sim$2~Mpc) as the stellar overdensity originally identified in \citet{janesh17a}, but with an even higher statistical significance (99.3\%) when compared to repeated experiments that randomly distribute the same number of stars around the field. 

\subsection{Putting Our Results in Context}

\subsubsection{Other Observational Searches for Dwarf Galaxy Counterparts to Compact \hi\ Clouds}


In the time since the discovery of Leo~P, we have had only limited success at finding candidate dwarf galaxy counterparts to the UCHVCs selected from the ALFALFA survey data, and we have certainly found no counterparts that are as obviously apparent as Leo~P.  Leo~P subtends roughly 90~arcsec on the sky, has a total absolute $V$-band magnitude of $M_V$$\sim$ $-$9.3, and has an underlying population of old ($\sim$12~Gyr) RGB stars, a young ($\leq$300~Myr) population of massive early-type stars, a prominent HII region
\citep{rhode13a,mcquinn15b}, and H-$\alpha$ emission that reveals the presence of extended ($\sim$100~pc) ringlike structures \citep{evans19a} in the galaxy's interstellar medium. Given its properties, Leo~P was readily visible in our modest-length exposures (20-30-minute integrations in each of the $B,V,R$ filters)
taken with WIYN 
\citep{rhode13a}. 

Other studies have looked for counterparts to \hi\ sources identified in ALFALFA and other neutral hydrogen surveys and have also met with limited or no success at finding Local Group dwarf galaxies, although they have identified new dwarf galaxies within the Local Volume and/or at Virgo Cluster distances.  
The survey that is most relevant to mention in this context is the 
Galactic Arecibo L-band Feed Array HI (GALFA-HI) survey.  GALFA-HI was aimed primarily at studying the neutral hydrogen component of the Galactic interstellar medium \citep{peek11a},
but in the process identified 27 clouds that were classified as "Galaxy Candidates" because of their kinematics and separation from known galaxies or high-velocity cloud complexes \citep{saul12a}.  Eleven of the ALFALFA UCHVCs from \citet{adams13a} also appear in the GALFA-HI catalog. 

Several groups carried out optical follow-up observations and archival imaging searches of objects chosen from the GALFA-HI and/or ALFALFA surveys. 
\citet{bellazzini15a,bellazzini15b} 
used the Large Binocular Telescope to conduct the SECCO Survey, which obtained deep optical images of 25 of the ALFALFA UCHVCs.  The survey resulted in the detection of a distant stellar counterpart to the UCHVC AGC~226067 that was dubbed SECCO~1 \citep{bellazzini15a}. \citet{bellazzini15b} suggested that the object was at least 3~Mpc away and was most likely a star-poor dwarf galaxy in the Virgo Cluster. 
\citet{adams15a}
subsequently published \hi\ synthesis maps and optical imaging of AGC~226067 and its putative optical counterpart. 
\citet{tollerud15a}
used the WIYN pODI instrument to obtain follow-up imaging of 22 of the 27 GALFA-HI Galaxy Candidates from the list published by \citet{saul12a}.  From this sample of 22 objects,
\citet{tollerud15a} identified two dwarf galaxies, dubbed Pisces~A and Pisces~B, that were visible in SDSS images and appeared to be nearby 
(at least within the Local Volume); they 
used spectroscopy with the Palomar 5-m to confirm that the dwarfs were genuinely associated with the \hi.  A subsequent HST imaging study by \citet{tollerud16a} placed 
Pisces~A and Pisces~B
at distances of 5.6 Mpc and 8.9 Mpc, respectively \citep[we note that these objects were also cataloged by ALFALFA and designated AGC~103722 and AGC~114843;][]{haynes18a}.

\citet{sand15a} 
searched archival UV and optical data from a number of different surveys and facilities (DSS, SDSS, Subaru
SMOKA, CFHT Megacam, GALEX, and Swift) to look for counterparts to a sample of UCHVCs selected from 
GALFA-HI and ALFALFA. They looked for
blue, diffuse emission similar to the faint blue emission that was visible in the SDSS images of Leo~P before its discovery 
\citep{giovanelli13a,rhode13a}.  They identified six possible counterparts to their sample of UCHVCs and used spectroscopic follow-up to confirm that five of the counterparts were genuinely associated with the HI gas. All five of these counterparts are dwarf galaxies that lie outside the Local Group. Two of the dwarfs were the above-mentioned galaxies Pisces~A and Pisces~B from \citet{tollerud15a}.   A third dwarf galaxy was SECCO~1, the object identified by \citet{bellazzini15a}; \citet{sand17a} followed up with an HST study that showed that this object is $\sim$17~Mpc away and is likely a remnant dwarf galaxy produced through a ram pressure stripping event that occurred within the M86 subgroup of the Virgo Cluster. The remaining two dwarf galaxies found by \citet{sand15a}, GALFA Dw3 and GALFA Dw4, were observed with HST to confirm their location outside the Local Group (at 7.6 Mpc and 3.1 Mpc, respectively) and note their isolated nature \citep[no other galaxies within 1.5~Mpc of either dwarf;][]{bennett22a}.

\citet{tollerud18a} utilized the Exploring the Local
Volume In Simulations (ELVIS) suite of simulations \citep{garrison14a}
to create mock \hi\ catalogs and explore how these compare to the results from GALFA-HI and the associated optical searches for Local Group dwarf galaxies. The ELVIS simulations were dark-matter-only simulations designed to model environments similar to the Local Group, with two massive galaxy halos in the same configuration (e.g., mass and position) as the dark matter halos that are thought to host M31 and the Milky Way. \citet{tollerud18a} utilized a simple empirical model to translate the dark-matter-only ELVIS simulations into GALFA-HI observables and then examine what the \hi\ survey would be expected to find.  Based on these mock catalogs, they found that they should potentially discover tens of Local Group dwarf galaxies (depending on the model), in contrast to the zero that were found in the actual observations.  They argued that this discrepancy could be explained if reionization inhibited star formation at mass scales below the mass of galaxies like Leo~T, which had an estimated virial mass at the time of reionization of 
$\sim$10$^{8.5}$ \msun.

The results shown in \citet{tollerud18a} were further investigated by \citet{defelippis19a}, 
who combined the data in the GALFA-HI catalog with archival optical imaging data from the Panoramic Survey Telescope and Rapid Response
System (Pan-STARRS; \citet{chambers16a}). 
\citet{defelippis19a}
began by searching the GALFA-HI data for \hi\ sources with the expected neutral gas properties of nearby dwarf galaxies and identifying 690 objects.  They then applied 
a filtering technique to the Pan-STARRS archival data to look for stellar populations with a range of ages and metallicities that might be spatially associated with the \hi.  They also tested their algorithm by searching for, and successfully finding, known dwarf galaxies (namely, Leo~T and Draco) at the appropriate distances. They found one potential dwarf galaxy candidate near the Galactic plane, but no objects that resemble Leo~T, and argued that the results rule out the existence of Leo~T-like objects within the GALFA footprint (which covers $\sim$one-third of the sky) at distances within 500~kpc.  They conclude that their results support 
the \citet{tollerud18a}
assertion that reionization caused the gas to be removed from Local Group objects with halo masses less than the $\sim$10$^{8.5}$ \msun\ threshold, and thereby prevented those objects from 
developing into full-fledged dwarf galaxies.

\subsubsection{Comparison of Our Results to Model Simulations}
\label{sec: model compare}

A useful next step for understanding our results in a larger context is to compare our most convincing dwarf galaxy candidates to the results from recent state-of-the-art simulations that attempt to model the formation of dwarf galaxies down to the lowest masses.  
\citet{applebaum21a} (hereafter A21) 
presented results from the DC Justice League suite of Milky Way-like zoom-in simulations, which were designed to probe galaxy formation down to the regime of ultra-faint dwarf galaxies (UFDs), i.e., down to absolute magnitudes of $M_V$ fainter than $-$8 and stellar masses \lapp10$^5$ \msun.  
The simulations were initially run to model large ($\sim$50~Mpc across), dark-matter-only volumes  and then small regions within those volumes were 
re-simulated at higher resolution with full hydrodynamical treatment.  
The smaller regions were specifically  chosen to contain Milky Way analogs 
to enable meaningful comparisons between the simulation results and the observed properties
of our local galaxy environment. 
Two Milky
Way analogs, nicknamed "Elena" and "Sandra", with masses of 7.5 x 10$^{11}$ \msun\ and and 2.4 x 10$^{12}$ \msun\ respectively, 
were selected for detailed simulations because they were isolated and had masses consistent with current observational estimates of the Milky Way’s virial mass.  The DC Justice League simulations included star and gas particles, feedback from supernovae and stellar winds, chemical evolution, star formation, and supermassive black hole (SMBH) formation and feedback.  
A21 cataloged the dwarf galaxies that were produced in each simulation and determined the satellite galaxies' stellar, gas, and virial masses, luminosities, half-light radii, metallicities, galactocentric distances, and star formation time scales.  They were able to resolve UFDs with absolute magnitudes as faint as M$_V$ $=$ $-$3 and found that they could match the range of dynamical properties and size-luminosity scaling relations seen in the known Local Group dwarf galaxies down to scales of $\sim$200 pc.  They also compared the properties of their simulated dwarf galaxies to the expected detection limits that the Legacy Survey of Space and Time (LSST) data from the Rubin Observatory will reach after 10 years.  They concluded that virtually all Local Group UFD galaxies will be detectable by LSST down to the luminosity limit probed by their simulations. 

\begin{figure*}[h!]
    \plotone{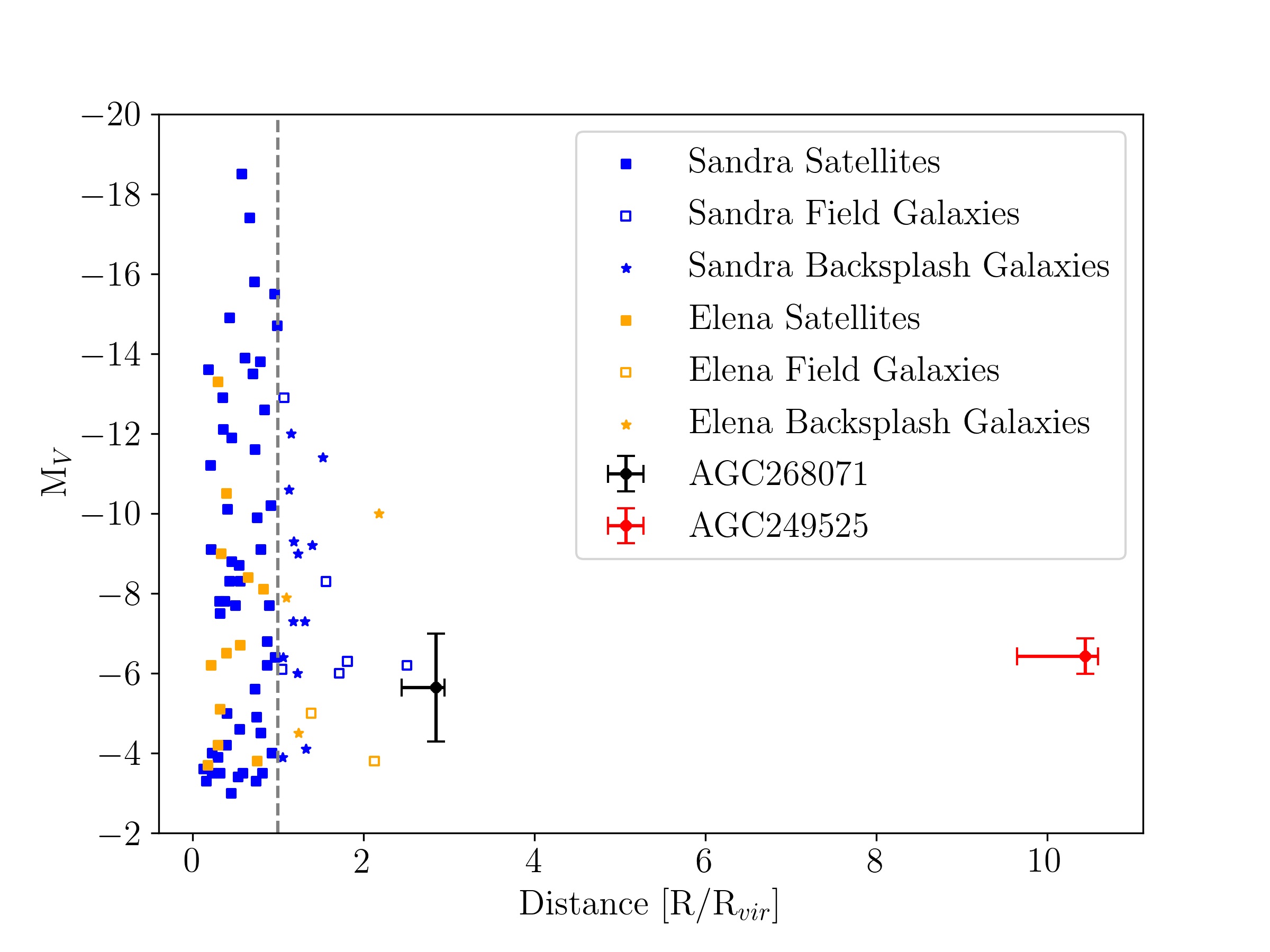}
    \caption{
A recreation of Figure 5 from \citet{applebaum21a}, which shows $V$ magnitude
as a function of normalized distance for the dwarf galaxies in the Sandra and Elena simulations (blue and orange points, respectively).  The dwarf galaxy candidates associated with the UCHVCs AGC~268071 and AGC~249525 are plotted on the figure (points with large error bars) for comparison. 
For both simulations, satellites are denoted
with filled squares, field galaxies are shown as open squares, and backsplash
galaxies are denoted with a star. For the optical counterparts to AGC268071 and AGC~249525, we plot the average of the
ratios calculated from the bright and faint estimates of their photometric properties and
use both estimates to bracket our uncertainty. The distance uncertainty for these two objects is determined by the distance over which each object is detected in our data (see Table~\ref{table: detections}). 
We note that the error bars representing the distance uncertainties for our dwarf galaxy candidates are not symmetric in this plot and in Figures~\ref{fig:ratio norm distance} and \ref{fig:LG dwarfs}. This  reflects the fact that the significance of a given overdensity falls below the 90\% or 80\% threshold relatively quickly after the peak significance is reached, as the tip of the RGB in the CMD filter is shifted downward to fainter magnitudes and the handful of stars that make up a typical detected overdensity become too bright to coincide with the filter. 
}
\label{fig:mag norm distance}
\end{figure*}

Because the DC Justice League simulations were able to produce dwarf galaxies
with realistic properties, it should be instructive to compare the properties of their dwarf galaxies with the properties of the two most convincing dwarf galaxy candidates we identified in our search,
namely the optical counterparts we find for the UCHVCs AGC~268071 and AGC~249525. We have recreated two of the figures from A21 that feature observable quantities that are also available for our dwarf galaxy candidates. In Figure~\ref{fig:mag norm distance}, which is a recreation of Figure~5 from A21, we show the absolute V-band magnitudes of the simulated dwarf galaxies as a function of the distance to their parent galaxy normalized by the parent galaxy's virial radius.  The figure makes it clear that 
the Sandra host galaxy was associated with many more satellite galaxies than the Elena host galaxy. A21 investigated this difference and attributed it to both the higher mass of the Sandra halo and to the fact that an analog to the Large Magellanic Cloud (which is associated with many of the ultra-faint dwarfs 
in the vicinity of the Milky Way) appears in the Sandra simulation but is not present in the Elena simulation.
The simulated galaxies are also separated into three categories: satellite galaxies that are within the virial radius of their parent galaxy, field galaxies that are outside the virial radius of the parent, and "backsplash" galaxies that are field galaxies that have experienced an infall that has brought them within their parent galaxy's virial radius. 

\begin{figure*}[h!]
    \plotone{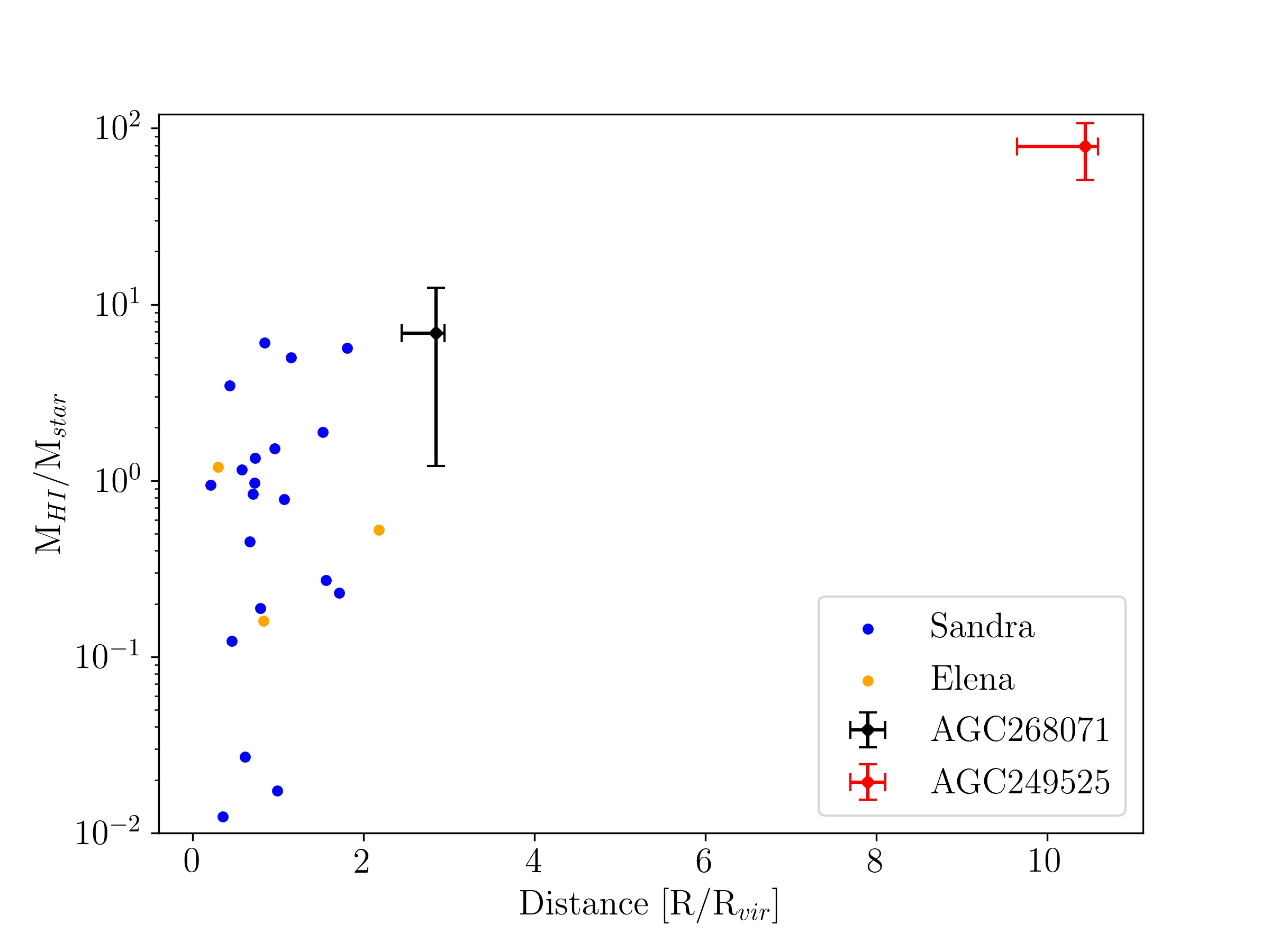}
    \caption{
A recreation of Figure 13 from A21, which shows the \hi-to-stellar mass ratio as a function of normalized distance from the parent galaxy for simulated dwarf satellite galaxies associated with Milky Way analogs.  We have added the dwarf galaxy candidates we found that are associated with the UCHVCs AGC~268071 and AGC~249525.  Blue points represent galaxies in the Sandra simulation, while orange points
represent galaxies in the Elena simulation. Note that many of the galaxies in the simulations do not have \hi\ gas and therefore are not plotted. For AGC~268071 and AGC~249525 (points with large error bars), we plot the average of the bright and faint magnitude estimates and use both estimates to set the uncertainty.
The distance uncertainty for these two objects is determined by 
the distance range over which we detect them (see Table~\ref{table: detections}). 
}
\label{fig:ratio norm distance}
\end{figure*}

\begin{figure*}[h!]
    \plotone{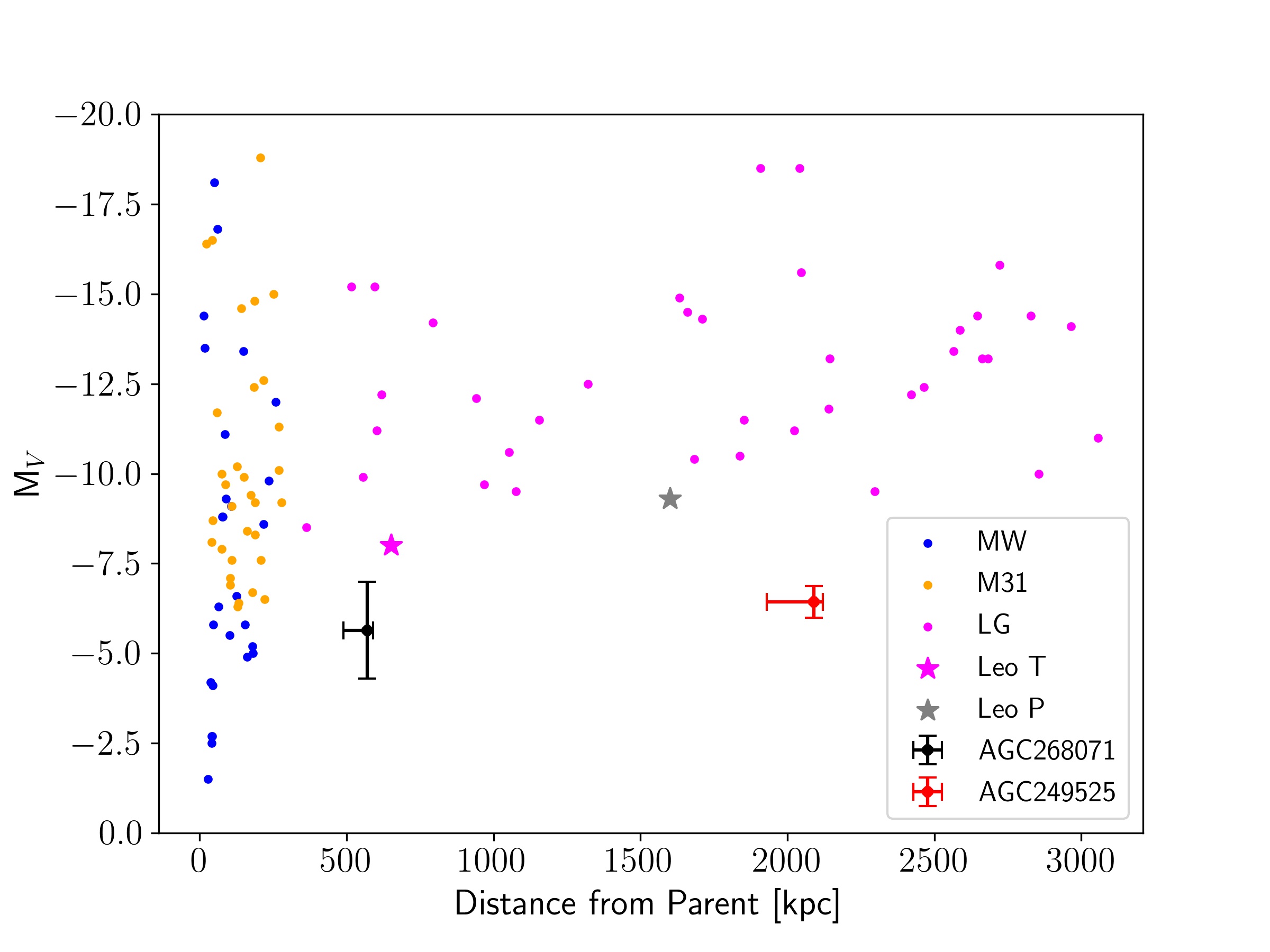}
    \caption{A plot of dwarf galaxy $M_V$ values, drawn from \citet{mcconnachie12a}, as a function of distance from
the parent galaxy for galaxies in and near the Local Group. Milky Way satellite galaxies are plotted in blue while M31 satellites are shown in orange. Dwarf galaxies that are not satellites of either the Milky Way or M31 appear in pink, including Leo~T (pink filled star symbol). For the latter set of objects, we plot the distance to the barycenter of the Local Group.  Leo~P is marked with a filled gray star and plotted using the distance from \citet{mcquinn15b}.  We also show our measurements for the dwarf galaxy candidate associated with AGC~268071 and AGC~249525 (black and red points with large error bars). For the two dwarf galaxy candidates, we plot the average of the bright and faint absolute magnitude estimates and use both estimates to set the uncertainty. The distance uncertainty is defined by the distance range over which we detect the stellar overdensity. We also note that the distances we are plotting for these two objects are the heliocentric distances.
}
\label{fig:LG dwarfs}
\end{figure*}

We have added our two dwarf galaxy candidates to Figure~\ref{fig:mag norm distance}. To include them in the plot, we calculated the distances of the two dwarf galaxy candidates in terms of virial radius of the Milky Way \citep[200~kpc;][]{dehnen06a}.
In  the figure,
AGC~268071 appears to fit within the population of field galaxies, which follow a weak trend of having fainter absolute magnitudes with increasing distance from the parent galaxy. One of the simulated field galaxies that is produced in the Sandra simulation has properties that overlap those of AGC~268071, given the uncertainties on our estimated absolute magnitude and distance for this dwarf galaxy candidate. We include AGC~249525 for completeness, although it lies in a region that is not populated with the dwarf galaxies included in the simulation; it has an estimated absolute magnitude that is similar to those of AGC~268071 and the other field dwarf galaxies, but a distance that corresponds to more than 10 times the virial radius of the Milky Way.

In Figure~\ref{fig:ratio norm distance}, we present a similar figure to Figure~13 in A21, and show the \hi-to-stellar mass ratios of the gas-bearing dwarf galaxies that appear in the Sandra and Elena simulations. Only the galaxies that A21 determined to have retained their gas (i.e., those that are not classified as "quenched" in their paper) are included in our version of the figure.  The ratios are shown as a function of the distance (again normalized by the virial radius) from the parent galaxy.  We also show the locations of the dwarf galaxy candidates we detect in the images of the UCHVCs AGC~249525 and AGC~268071.  Many of the dwarf galaxies produced by the DC Justice League simulations do not have any neutral gas and therefore are not plotted. The dwarf galaxy candidate associated with AGC~268071 exists much farther out in distance from the parent galaxy than the majority of the simulated dwarfs 
and lies on the high end of the range of \hi-to-stellar mass ratios. The dwarf galaxy candidate corresponding to AGC~249525 is even more extreme, in terms of both gas-to-stellar ratio and distance from the massive galaxy.  Both dwarf candidates appear generally consistent with a mild trend for higher gas-to-stellar ratios with larger distances that is present in the simulated sample.  The low stellar content (and/or high \hi-to-stellar mass ratio) of the dwarfs at large distances may be at least partially explained by their isolated nature, as they would presumably be less subject to tidal effects from their parent galaxy or other nearby satellites and therefore perhaps less likely to either be stripped of their gas and/or to have undergone starbursts triggered by close interactions. 

\subsubsection{Comparison of Our Results to the Population of Local Group Dwarf Galaxies}
\label{sec:LG dwarfs}

In addition to comparing to simulated data, it is worthwhile to compare the dwarf galaxy candidates we detect to the actual dwarf galaxy population in and around the Local Group.  In Figure~\ref{fig:LG dwarfs}, we plot the absolute $V$-band magnitude as a function of distance for dwarf galaxies within $\sim$3~Mpc.  The values we show are taken from the sample included in \citet{mcconnachie12a}, who compiled the measured properties (e.g., distances, magnitudes, metallicities, structural and dynamical characteristics) of galaxies in the Local Group and its environs.  In the figure, Milky Way satellite galaxies are shown in blue, M31 satellites are shown in orange, other nearby dwarf galaxies are shown in pink, and our dwarf galaxy candidates are shown in black and red.  The positions of Leo~T and Leo~P are plotted with filled star symbols that are pink and gray, respectively.  
Since the other nearby dwarf galaxies are not bound to either the Milky Way or M31, we use their distance from the barycenter of the Local Group as the distance value on the x-axis of the figure, except that for Leo~P we use the distance from \citet{mcquinn15b}. For the counterparts to AGC~268071 and AGC~249525, we simply use the estimated heliocentric distance determined from our detection and analysis process. 

Our estimate of the absolute $V$-band magnitude of the dwarf galaxy candidate associated with AGC~268071 falls toward the faint end of the range of absolute magnitudes for Milky Way satellites and at the very bottom end of the range of luminosities of M31 satellites.  The main difference between the AGC~268071 dwarf galaxy candidate and the known satellites of the Milky Way and M31 is its large distance from the Milky Way (which we estimate at $\sim$570~kpc).  This distance places it substantially farther from the Milky Way or from M31 than any of the other dwarf satellite galaxies that appear on the plot.  This suggests that it may not be a satellite of the Milky Way, but may instead be simply a Local Group dwarf galaxy (if it is a genuine galaxy at all); in that case, the most relevant comparison sample for this object would be the other nearby dwarf galaxies that are not satellites (pink points with distances extending to $\sim$3~Mpc on Figure~\ref{fig:LG dwarfs}). Compared to this sample of objects, the dwarf galaxy candidate associated with AGC~268071 is roughly 2 magnitudes less luminous, and is also closer in distance than all but a few of the dwarf galaxies in the figure.  Overall it seems that the detection for AGC~268071 makes a reasonable dwarf galaxy candidate in the sense that it at least lies within a similar parameter space as the other dwarf galaxies in and around the Local Group, although on the low-luminosity edge of the distribution. 

The estimated $M_V$ magnitude and distance for the dwarf galaxy candidate associated with AGC~249525 puts this galaxy $\sim$2.5 magnitudes below that same sample of nearby dwarf galaxies. This object would thus be a gas-rich ultra-faint dwarf galaxy that is substantially fainter than the other currently-known galaxies at this distance ($\sim$2~Mpc).  As noted in \citet{janesh17a}, a potential close neighbor to AGC~249525 in terms of distance and sky position is UGC~9128, which is 10 degrees away on the sky and has a heliocentric velocity $cz$ of 152~\kms \citep{mcconnachie12a}, compared to 48~\kms\ for AGC~249525.

%

\subsection{Next Steps and Future Observing Capabilities}
\label{sec:future work}

In our analysis of the ALFALFA UCHVCs, we found six possible dwarf galaxy candidates.
However, we again note that -- even with our deep optical multi-band imaging data, with typical seeing $\sim$1 arcsec or better -- 
there was no clearly visible diffuse optical emission, nor any obvious (to the eye) stellar association, 
at the locations of the dwarf galaxy candidates detected via the CMD filtering process. Without such additional evidence,
it is appropriate to question whether the objects we found are indeed genuine dwarf
galaxies.  
The statistically-significant stellar overdensities we have detected may actually be made up of some combination of compact background galaxies and Galactic foreground stars (that are below the detection limits for Gaia EDR3 and/or SDSS and therefore too faint to be eliminated in 
the cross-matching step) 
which happen to 
coincide with the rough location of the \hi\ source. 
To confirm that the stars in the detected overdensities are truly associated with the corresponding UCHVC, we would need spectroscopic follow-up observations that show that the stars' velocities are similar to that of the \hi. 
The brightest stars that make up the detections in our best two candidates, AGC~268071 and AGC~249525, have $i$ magnitudes of 
20.4 and 23.1~mag, respectively.
%
Deriving accurate radial velocities of stars of these magnitude, even with time on an $\sim$8$-$10-meter telescope, is challenging; for example, our initial attempt to obtain a spectrum of a star in the AGC~268071 dwarf galaxy candidate 
using the LRS2 instrument \citep{chonis16a} on the 10-meter Hobby-Eberly Telescope\footnote{Based on observations obtained with the Hobby-Eberly Telescope (HET), which is a joint project of the University of Texas at Austin, the Pennsylvania State University, Ludwig-Maximillians-Universitaet Muenchen, and Georg-August Universitaet Goettingen. The HET is named in honor of its principal benefactors, William P. Hobby and Robert E. Eberly.} yielded a spectrum with insufficient signal-to-noise to measure a radial velocity. 
The rest of the stars that make up the overdensities for our two best candidates are 
even fainter (see Figures~\ref{fig:four plot agc268071} and \ref{fig:four plot agc249525}). Thus, follow-up spectroscopy is extremely challenging
with current ground-based optical spectroscopy capabilities.  

More generally, further progress toward completing the census of low-mass galaxies in the Local Group and its environs will be made with the advent of the next generation of large-scale systematic surveys like the Rubin Observatory’s Legacy Survey of Space and Time 
(LSST; \citealt{ivezic19a}).
LSST will uniformly survey the Southern sky in six filters ($ugrizy$) over the course of 10 years; the median 5-$\sigma$ point source detection limit should be $r = 24$ mag in the single-visit images and $r$ $\sim$27 mag in the 10-year co-added images, with median effective seeing of 1.0 arcsec.  
With this wide-field, multi-color, high-quality imaging data, LSST should be able to resolve RGB stars out to $\sim$10~Mpc within its survey area. 
The survey should be sensitive to galaxies as faint as the candidate dwarf galaxy we detected in the AGC~268071 field (i.e., $M_V$ $\sim$ $-$6) in galaxy groups out to $\sim$3$-$4~Mpc \citep{lsst09a,simon19a}. LSST should also be able to probe even fainter galaxies with resolved stellar photometry out to 1$-$2 Mpc and should be complete for galaxies with stellar luminosities 
\gapp 2 x 10$^3$ \lsun\ 
that are within $\sim$1 Mpc of the Milky Way \citep{bullock17a}.
Comparing the catalogs of resolved stars and newly-discovered dwarf galaxies that will likely be detected by LSST to \hi\ observations from sensitive radio facilities like the Five Hundred Meter Spherical Telescope (FAST), the MeerKAT telescope, the ASKAP telescope, and the planned Square Kilometre Array  Observatory (SKAO)
may help reveal the nature of objects like the UCHVCs or the Galaxy Candidates identified in GALFA-HI data by \citet{saul12a} and whether these are genuine gas-rich ultra-faint dwarf galaxies or simply isolated gas clouds with no associated stellar population. 

%
%

\begin{acknowledgments}
The authors wish to thank the anonymous referee for providing helpful suggestions that improved the manuscript.  We also thank the staff of the WIYN Observatory and Kitt Peak National Observatory for their assistance during our pODI and ODI observing runs.  We acknowledge the work of the ALFALFA collaboration in observing, processing, and extracting \hi\ sources. We are grateful to the staff members at WIYN, NOIRLab, and Indiana University Pervasive Technology Institute for designing and implementing the ODI Pipeline, Portal, and Archive (ODI-PPA) and assisting us with the pODI and ODI image reduction. 
K.L.R., N.J.S., and W.F.J. were supported by NSF grant AST-1615483. We thank the Indiana University (IU) College of Arts and Sciences for funding IU's share of the WIYN telescope.
This research has made use of the NASA/IPAC Extragalactic Database (NED), which is funded by the National Aeronautics and Space
Administration and operated by the California Institute of Technology.
\end{acknowledgments}

%

\vspace{5mm}
\facilities{WIYN (pODI and ODI), Arecibo (ALFA), WSRT}

\software{ODI-PPA \citep{gopu14a}, QuickReduce pipeline \citep{kotulla14a}, IRAF \citep{tody86a,tody93a}, Python (https://www.python.org), Astropy \citep{astropy13a,astropy18a}}


\bibliography{uchvcs_odi}{}

\begin{thebibliography}{}
\expandafter\ifx\csname natexlab\endcsname\relax\def\natexlab#1{#1}\fi
\providecommand{\url}[1]{\href{#1}{#1}}
\providecommand{\dodoi}[1]{doi:~\href{http://doi.org/#1}{\nolinkurl{#1}}}
\providecommand{\doeprint}[1]{\href{http://ascl.net/#1}{\nolinkurl{http://ascl.net/#1}}}
\providecommand{\doarXiv}[1]{\href{https://arxiv.org/abs/#1}{\nolinkurl{https://arxiv.org/abs/#1}}}

\bibitem[{{Adams} {et~al.}(2013){Adams}, {Giovanelli}, \& {Haynes}}]{adams13a}
{Adams}, E. A.~K., {Giovanelli}, R., \& {Haynes}, M.~P. 2013, \apj, 768, 77,
  \dodoi{10.1088/0004-637X/768/1/77}

\bibitem[{{Adams} {et~al.}(2016){Adams}, {Oosterloo}, {Cannon}, {Giovanelli},
  \& {Haynes}}]{adams16a}
{Adams}, E. A.~K., {Oosterloo}, T.~A., {Cannon}, J.~M., {Giovanelli}, R., \&
  {Haynes}, M.~P. 2016, \aap, 596, A117, \dodoi{10.1051/0004-6361/201629262}

\bibitem[{{Adams} {et~al.}(2015){Adams}, {Cannon}, {Rhode}, {Janesh},
  {Janowiecki}, {Leisman}, {Giovanelli}, {Haynes}, {Oosterloo}, {Salzer}, \&
  {Zaidi}}]{adams15a}
{Adams}, E.~A.~K., {Cannon}, J.~M., {Rhode}, K.~L., {et~al.} 2015, \aap, 580,
  A134, \dodoi{10.1051/0004-6361/201526857}

\bibitem[{{Applebaum} {et~al.}(2021){Applebaum}, {Brooks}, {Christensen},
  {Munshi}, {Quinn}, {Shen}, \& {Tremmel}}]{applebaum21a}
{Applebaum}, E., {Brooks}, A.~M., {Christensen}, C.~R., {et~al.} 2021, \apj,
  906, 96, \dodoi{10.3847/1538-4357/abcafa}

\bibitem[{{Astropy Collaboration} {et~al.}(2013){Astropy Collaboration},
  {Robitaille}, {Tollerud}, {Greenfield}, {Droettboom}, {Bray}, {Aldcroft},
  {Davis}, {Ginsburg}, {Price-Whelan}, {Kerzendorf}, {Conley}, {Crighton},
  {Barbary}, {Muna}, {Ferguson}, {Grollier}, {Parikh}, {Nair}, {Unther},
  {Deil}, {Woillez}, {Conseil}, {Kramer}, {Turner}, {Singer}, {Fox}, {Weaver},
  {Zabalza}, {Edwards}, {Azalee Bostroem}, {Burke}, {Casey}, {Crawford},
  {Dencheva}, {Ely}, {Jenness}, {Labrie}, {Lim}, {Pierfederici}, {Pontzen},
  {Ptak}, {Refsdal}, {Servillat}, \& {Streicher}}]{astropy13a}
{Astropy Collaboration}, {Robitaille}, T.~P., {Tollerud}, E.~J., {et~al.} 2013,
  \aap, 558, A33, \dodoi{10.1051/0004-6361/201322068}

\bibitem[{{Astropy Collaboration} {et~al.}(2018){Astropy Collaboration},
  {Price-Whelan}, {Sip{\H{o}}cz}, {G{\"u}nther}, {Lim}, {Crawford}, {Conseil},
  {Shupe}, {Craig}, {Dencheva}, {Ginsburg}, {VanderPlas}, {Bradley},
  {P{\'e}rez-Su{\'a}rez}, {de Val-Borro}, {Aldcroft}, {Cruz}, {Robitaille},
  {Tollerud}, {Ardelean}, {Babej}, {Bach}, {Bachetti}, {Bakanov}, {Bamford},
  {Barentsen}, {Barmby}, {Baumbach}, {Berry}, {Biscani}, {Boquien}, {Bostroem},
  {Bouma}, {Brammer}, {Bray}, {Breytenbach}, {Buddelmeijer}, {Burke},
  {Calderone}, {Cano Rodr{\'\i}guez}, {Cara}, {Cardoso}, {Cheedella}, {Copin},
  {Corrales}, {Crichton}, {D'Avella}, {Deil}, {Depagne}, {Dietrich}, {Donath},
  {Droettboom}, {Earl}, {Erben}, {Fabbro}, {Ferreira}, {Finethy}, {Fox},
  {Garrison}, {Gibbons}, {Goldstein}, {Gommers}, {Greco}, {Greenfield},
  {Groener}, {Grollier}, {Hagen}, {Hirst}, {Homeier}, {Horton}, {Hosseinzadeh},
  {Hu}, {Hunkeler}, {Ivezi{\'c}}, {Jain}, {Jenness}, {Kanarek}, {Kendrew},
  {Kern}, {Kerzendorf}, {Khvalko}, {King}, {Kirkby}, {Kulkarni}, {Kumar},
  {Lee}, {Lenz}, {Littlefair}, {Ma}, {Macleod}, {Mastropietro}, {McCully},
  {Montagnac}, {Morris}, {Mueller}, {Mumford}, {Muna}, {Murphy}, {Nelson},
  {Nguyen}, {Ninan}, {N{\"o}the}, {Ogaz}, {Oh}, {Parejko}, {Parley}, {Pascual},
  {Patil}, {Patil}, {Plunkett}, {Prochaska}, {Rastogi}, {Reddy Janga},
  {Sabater}, {Sakurikar}, {Seifert}, {Sherbert}, {Sherwood-Taylor}, {Shih},
  {Sick}, {Silbiger}, {Singanamalla}, {Singer}, {Sladen}, {Sooley},
  {Sornarajah}, {Streicher}, {Teuben}, {Thomas}, {Tremblay}, {Turner},
  {Terr{\'o}n}, {van Kerkwijk}, {de la Vega}, {Watkins}, {Weaver}, {Whitmore},
  {Woillez}, {Zabalza}, \& {Astropy Contributors}}]{astropy18a}
{Astropy Collaboration}, {Price-Whelan}, A.~M., {Sip{\H{o}}cz}, B.~M., {et~al.}
  2018, \aj, 156, 123, \dodoi{10.3847/1538-3881/aabc4f}

\bibitem[{{Baumgardt} \& {Hilker}(2018)}]{baumgardt18a}
{Baumgardt}, H., \& {Hilker}, M. 2018, \mnras, 478, 1520,
  \dodoi{10.1093/mnras/sty1057}

\bibitem[{{Bell} {et~al.}(2003){Bell}, {McIntosh}, {Katz}, \&
  {Weinberg}}]{bell03a}
{Bell}, E.~F., {McIntosh}, D.~H., {Katz}, N., \& {Weinberg}, M.~D. 2003, \apjs,
  149, 289, \dodoi{10.1086/378847}

\bibitem[{{Bellazzini} {et~al.}(2015{\natexlab{a}}){Bellazzini}, {Magrini},
  {Mucciarelli}, {Beccari}, {Ibata}, {Battaglia}, {Martin}, {Testa}, {Fumana},
  {Marchetti}, {Correnti}, \& {Fraternali}}]{bellazzini15a}
{Bellazzini}, M., {Magrini}, L., {Mucciarelli}, A., {et~al.}
  2015{\natexlab{a}}, \apjl, 800, L15, \dodoi{10.1088/2041-8205/800/1/L15}

\bibitem[{{Bellazzini} {et~al.}(2015{\natexlab{b}}){Bellazzini}, {Beccari},
  {Battaglia}, {Martin}, {Testa}, {Ibata}, {Correnti}, {Cusano}, \&
  {Sani}}]{bellazzini15b}
{Bellazzini}, M., {Beccari}, G., {Battaglia}, G., {et~al.} 2015{\natexlab{b}},
  \aap, 575, A126, \dodoi{10.1051/0004-6361/201425262}

\bibitem[{{Bennet} {et~al.}(2022){Bennet}, {Sand}, {Crnojevi{\'c}}, {Weisz},
  {Caldwell}, {Guhathakurta}, {Hargis}, {Karunakaran}, {Mutlu-Pakdil},
  {Olszewski}, {Salzer}, {Seth}, {Simon}, {Spekkens}, {Stark}, {Strader},
  {Tollerud}, {Toloba}, \& {Willman}}]{bennett22a}
{Bennet}, P., {Sand}, D.~J., {Crnojevi{\'c}}, D., {et~al.} 2022, \apj, 924, 98,
  \dodoi{10.3847/1538-4357/ac356c}

\bibitem[{{Birkinshaw} {et~al.}(1983){Birkinshaw}, {Ho}, \&
  {Baud}}]{birkinshaw83a}
{Birkinshaw}, M., {Ho}, P.~T.~P., \& {Baud}, B. 1983, \aap, 125, 271

\bibitem[{{Bralts-Kelly} {et~al.}(2020){Bralts-Kelly}, {Paine}, {Adams},
  {Cannon}, {Giovanelli}, {Haynes}, {Janesh}, {Janowiecki}, {Oosterloo},
  {Rhode}, \& {Salzer}}]{bralts20a}
{Bralts-Kelly}, L., {Paine}, S., {Adams}, E.~A., {et~al.} 2020, in American
  Astronomical Society Meeting Abstracts, Vol. 235, American Astronomical
  Society Meeting Abstracts \#235, 168.01

\bibitem[{{Bullock} \& {Boylan-Kolchin}(2017)}]{bullock17a}
{Bullock}, J.~S., \& {Boylan-Kolchin}, M. 2017, \araa, 55, 343,
  \dodoi{10.1146/annurev-astro-091916-055313}

\bibitem[{{Chambers} {et~al.}(2016){Chambers}, {Magnier}, {Metcalfe},
  {Flewelling}, {Huber}, {Waters}, {Denneau}, {Draper}, {Farrow}, {Finkbeiner},
  {Holmberg}, {Koppenhoefer}, {Price}, {Rest}, {Saglia}, {Schlafly}, {Smartt},
  {Sweeney}, {Wainscoat}, {Burgett}, {Chastel}, {Grav}, {Heasley}, {Hodapp},
  {Jedicke}, {Kaiser}, {Kudritzki}, {Luppino}, {Lupton}, {Monet}, {Morgan},
  {Onaka}, {Shiao}, {Stubbs}, {Tonry}, {White}, {Ba{\~n}ados}, {Bell},
  {Bender}, {Bernard}, {Boegner}, {Boffi}, {Botticella}, {Calamida},
  {Casertano}, {Chen}, {Chen}, {Cole}, {Deacon}, {Frenk}, {Fitzsimmons},
  {Gezari}, {Gibbs}, {Goessl}, {Goggia}, {Gourgue}, {Goldman}, {Grant},
  {Grebel}, {Hambly}, {Hasinger}, {Heavens}, {Heckman}, {Henderson}, {Henning},
  {Holman}, {Hopp}, {Ip}, {Isani}, {Jackson}, {Keyes}, {Koekemoer}, {Kotak},
  {Le}, {Liska}, {Long}, {Lucey}, {Liu}, {Martin}, {Masci}, {McLean}, {Mindel},
  {Misra}, {Morganson}, {Murphy}, {Obaika}, {Narayan}, {Nieto-Santisteban},
  {Norberg}, {Peacock}, {Pier}, {Postman}, {Primak}, {Rae}, {Rai}, {Riess},
  {Riffeser}, {Rix}, {R{\"o}ser}, {Russel}, {Rutz}, {Schilbach}, {Schultz},
  {Scolnic}, {Strolger}, {Szalay}, {Seitz}, {Small}, {Smith}, {Soderblom},
  {Taylor}, {Thomson}, {Taylor}, {Thakar}, {Thiel}, {Thilker}, {Unger},
  {Urata}, {Valenti}, {Wagner}, {Walder}, {Walter}, {Watters}, {Werner},
  {Wood-Vasey}, \& {Wyse}}]{chambers16a}
{Chambers}, K.~C., {Magnier}, E.~A., {Metcalfe}, N., {et~al.} 2016, arXiv
  e-prints, arXiv:1612.05560, \dodoi{10.48550/arXiv.1612.05560}

\bibitem[{{Chonis} {et~al.}(2016){Chonis}, {Hill}, {Lee}, {Tuttle}, {Vattiat},
  {Drory}, {Indahl}, {Peterson}, \& {Ramsey}}]{chonis16a}
{Chonis}, T.~S., {Hill}, G.~J., {Lee}, H., {et~al.} 2016, in Society of
  Photo-Optical Instrumentation Engineers (SPIE) Conference Series, Vol. 9908,
  Ground-based and Airborne Instrumentation for Astronomy VI, ed. C.~J.
  {Evans}, L.~{Simard}, \& H.~{Takami}, 99084C, \dodoi{10.1117/12.2232209}

\bibitem[{{DeFelippis} {et~al.}(2019){DeFelippis}, {Putman}, \&
  {Tollerud}}]{defelippis19a}
{DeFelippis}, D., {Putman}, M., \& {Tollerud}, E. 2019, \apj, 879, 22,
  \dodoi{10.3847/1538-4357/ab1e57}

\bibitem[{{Dehnen} {et~al.}(2006){Dehnen}, {McLaughlin}, \&
  {Sachania}}]{dehnen06a}
{Dehnen}, W., {McLaughlin}, D.~E., \& {Sachania}, J. 2006, \mnras, 369, 1688,
  \dodoi{10.1111/j.1365-2966.2006.10404.x}

\bibitem[{{Eisenstein} {et~al.}(2011){Eisenstein}, {Weinberg}, {Agol},
  {Aihara}, {Allende Prieto}, {Anderson}, {Arns}, {Aubourg}, {Bailey},
  {Balbinot}, {Barkhouser}, {Beers}, {Berlind}, {Bickerton}, {Bizyaev},
  {Blanton}, {Bochanski}, {Bolton}, {Bosman}, {Bovy}, {Brandt}, {Breslauer},
  {Brewington}, {Brinkmann}, {Brown}, {Brownstein}, {Burger}, {Busca},
  {Campbell}, {Cargile}, {Carithers}, {Carlberg}, {Carr}, {Chang}, {Chen},
  {Chiappini}, {Comparat}, {Connolly}, {Cortes}, {Croft}, {Cunha}, {da Costa},
  {Davenport}, {Dawson}, {De Lee}, {Porto de Mello}, {de Simoni}, {Dean},
  {Dhital}, {Ealet}, {Ebelke}, {Edmondson}, {Eiting}, {Escoffier}, {Esposito},
  {Evans}, {Fan}, {Femen{\'\i}a Castell{\'a}}, {Dutra Ferreira}, {Fitzgerald},
  {Fleming}, {Font-Ribera}, {Ford}, {Frinchaboy}, {Garc{\'\i}a P{\'e}rez},
  {Gaudi}, {Ge}, {Ghezzi}, {Gillespie}, {Gilmore}, {Girardi}, {Gott}, {Gould},
  {Grebel}, {Gunn}, {Hamilton}, {Harding}, {Harris}, {Hawley}, {Hearty},
  {Hennawi}, {Gonz{\'a}lez Hern{\'a}ndez}, {Ho}, {Hogg}, {Holtzman},
  {Honscheid}, {Inada}, {Ivans}, {Jiang}, {Jiang}, {Johnson}, {Jordan},
  {Jordan}, {Kauffmann}, {Kazin}, {Kirkby}, {Klaene}, {Knapp}, {Kneib},
  {Kochanek}, {Koesterke}, {Kollmeier}, {Kron}, {Lampeitl}, {Lang}, {Lawler},
  {Le Goff}, {Lee}, {Lee}, {Leisenring}, {Lin}, {Liu}, {Long}, {Loomis},
  {Lucatello}, {Lundgren}, {Lupton}, {Ma}, {Ma}, {MacDonald}, {Mack},
  {Mahadevan}, {Maia}, {Majewski}, {Makler}, {Malanushenko}, {Malanushenko},
  {Mandelbaum}, {Maraston}, {Margala}, {Maseman}, {Masters}, {McBride},
  {McDonald}, {McGreer}, {McMahon}, {Mena Requejo}, {M{\'e}nard},
  {Miralda-Escud{\'e}}, {Morrison}, {Mullally}, {Muna}, {Murayama}, {Myers},
  {Naugle}, {Neto}, {Nguyen}, {Nichol}, {Nidever}, {O'Connell}, {Ogando},
  {Olmstead}, {Oravetz}, {Padmanabhan}, {Paegert}, {Palanque-Delabrouille},
  {Pan}, {Pandey}, {Parejko}, {P{\^a}ris}, {Pellegrini}, {Pepper}, {Percival},
  {Petitjean}, {Pfaffenberger}, {Pforr}, {Phleps}, {Pichon}, {Pieri}, {Prada},
  {Price-Whelan}, {Raddick}, {Ramos}, {Reid}, {Reyle}, {Rich}, {Richards},
  {Rieke}, {Rieke}, {Rix}, {Robin}, {Rocha-Pinto}, {Rockosi}, {Roe},
  {Rollinde}, {Ross}, {Ross}, {Rossetto}, {S{\'a}nchez}, {Santiago}, {Sayres},
  {Schiavon}, {Schlegel}, {Schlesinger}, {Schmidt}, {Schneider}, {Sellgren},
  {Shelden}, {Sheldon}, {Shetrone}, {Shu}, {Silverman}, {Simmerer}, {Simmons},
  {Sivarani}, {Skrutskie}, {Slosar}, {Smee}, {Smith}, {Snedden}, {Stassun},
  {Steele}, {Steinmetz}, {Stockett}, {Stollberg}, {Strauss}, {Szalay},
  {Tanaka}, {Thakar}, {Thomas}, {Tinker}, {Tofflemire}, {Tojeiro}, {Tremonti},
  {Vargas Maga{\~n}a}, {Verde}, {Vogt}, {Wake}, {Wan}, {Wang}, {Weaver},
  {White}, {White}, {Wilson}, {Wisniewski}, {Wood-Vasey}, {Yanny}, {Yasuda},
  {Y{\`e}che}, {York}, {Young}, {Zasowski}, {Zehavi}, \&
  {Zhao}}]{eisenstein11a}
{Eisenstein}, D.~J., {Weinberg}, D.~H., {Agol}, E., {et~al.} 2011, \aj, 142,
  72, \dodoi{10.1088/0004-6256/142/3/72}

\bibitem[{{Erkes} \& {Philip}(1975)}]{erkes75a}
{Erkes}, J.~W., \& {Philip}, A.~G.~D. 1975, \apj, 197, 533,
  \dodoi{10.1086/153540}

\bibitem[{{Evans} {et~al.}(2019){Evans}, {Castro}, {Gonzalez}, {Garcia},
  {Bastian}, {Cioni}, {Clark}, {Davies}, {Ferguson}, {Kamann}, {Lennon},
  {Patrick}, {Vink}, \& {Weisz}}]{evans19a}
{Evans}, C.~J., {Castro}, N., {Gonzalez}, O.~A., {et~al.} 2019, \aap, 622,
  A129, \dodoi{10.1051/0004-6361/201834145}

\bibitem[{{Faulkner} {et~al.}(1991){Faulkner}, {Scott}, {Wood}, \&
  {Wright}}]{faulkner91a}
{Faulkner}, D.~J., {Scott}, T.~R., {Wood}, P.~R., \& {Wright}, A.~E. 1991,
  \apjl, 374, L45, \dodoi{10.1086/186068}

\bibitem[{{Frank} \& {Gisler}(1976)}]{frank76a}
{Frank}, J., \& {Gisler}, G. 1976, \mnras, 176, 533,
  \dodoi{10.1093/mnras/176.3.533}

\bibitem[{{Freire} {et~al.}(2001){Freire}, {Kramer}, {Lyne}, {Camilo},
  {Manchester}, \& {D'Amico}}]{freire01a}
{Freire}, P.~C., {Kramer}, M., {Lyne}, A.~G., {et~al.} 2001, \apjl, 557, L105,
  \dodoi{10.1086/323248}

\bibitem[{{Gaia Collaboration} {et~al.}(2016){Gaia Collaboration}, {Prusti},
  {de Bruijne}, {Brown}, {Vallenari}, {Babusiaux}, {Bailer-Jones}, {Bastian},
  {Biermann}, {Evans}, {Eyer}, {Jansen}, {Jordi}, {Klioner}, {Lammers},
  {Lindegren}, {Luri}, {Mignard}, {Milligan}, {Panem}, {Poinsignon},
  {Pourbaix}, {Randich}, {Sarri}, {Sartoretti}, {Siddiqui}, {Soubiran},
  {Valette}, {van Leeuwen}, {Walton}, {Aerts}, {Arenou}, {Cropper}, {Drimmel},
  {H{\o}g}, {Katz}, {Lattanzi}, {O'Mullane}, {Grebel}, {Holland}, {Huc},
  {Passot}, {Bramante}, {Cacciari}, {Casta{\~n}eda}, {Chaoul}, {Cheek}, {De
  Angeli}, {Fabricius}, {Guerra}, {Hern{\'a}ndez}, {Jean-Antoine-Piccolo},
  {Masana}, {Messineo}, {Mowlavi}, {Nienartowicz}, {Ord{\'o}{\~n}ez-Blanco},
  {Panuzzo}, {Portell}, {Richards}, {Riello}, {Seabroke}, {Tanga},
  {Th{\'e}venin}, {Torra}, {Els}, {Gracia-Abril}, {Comoretto},
  {Garcia-Reinaldos}, {Lock}, {Mercier}, {Altmann}, {Andrae}, {Astraatmadja},
  {Bellas-Velidis}, {Benson}, {Berthier}, {Blomme}, {Busso}, {Carry},
  {Cellino}, {Clementini}, {Cowell}, {Creevey}, {Cuypers}, {Davidson}, {De
  Ridder}, {de Torres}, {Delchambre}, {Dell'Oro}, {Ducourant}, {Fr{\'e}mat},
  {Garc{\'\i}a-Torres}, {Gosset}, {Halbwachs}, {Hambly}, {Harrison}, {Hauser},
  {Hestroffer}, {Hodgkin}, {Huckle}, {Hutton}, {Jasniewicz}, {Jordan},
  {Kontizas}, {Korn}, {Lanzafame}, {Manteiga}, {Moitinho}, {Muinonen},
  {Osinde}, {Pancino}, {Pauwels}, {Petit}, {Recio-Blanco}, {Robin}, {Sarro},
  {Siopis}, {Smith}, {Smith}, {Sozzetti}, {Thuillot}, {van Reeven}, {Viala},
  {Abbas}, {Abreu Aramburu}, {Accart}, {Aguado}, {Allan}, {Allasia},
  {Altavilla}, {{\'A}lvarez}, {Alves}, {Anderson}, {Andrei}, {Anglada Varela},
  {Antiche}, {Antoja}, {Ant{\'o}n}, {Arcay}, {Atzei}, {Ayache}, {Bach},
  {Baker}, {Balaguer-N{\'u}{\~n}ez}, {Barache}, {Barata}, {Barbier}, {Barblan},
  {Baroni}, {Barrado y Navascu{\'e}s}, {Barros}, {Barstow}, {Becciani},
  {Bellazzini}, {Bellei}, {Bello Garc{\'\i}a}, {Belokurov}, {Bendjoya},
  {Berihuete}, {Bianchi}, {Bienaym{\'e}}, {Billebaud}, {Blagorodnova},
  {Blanco-Cuaresma}, {Boch}, {Bombrun}, {Borrachero}, {Bouquillon}, {Bourda},
  {Bouy}, {Bragaglia}, {Breddels}, {Brouillet}, {Br{\"u}semeister},
  {Bucciarelli}, {Budnik}, {Burgess}, {Burgon}, {Burlacu}, {Busonero}, {Buzzi},
  {Caffau}, {Cambras}, {Campbell}, {Cancelliere}, {Cantat-Gaudin}, {Carlucci},
  {Carrasco}, {Castellani}, {Charlot}, {Charnas}, {Charvet}, {Chassat},
  {Chiavassa}, {Clotet}, {Cocozza}, {Collins}, {Collins}, {Costigan}, {Crifo},
  {Cross}, {Crosta}, {Crowley}, {Dafonte}, {Damerdji}, {Dapergolas}, {David},
  {David}, {De Cat}, {de Felice}, {de Laverny}, {De Luise}, {De March}, {de
  Martino}, {de Souza}, {Debosscher}, {del Pozo}, {Delbo}, {Delgado},
  {Delgado}, {di Marco}, {Di Matteo}, {Diakite}, {Distefano}, {Dolding}, {Dos
  Anjos}, {Drazinos}, {Dur{\'a}n}, {Dzigan}, {Ecale}, {Edvardsson}, {Enke},
  {Erdmann}, {Escolar}, {Espina}, {Evans}, {Eynard Bontemps}, {Fabre},
  {Fabrizio}, {Faigler}, {Falc{\~a}o}, {Farr{\`a}s Casas}, {Faye}, {Federici},
  {Fedorets}, {Fern{\'a}ndez-Hern{\'a}ndez}, {Fernique}, {Fienga}, {Figueras},
  {Filippi}, {Findeisen}, {Fonti}, {Fouesneau}, {Fraile}, {Fraser}, {Fuchs},
  {Furnell}, {Gai}, {Galleti}, {Galluccio}, {Garabato}, {Garc{\'\i}a-Sedano},
  {Gar{\'e}}, {Garofalo}, {Garralda}, {Gavras}, {Gerssen}, {Geyer}, {Gilmore},
  {Girona}, {Giuffrida}, {Gomes}, {Gonz{\'a}lez-Marcos},
  {Gonz{\'a}lez-N{\'u}{\~n}ez}, {Gonz{\'a}lez-Vidal}, {Granvik}, {Guerrier},
  {Guillout}, {Guiraud}, {G{\'u}rpide}, {Guti{\'e}rrez-S{\'a}nchez}, {Guy},
  {Haigron}, {Hatzidimitriou}, {Haywood}, {Heiter}, {Helmi}, {Hobbs},
  {Hofmann}, {Holl}, {Holland}, {Hunt}, {Hypki}, {Icardi}, {Irwin}, {Jevardat
  de Fombelle}, {Jofr{\'e}}, {Jonker}, {Jorissen}, {Julbe}, {Karampelas},
  {Kochoska}, {Kohley}, {Kolenberg}, {Kontizas}, {Koposov}, {Kordopatis},
  {Koubsky}, {Kowalczyk}, {Krone-Martins}, {Kudryashova}, {Kull}, {Bachchan},
  {Lacoste-Seris}, {Lanza}, {Lavigne}, {Le Poncin-Lafitte}, {Lebreton},
  {Lebzelter}, {Leccia}, {Leclerc}, {Lecoeur-Taibi}, {Lemaitre}, {Lenhardt},
  {Leroux}, {Liao}, {Licata}, {Lindstr{\o}m}, {Lister}, {Livanou}, {Lobel},
  {L{\"o}ffler}, {L{\'o}pez}, {Lopez-Lozano}, {Lorenz}, {Loureiro},
  {MacDonald}, {Magalh{\~a}es Fernandes}, {Managau}, {Mann}, {Mantelet},
  {Marchal}, {Marchant}, {Marconi}, {Marie}, {Marinoni}, {Marrese},
  {Marschalk{\'o}}, {Marshall}, {Mart{\'\i}n-Fleitas}, {Martino}, {Mary},
  {Matijevi{\v{c}}}, {Mazeh}, {McMillan}, {Messina}, {Mestre}, {Michalik},
  {Millar}, {Miranda}, {Molina}, {Molinaro}, {Molinaro}, {Moln{\'a}r},
  {Moniez}, {Montegriffo}, {Monteiro}, {Mor}, {Mora}, {Morbidelli}, {Morel},
  {Morgenthaler}, {Morley}, {Morris}, {Mulone}, {Muraveva}, {Musella},
  {Narbonne}, {Nelemans}, {Nicastro}, {Noval}, {Ord{\'e}novic},
  {Ordieres-Mer{\'e}}, {Osborne}, {Pagani}, {Pagano}, {Pailler}, {Palacin},
  {Palaversa}, {Parsons}, {Paulsen}, {Pecoraro}, {Pedrosa}, {Pentik{\"a}inen},
  {Pereira}, {Pichon}, {Piersimoni}, {Pineau}, {Plachy}, {Plum}, {Poujoulet},
  {Pr{\v{s}}a}, {Pulone}, {Ragaini}, {Rago}, {Rambaux}, {Ramos-Lerate},
  {Ranalli}, {Rauw}, {Read}, {Regibo}, {Renk}, {Reyl{\'e}}, {Ribeiro},
  {Rimoldini}, {Ripepi}, {Riva}, {Rixon}, {Roelens}, {Romero-G{\'o}mez},
  {Rowell}, {Royer}, {Rudolph}, {Ruiz-Dern}, {Sadowski}, {Sagrist{\`a}
  Sell{\'e}s}, {Sahlmann}, {Salgado}, {Salguero}, {Sarasso}, {Savietto},
  {Schnorhk}, {Schultheis}, {Sciacca}, {Segol}, {Segovia}, {Segransan},
  {Serpell}, {Shih}, {Smareglia}, {Smart}, {Smith}, {Solano}, {Solitro},
  {Sordo}, {Soria Nieto}, {Souchay}, {Spagna}, {Spoto}, {Stampa}, {Steele},
  {Steidelm{\"u}ller}, {Stephenson}, {Stoev}, {Suess}, {S{\"u}veges}, {Surdej},
  {Szabados}, {Szegedi-Elek}, {Tapiador}, {Taris}, {Tauran}, {Taylor},
  {Teixeira}, {Terrett}, {Tingley}, {Trager}, {Turon}, {Ulla}, {Utrilla},
  {Valentini}, {van Elteren}, {Van Hemelryck}, {van Leeuwen}, {Varadi},
  {Vecchiato}, {Veljanoski}, {Via}, {Vicente}, {Vogt}, {Voss}, {Votruba},
  {Voutsinas}, {Walmsley}, {Weiler}, {Weingrill}, {Werner}, {Wevers},
  {Whitehead}, {Wyrzykowski}, {Yoldas}, {{\v{Z}}erjal}, {Zucker}, {Zurbach},
  {Zwitter}, {Alecu}, {Allen}, {Allende Prieto}, {Amorim},
  {Anglada-Escud{\'e}}, {Arsenijevic}, {Azaz}, {Balm}, {Beck}, {Bernstein},
  {Bigot}, {Bijaoui}, {Blasco}, {Bonfigli}, {Bono}, {Boudreault}, {Bressan},
  {Brown}, {Brunet}, {Bunclark}, {Buonanno}, {Butkevich}, {Carret}, {Carrion},
  {Chemin}, {Ch{\'e}reau}, {Corcione}, {Darmigny}, {de Boer}, {de Teodoro}, {de
  Zeeuw}, {Delle Luche}, {Domingues}, {Dubath}, {Fodor}, {Fr{\'e}zouls},
  {Fries}, {Fustes}, {Fyfe}, {Gallardo}, {Gallegos}, {Gardiol}, {Gebran},
  {Gomboc}, {G{\'o}mez}, {Grux}, {Gueguen}, {Heyrovsky}, {Hoar}, {Iannicola},
  {Isasi Parache}, {Janotto}, {Joliet}, {Jonckheere}, {Keil}, {Kim},
  {Klagyivik}, {Klar}, {Knude}, {Kochukhov}, {Kolka}, {Kos}, {Kutka}, {Lainey},
  {LeBouquin}, {Liu}, {Loreggia}, {Makarov}, {Marseille}, {Martayan},
  {Martinez-Rubi}, {Massart}, {Meynadier}, {Mignot}, {Munari}, {Nguyen},
  {Nordlander}, {Ocvirk}, {O'Flaherty}, {Olias Sanz}, {Ortiz}, {Osorio},
  {Oszkiewicz}, {Ouzounis}, {Palmer}, {Park}, {Pasquato}, {Peltzer}, {Peralta},
  {P{\'e}turaud}, {Pieniluoma}, {Pigozzi}, {Poels}, {Prat}, {Prod'homme},
  {Raison}, {Rebordao}, {Risquez}, {Rocca-Volmerange}, {Rosen}, {Ruiz-Fuertes},
  {Russo}, {Sembay}, {Serraller Vizcaino}, {Short}, {Siebert}, {Silva},
  {Sinachopoulos}, {Slezak}, {Soffel}, {Sosnowska}, {Strai{\v{z}}ys}, {ter
  Linden}, {Terrell}, {Theil}, {Tiede}, {Troisi}, {Tsalmantza}, {Tur},
  {Vaccari}, {Vachier}, {Valles}, {Van Hamme}, {Veltz}, {Virtanen}, {Wallut},
  {Wichmann}, {Wilkinson}, {Ziaeepour}, \& {Zschocke}}]{gaia16}
{Gaia Collaboration}, {Prusti}, T., {de Bruijne}, J.~H.~J., {et~al.} 2016,
  \aap, 595, A1, \dodoi{10.1051/0004-6361/201629272}

\bibitem[{{Gaia Collaboration} {et~al.}(2021){Gaia Collaboration}, {Brown},
  {Vallenari}, {Prusti}, {de Bruijne}, {Babusiaux}, {Biermann}, {Creevey},
  {Evans}, {Eyer}, {Hutton}, {Jansen}, {Jordi}, {Klioner}, {Lammers},
  {Lindegren}, {Luri}, {Mignard}, {Panem}, {Pourbaix}, {Randich}, {Sartoretti},
  {Soubiran}, {Walton}, {Arenou}, {Bailer-Jones}, {Bastian}, {Cropper},
  {Drimmel}, {Katz}, {Lattanzi}, {van Leeuwen}, {Bakker}, {Cacciari},
  {Casta{\~n}eda}, {De Angeli}, {Ducourant}, {Fabricius}, {Fouesneau},
  {Fr{\'e}mat}, {Guerra}, {Guerrier}, {Guiraud}, {Jean-Antoine Piccolo},
  {Masana}, {Messineo}, {Mowlavi}, {Nicolas}, {Nienartowicz}, {Pailler},
  {Panuzzo}, {Riclet}, {Roux}, {Seabroke}, {Sordo}, {Tanga}, {Th{\'e}venin},
  {Gracia-Abril}, {Portell}, {Teyssier}, {Altmann}, {Andrae}, {Bellas-Velidis},
  {Benson}, {Berthier}, {Blomme}, {Brugaletta}, {Burgess}, {Busso}, {Carry},
  {Cellino}, {Cheek}, {Clementini}, {Damerdji}, {Davidson}, {Delchambre},
  {Dell'Oro}, {Fern{\'a}ndez-Hern{\'a}ndez}, {Galluccio}, {Garc{\'\i}a-Lario},
  {Garcia-Reinaldos}, {Gonz{\'a}lez-N{\'u}{\~n}ez}, {Gosset}, {Haigron},
  {Halbwachs}, {Hambly}, {Harrison}, {Hatzidimitriou}, {Heiter},
  {Hern{\'a}ndez}, {Hestroffer}, {Hodgkin}, {Holl}, {Jan{\ss}en}, {Jevardat de
  Fombelle}, {Jordan}, {Krone-Martins}, {Lanzafame}, {L{\"o}ffler}, {Lorca},
  {Manteiga}, {Marchal}, {Marrese}, {Moitinho}, {Mora}, {Muinonen}, {Osborne},
  {Pancino}, {Pauwels}, {Petit}, {Recio-Blanco}, {Richards}, {Riello},
  {Rimoldini}, {Robin}, {Roegiers}, {Rybizki}, {Sarro}, {Siopis}, {Smith},
  {Sozzetti}, {Ulla}, {Utrilla}, {van Leeuwen}, {van Reeven}, {Abbas}, {Abreu
  Aramburu}, {Accart}, {Aerts}, {Aguado}, {Ajaj}, {Altavilla}, {{\'A}lvarez},
  {{\'A}lvarez Cid-Fuentes}, {Alves}, {Anderson}, {Anglada Varela}, {Antoja},
  {Audard}, {Baines}, {Baker}, {Balaguer-N{\'u}{\~n}ez}, {Balbinot}, {Balog},
  {Barache}, {Barbato}, {Barros}, {Barstow}, {Bartolom{\'e}}, {Bassilana},
  {Bauchet}, {Baudesson-Stella}, {Becciani}, {Bellazzini}, {Bernet}, {Bertone},
  {Bianchi}, {Blanco-Cuaresma}, {Boch}, {Bombrun}, {Bossini}, {Bouquillon},
  {Bragaglia}, {Bramante}, {Breedt}, {Bressan}, {Brouillet}, {Bucciarelli},
  {Burlacu}, {Busonero}, {Butkevich}, {Buzzi}, {Caffau}, {Cancelliere},
  {C{\'a}novas}, {Cantat-Gaudin}, {Carballo}, {Carlucci}, {Carnerero},
  {Carrasco}, {Casamiquela}, {Castellani}, {Castro-Ginard}, {Castro Sampol},
  {Chaoul}, {Charlot}, {Chemin}, {Chiavassa}, {Cioni}, {Comoretto}, {Cooper},
  {Cornez}, {Cowell}, {Crifo}, {Crosta}, {Crowley}, {Dafonte}, {Dapergolas},
  {David}, {David}, {de Laverny}, {De Luise}, {De March}, {De Ridder}, {de
  Souza}, {de Teodoro}, {de Torres}, {del Peloso}, {del Pozo}, {Delbo},
  {Delgado}, {Delgado}, {Delisle}, {Di Matteo}, {Diakite}, {Diener},
  {Distefano}, {Dolding}, {Eappachen}, {Edvardsson}, {Enke}, {Esquej}, {Fabre},
  {Fabrizio}, {Faigler}, {Fedorets}, {Fernique}, {Fienga}, {Figueras},
  {Fouron}, {Fragkoudi}, {Fraile}, {Franke}, {Gai}, {Garabato},
  {Garcia-Gutierrez}, {Garc{\'\i}a-Torres}, {Garofalo}, {Gavras}, {Gerlach},
  {Geyer}, {Giacobbe}, {Gilmore}, {Girona}, {Giuffrida}, {Gomel}, {Gomez},
  {Gonzalez-Santamaria}, {Gonz{\'a}lez-Vidal}, {Granvik},
  {Guti{\'e}rrez-S{\'a}nchez}, {Guy}, {Hauser}, {Haywood}, {Helmi}, {Hidalgo},
  {Hilger}, {H{\l}adczuk}, {Hobbs}, {Holland}, {Huckle}, {Jasniewicz},
  {Jonker}, {Juaristi Campillo}, {Julbe}, {Karbevska}, {Kervella}, {Khanna},
  {Kochoska}, {Kontizas}, {Kordopatis}, {Korn}, {Kostrzewa-Rutkowska},
  {Kruszy{\'n}ska}, {Lambert}, {Lanza}, {Lasne}, {Le Campion}, {Le Fustec},
  {Lebreton}, {Lebzelter}, {Leccia}, {Leclerc}, {Lecoeur-Taibi}, {Liao},
  {Licata}, {Lindstr{\o}m}, {Lister}, {Livanou}, {Lobel}, {Madrero Pardo},
  {Managau}, {Mann}, {Marchant}, {Marconi}, {Marcos Santos}, {Marinoni},
  {Marocco}, {Marshall}, {Martin Polo}, {Mart{\'\i}n-Fleitas}, {Masip},
  {Massari}, {Mastrobuono-Battisti}, {Mazeh}, {McMillan}, {Messina},
  {Michalik}, {Millar}, {Mints}, {Molina}, {Molinaro}, {Moln{\'a}r},
  {Montegriffo}, {Mor}, {Morbidelli}, {Morel}, {Morris}, {Mulone}, {Munoz},
  {Muraveva}, {Murphy}, {Musella}, {Noval}, {Ord{\'e}novic}, {Orr{\`u}},
  {Osinde}, {Pagani}, {Pagano}, {Palaversa}, {Palicio}, {Panahi}, {Pawlak},
  {Pe{\~n}alosa Esteller}, {Penttil{\"a}}, {Piersimoni}, {Pineau}, {Plachy},
  {Plum}, {Poggio}, {Poretti}, {Poujoulet}, {Pr{\v{s}}a}, {Pulone}, {Racero},
  {Ragaini}, {Rainer}, {Raiteri}, {Rambaux}, {Ramos}, {Ramos-Lerate}, {Re
  Fiorentin}, {Regibo}, {Reyl{\'e}}, {Ripepi}, {Riva}, {Rixon}, {Robichon},
  {Robin}, {Roelens}, {Rohrbasser}, {Romero-G{\'o}mez}, {Rowell}, {Royer},
  {Rybicki}, {Sadowski}, {Sagrist{\`a} Sell{\'e}s}, {Sahlmann}, {Salgado},
  {Salguero}, {Samaras}, {Sanchez Gimenez}, {Sanna}, {Santove{\~n}a},
  {Sarasso}, {Schultheis}, {Sciacca}, {Segol}, {Segovia}, {S{\'e}gransan},
  {Semeux}, {Shahaf}, {Siddiqui}, {Siebert}, {Siltala}, {Slezak}, {Smart},
  {Solano}, {Solitro}, {Souami}, {Souchay}, {Spagna}, {Spoto}, {Steele},
  {Steidelm{\"u}ller}, {Stephenson}, {S{\"u}veges}, {Szabados}, {Szegedi-Elek},
  {Taris}, {Tauran}, {Taylor}, {Teixeira}, {Thuillot}, {Tonello}, {Torra},
  {Torra}, {Turon}, {Unger}, {Vaillant}, {van Dillen}, {Vanel}, {Vecchiato},
  {Viala}, {Vicente}, {Voutsinas}, {Weiler}, {Wevers}, {Wyrzykowski}, {Yoldas},
  {Yvard}, {Zhao}, {Zorec}, {Zucker}, {Zurbach}, \& {Zwitter}}]{gaia21}
{Gaia Collaboration}, {Brown}, A.~G.~A., {Vallenari}, A., {et~al.} 2021, \aap,
  649, A1, \dodoi{10.1051/0004-6361/202039657}

\bibitem[{{Garrison-Kimmel} {et~al.}(2014){Garrison-Kimmel}, {Boylan-Kolchin},
  {Bullock}, \& {Lee}}]{garrison14a}
{Garrison-Kimmel}, S., {Boylan-Kolchin}, M., {Bullock}, J.~S., \& {Lee}, K.
  2014, \mnras, 438, 2578, \dodoi{10.1093/mnras/stt2377}

\bibitem[{{Giovanelli} {et~al.}(2010){Giovanelli}, {Haynes}, {Kent}, \&
  {Adams}}]{giovanelli10a}
{Giovanelli}, R., {Haynes}, M.~P., {Kent}, B.~R., \& {Adams}, E. A.~K. 2010,
  \apjl, 708, L22, \dodoi{10.1088/2041-8205/708/1/L22}

\bibitem[{{Giovanelli} {et~al.}(2005){Giovanelli}, {Haynes}, {Kent},
  {Perillat}, {Saintonge}, {Brosch}, {Catinella}, {Hoffman}, {Stierwalt},
  {Spekkens}, {Lerner}, {Masters}, {Momjian}, {Rosenberg}, {Springob},
  {Boselli}, {Charmandaris}, {Darling}, {Davies}, {Garcia Lambas}, {Gavazzi},
  {Giovanardi}, {Hardy}, {Hunt}, {Iovino}, {Karachentsev}, {Karachentseva},
  {Koopmann}, {Marinoni}, {Minchin}, {Muller}, {Putman}, {Pantoja}, {Salzer},
  {Scodeggio}, {Skillman}, {Solanes}, {Valotto}, {van Driel}, \& {van
  Zee}}]{giovanelli05a}
{Giovanelli}, R., {Haynes}, M.~P., {Kent}, B.~R., {et~al.} 2005, \aj, 130,
  2598, \dodoi{10.1086/497431}

\bibitem[{{Giovanelli} {et~al.}(2013){Giovanelli}, {Haynes}, {Adams}, {Cannon},
  {Rhode}, {Salzer}, {Skillman}, {Bernstein-Cooper}, \&
  {McQuinn}}]{giovanelli13a}
{Giovanelli}, R., {Haynes}, M.~P., {Adams}, E. A.~K., {et~al.} 2013, \aj, 146,
  15, \dodoi{10.1088/0004-6256/146/1/15}

\bibitem[{{Girardi} {et~al.}(2004){Girardi}, {Grebel}, {Odenkirchen}, \&
  {Chiosi}}]{girardi04a}
{Girardi}, L., {Grebel}, E.~K., {Odenkirchen}, M., \& {Chiosi}, C. 2004, \aap,
  422, 205, \dodoi{10.1051/0004-6361:20040250}

\bibitem[{{Gopu} {et~al.}(2014){Gopu}, {Hayashi}, {Young}, {Harbeck},
  {Boroson}, {Liu}, {Kotulla}, {Shaw}, {Henschel}, {Rajagopal}, {Stobie},
  {Knezek}, {Martin}, \& {Archbold}}]{gopu14a}
{Gopu}, A., {Hayashi}, S., {Young}, M.~D., {et~al.} 2014, in Society of
  Photo-Optical Instrumentation Engineers (SPIE) Conference Series, Vol. 9152,
  Software and Cyberinfrastructure for Astronomy III, ed. G.~{Chiozzi} \& N.~M.
  {Radziwill}, 91520E, \dodoi{10.1117/12.2057123}

\bibitem[{{Grogin} {et~al.}(2011){Grogin}, {Kocevski}, {Faber}, {Ferguson},
  {Koekemoer}, {Riess}, {Acquaviva}, {Alexander}, {Almaini}, {Ashby}, {Barden},
  {Bell}, {Bournaud}, {Brown}, {Caputi}, {Casertano}, {Cassata}, {Castellano},
  {Challis}, {Chary}, {Cheung}, {Cirasuolo}, {Conselice}, {Roshan Cooray},
  {Croton}, {Daddi}, {Dahlen}, {Dav{\'e}}, {de Mello}, {Dekel}, {Dickinson},
  {Dolch}, {Donley}, {Dunlop}, {Dutton}, {Elbaz}, {Fazio}, {Filippenko},
  {Finkelstein}, {Fontana}, {Gardner}, {Garnavich}, {Gawiser}, {Giavalisco},
  {Grazian}, {Guo}, {Hathi}, {H{\"a}ussler}, {Hopkins}, {Huang}, {Huang},
  {Jha}, {Kartaltepe}, {Kirshner}, {Koo}, {Lai}, {Lee}, {Li}, {Lotz}, {Lucas},
  {Madau}, {McCarthy}, {McGrath}, {McIntosh}, {McLure}, {Mobasher},
  {Moustakas}, {Mozena}, {Nandra}, {Newman}, {Niemi}, {Noeske}, {Papovich},
  {Pentericci}, {Pope}, {Primack}, {Rajan}, {Ravindranath}, {Reddy}, {Renzini},
  {Rix}, {Robaina}, {Rodney}, {Rosario}, {Rosati}, {Salimbeni}, {Scarlata},
  {Siana}, {Simard}, {Smidt}, {Somerville}, {Spinrad}, {Straughn}, {Strolger},
  {Telford}, {Teplitz}, {Trump}, {van der Wel}, {Villforth}, {Wechsler},
  {Weiner}, {Wiklind}, {Wild}, {Wilson}, {Wuyts}, {Yan}, \& {Yun}}]{grogin11a}
{Grogin}, N.~A., {Kocevski}, D.~D., {Faber}, S.~M., {et~al.} 2011, \apjs, 197,
  35, \dodoi{10.1088/0067-0049/197/2/35}

\bibitem[{{Harris}(1996)}]{harris96a}
{Harris}, W.~E. 1996, \aj, 112, 1487, \dodoi{10.1086/118116}

\bibitem[{{Haynes} {et~al.}(2011){Haynes}, {Giovanelli}, {Martin}, {Hess},
  {Saintonge}, {Adams}, {Hallenbeck}, {Hoffman}, {Huang}, {Kent}, {Koopmann},
  {Papastergis}, {Stierwalt}, {Balonek}, {Craig}, {Higdon}, {Kornreich},
  {Miller}, {O'Donoghue}, {Olowin}, {Rosenberg}, {Spekkens}, {Troischt}, \&
  {Wilcots}}]{haynes11a}
{Haynes}, M.~P., {Giovanelli}, R., {Martin}, A.~M., {et~al.} 2011, \aj, 142,
  170, \dodoi{10.1088/0004-6256/142/5/170}

\bibitem[{{Haynes} {et~al.}(2018){Haynes}, {Giovanelli}, {Kent}, {Adams},
  {Balonek}, {Craig}, {Fertig}, {Finn}, {Giovanardi}, {Hallenbeck}, {Hess},
  {Hoffman}, {Huang}, {Jones}, {Koopmann}, {Kornreich}, {Leisman}, {Miller},
  {Moorman}, {O'Connor}, {O'Donoghue}, {Papastergis}, {Troischt}, {Stark}, \&
  {Xiao}}]{haynes18a}
{Haynes}, M.~P., {Giovanelli}, R., {Kent}, B.~R., {et~al.} 2018, \apj, 861, 49,
  \dodoi{10.3847/1538-4357/aac956}

\bibitem[{{Heiles} \& {Henry}(1966)}]{heiles66a}
{Heiles}, C., \& {Henry}, R.~C. 1966, \apj, 146, 953, \dodoi{10.1086/148971}

\bibitem[{{Hilker}(2006)}]{hilker06a}
{Hilker}, M. 2006, \aap, 448, 171, \dodoi{10.1051/0004-6361:20054327}

\bibitem[{{Irwin} {et~al.}(2007){Irwin}, {Belokurov}, {Evans}, {Ryan-Weber},
  {de Jong}, {Koposov}, {Zucker}, {Hodgkin}, {Gilmore}, {Prema}, {Hebb},
  {Begum}, {Fellhauer}, {Hewett}, {Kennicutt}, {Wilkinson}, {Bramich},
  {Vidrih}, {Rix}, {Beers}, {Barentine}, {Brewington}, {Harvanek},
  {Krzesinski}, {Long}, {Nitta}, \& {Snedden}}]{irwin07a}
{Irwin}, M.~J., {Belokurov}, V., {Evans}, N.~W., {et~al.} 2007, \apjl, 656,
  L13, \dodoi{10.1086/512183}

\bibitem[{{Ivezi{\'c}} {et~al.}(2019){Ivezi{\'c}}, {Kahn}, {Tyson}, {Abel},
  {Acosta}, {Allsman}, {Alonso}, {AlSayyad}, {Anderson}, {Andrew}, {Angel},
  {Angeli}, {Ansari}, {Antilogus}, {Araujo}, {Armstrong}, {Arndt}, {Astier},
  {Aubourg}, {Auza}, {Axelrod}, {Bard}, {Barr}, {Barrau}, {Bartlett}, {Bauer},
  {Bauman}, {Baumont}, {Bechtol}, {Bechtol}, {Becker}, {Becla}, {Beldica},
  {Bellavia}, {Bianco}, {Biswas}, {Blanc}, {Blazek}, {Blandford}, {Bloom},
  {Bogart}, {Bond}, {Booth}, {Borgland}, {Borne}, {Bosch}, {Boutigny},
  {Brackett}, {Bradshaw}, {Brandt}, {Brown}, {Bullock}, {Burchat}, {Burke},
  {Cagnoli}, {Calabrese}, {Callahan}, {Callen}, {Carlin}, {Carlson},
  {Chandrasekharan}, {Charles-Emerson}, {Chesley}, {Cheu}, {Chiang}, {Chiang},
  {Chirino}, {Chow}, {Ciardi}, {Claver}, {Cohen-Tanugi}, {Cockrum}, {Coles},
  {Connolly}, {Cook}, {Cooray}, {Covey}, {Cribbs}, {Cui}, {Cutri}, {Daly},
  {Daniel}, {Daruich}, {Daubard}, {Daues}, {Dawson}, {Delgado}, {Dellapenna},
  {de Peyster}, {de Val-Borro}, {Digel}, {Doherty}, {Dubois},
  {Dubois-Felsmann}, {Durech}, {Economou}, {Eifler}, {Eracleous}, {Emmons},
  {Fausti Neto}, {Ferguson}, {Figueroa}, {Fisher-Levine}, {Focke}, {Foss},
  {Frank}, {Freemon}, {Gangler}, {Gawiser}, {Geary}, {Gee}, {Geha}, {Gessner},
  {Gibson}, {Gilmore}, {Glanzman}, {Glick}, {Goldina}, {Goldstein}, {Goodenow},
  {Graham}, {Gressler}, {Gris}, {Guy}, {Guyonnet}, {Haller}, {Harris},
  {Hascall}, {Haupt}, {Hernandez}, {Herrmann}, {Hileman}, {Hoblitt}, {Hodgson},
  {Hogan}, {Howard}, {Huang}, {Huffer}, {Ingraham}, {Innes}, {Jacoby}, {Jain},
  {Jammes}, {Jee}, {Jenness}, {Jernigan}, {Jevremovi{\'c}}, {Johns}, {Johnson},
  {Johnson}, {Jones}, {Juramy-Gilles}, {Juri{\'c}}, {Kalirai}, {Kallivayalil},
  {Kalmbach}, {Kantor}, {Karst}, {Kasliwal}, {Kelly}, {Kessler}, {Kinnison},
  {Kirkby}, {Knox}, {Kotov}, {Krabbendam}, {Krughoff}, {Kub{\'a}nek},
  {Kuczewski}, {Kulkarni}, {Ku}, {Kurita}, {Lage}, {Lambert}, {Lange},
  {Langton}, {Le Guillou}, {Levine}, {Liang}, {Lim}, {Lintott}, {Long},
  {Lopez}, {Lotz}, {Lupton}, {Lust}, {MacArthur}, {Mahabal}, {Mandelbaum},
  {Markiewicz}, {Marsh}, {Marshall}, {Marshall}, {May}, {McKercher}, {McQueen},
  {Meyers}, {Migliore}, {Miller}, {Mills}, {Miraval}, {Moeyens}, {Moolekamp},
  {Monet}, {Moniez}, {Monkewitz}, {Montgomery}, {Morrison}, {Mueller},
  {Muller}, {Mu{\~n}oz Arancibia}, {Neill}, {Newbry}, {Nief}, {Nomerotski},
  {Nordby}, {O'Connor}, {Oliver}, {Olivier}, {Olsen}, {O'Mullane}, {Ortiz},
  {Osier}, {Owen}, {Pain}, {Palecek}, {Parejko}, {Parsons}, {Pease},
  {Peterson}, {Peterson}, {Petravick}, {Libby Petrick}, {Petry},
  {Pierfederici}, {Pietrowicz}, {Pike}, {Pinto}, {Plante}, {Plate}, {Plutchak},
  {Price}, {Prouza}, {Radeka}, {Rajagopal}, {Rasmussen}, {Regnault}, {Reil},
  {Reiss}, {Reuter}, {Ridgway}, {Riot}, {Ritz}, {Robinson}, {Roby}, {Roodman},
  {Rosing}, {Roucelle}, {Rumore}, {Russo}, {Saha}, {Sassolas}, {Schalk},
  {Schellart}, {Schindler}, {Schmidt}, {Schneider}, {Schneider}, {Schoening},
  {Schumacher}, {Schwamb}, {Sebag}, {Selvy}, {Sembroski}, {Seppala}, {Serio},
  {Serrano}, {Shaw}, {Shipsey}, {Sick}, {Silvestri}, {Slater}, {Smith},
  {Smith}, {Sobhani}, {Soldahl}, {Storrie-Lombardi}, {Stover}, {Strauss},
  {Street}, {Stubbs}, {Sullivan}, {Sweeney}, {Swinbank}, {Szalay}, {Takacs},
  {Tether}, {Thaler}, {Thayer}, {Thomas}, {Thornton}, {Thukral}, {Tice},
  {Trilling}, {Turri}, {Van Berg}, {Vanden Berk}, {Vetter}, {Virieux},
  {Vucina}, {Wahl}, {Walkowicz}, {Walsh}, {Walter}, {Wang}, {Wang}, {Warner},
  {Wiecha}, {Willman}, {Winters}, {Wittman}, {Wolff}, {Wood-Vasey}, {Wu},
  {Xin}, {Yoachim}, \& {Zhan}}]{ivezic19a}
{Ivezi{\'c}}, {\v{Z}}., {Kahn}, S.~M., {Tyson}, J.~A., {et~al.} 2019, \apj,
  873, 111, \dodoi{10.3847/1538-4357/ab042c}

\bibitem[{{Janesh} {et~al.}(2017){Janesh}, {Rhode}, {Salzer}, {Janowiecki},
  {Adams}, {Haynes}, {Giovanelli}, \& {Cannon}}]{janesh17a}
{Janesh}, W., {Rhode}, K.~L., {Salzer}, J.~J., {et~al.} 2017, \apjl, 837, L16,
  \dodoi{10.3847/2041-8213/aa62a1}

\bibitem[{{Janesh} {et~al.}(2019){Janesh}, {Rhode}, {Salzer}, {Janowiecki},
  {Adams}, {Haynes}, {Giovanelli}, \& {Cannon}}]{janesh19a}
---. 2019, \aj, 157, 183, \dodoi{10.3847/1538-3881/ab12d3}

\bibitem[{{Janesh} {et~al.}(2015){Janesh}, {Rhode}, {Salzer}, {Janowiecki},
  {Adams}, {Haynes}, {Giovanelli}, {Cannon}, \& {Mu{\~n}oz}}]{janesh15a}
---. 2015, \apj, 811, 35, \dodoi{10.1088/0004-637X/811/1/35}

\bibitem[{{Janesh}(2018)}]{janesh18a}
{Janesh}, W.~F. 2018, PhD thesis, Indiana University, Bloomington

\bibitem[{{Jester} {et~al.}(2005){Jester}, {Schneider}, {Richards}, {Green},
  {Schmidt}, {Hall}, {Strauss}, {Vanden Berk}, {Stoughton}, {Gunn},
  {Brinkmann}, {Kent}, {Smith}, {Tucker}, \& {Yanny}}]{jester05a}
{Jester}, S., {Schneider}, D.~P., {Richards}, G.~T., {et~al.} 2005, \aj, 130,
  873, \dodoi{10.1086/432466}

\bibitem[{{Kerr} \& {Knapp}(1972)}]{kerr72a}
{Kerr}, F.~J., \& {Knapp}, G.~R. 1972, \aj, 77, 573, \dodoi{10.1086/111320}

\bibitem[{{Knapp} {et~al.}(1973){Knapp}, {Rose}, \& {Kerr}}]{knapp73a}
{Knapp}, G.~R., {Rose}, W.~K., \& {Kerr}, F.~J. 1973, \apj, 186, 831,
  \dodoi{10.1086/152550}

\bibitem[{{Koch} {et~al.}(2009){Koch}, {C{\^o}t{\'e}}, \&
  {McWilliam}}]{koch09a}
{Koch}, A., {C{\^o}t{\'e}}, P., \& {McWilliam}, A. 2009, \aap, 506, 729,
  \dodoi{10.1051/0004-6361/200912819}

\bibitem[{{Koekemoer} {et~al.}(2011){Koekemoer}, {Faber}, {Ferguson}, {Grogin},
  {Kocevski}, {Koo}, {Lai}, {Lotz}, {Lucas}, {McGrath}, {Ogaz}, {Rajan},
  {Riess}, {Rodney}, {Strolger}, {Casertano}, {Castellano}, {Dahlen},
  {Dickinson}, {Dolch}, {Fontana}, {Giavalisco}, {Grazian}, {Guo}, {Hathi},
  {Huang}, {van der Wel}, {Yan}, {Acquaviva}, {Alexander}, {Almaini}, {Ashby},
  {Barden}, {Bell}, {Bournaud}, {Brown}, {Caputi}, {Cassata}, {Challis},
  {Chary}, {Cheung}, {Cirasuolo}, {Conselice}, {Roshan Cooray}, {Croton},
  {Daddi}, {Dav{\'e}}, {de Mello}, {de Ravel}, {Dekel}, {Donley}, {Dunlop},
  {Dutton}, {Elbaz}, {Fazio}, {Filippenko}, {Finkelstein}, {Frazer}, {Gardner},
  {Garnavich}, {Gawiser}, {Gruetzbauch}, {Hartley}, {H{\"a}ussler},
  {Herrington}, {Hopkins}, {Huang}, {Jha}, {Johnson}, {Kartaltepe},
  {Khostovan}, {Kirshner}, {Lani}, {Lee}, {Li}, {Madau}, {McCarthy},
  {McIntosh}, {McLure}, {McPartland}, {Mobasher}, {Moreira}, {Mortlock},
  {Moustakas}, {Mozena}, {Nandra}, {Newman}, {Nielsen}, {Niemi}, {Noeske},
  {Papovich}, {Pentericci}, {Pope}, {Primack}, {Ravindranath}, {Reddy},
  {Renzini}, {Rix}, {Robaina}, {Rosario}, {Rosati}, {Salimbeni}, {Scarlata},
  {Siana}, {Simard}, {Smidt}, {Snyder}, {Somerville}, {Spinrad}, {Straughn},
  {Telford}, {Teplitz}, {Trump}, {Vargas}, {Villforth}, {Wagner}, {Wandro},
  {Wechsler}, {Weiner}, {Wiklind}, {Wild}, {Wilson}, {Wuyts}, \&
  {Yun}}]{koekemoer11a}
{Koekemoer}, A.~M., {Faber}, S.~M., {Ferguson}, H.~C., {et~al.} 2011, \apjs,
  197, 36, \dodoi{10.1088/0067-0049/197/2/36}

\bibitem[{{Kotulla}(2014)}]{kotulla14a}
{Kotulla}, R. 2014, in Astronomical Society of the Pacific Conference Series,
  Vol. 485, Astronomical Data Analysis Software and Systems XXIII, ed.
  N.~{Manset} \& P.~{Forshay}, 375

\bibitem[{{LSST Science Collaboration} {et~al.}(2009){LSST Science
  Collaboration}, {Abell}, {Allison}, {Anderson}, {Andrew}, {Angel}, {Armus},
  {Arnett}, {Asztalos}, {Axelrod}, {Bailey}, {Ballantyne}, {Bankert},
  {Barkhouse}, {Barr}, {Barrientos}, {Barth}, {Bartlett}, {Becker}, {Becla},
  {Beers}, {Bernstein}, {Biswas}, {Blanton}, {Bloom}, {Bochanski}, {Boeshaar},
  {Borne}, {Bradac}, {Brandt}, {Bridge}, {Brown}, {Brunner}, {Bullock},
  {Burgasser}, {Burge}, {Burke}, {Cargile}, {Chandrasekharan}, {Chartas},
  {Chesley}, {Chu}, {Cinabro}, {Claire}, {Claver}, {Clowe}, {Connolly}, {Cook},
  {Cooke}, {Cooray}, {Covey}, {Culliton}, {de Jong}, {de Vries}, {Debattista},
  {Delgado}, {Dell'Antonio}, {Dhital}, {Di Stefano}, {Dickinson}, {Dilday},
  {Djorgovski}, {Dobler}, {Donalek}, {Dubois-Felsmann}, {Durech},
  {Eliasdottir}, {Eracleous}, {Eyer}, {Falco}, {Fan}, {Fassnacht}, {Ferguson},
  {Fernandez}, {Fields}, {Finkbeiner}, {Figueroa}, {Fox}, {Francke}, {Frank},
  {Frieman}, {Fromenteau}, {Furqan}, {Galaz}, {Gal-Yam}, {Garnavich},
  {Gawiser}, {Geary}, {Gee}, {Gibson}, {Gilmore}, {Grace}, {Green}, {Gressler},
  {Grillmair}, {Habib}, {Haggerty}, {Hamuy}, {Harris}, {Hawley}, {Heavens},
  {Hebb}, {Henry}, {Hileman}, {Hilton}, {Hoadley}, {Holberg}, {Holman},
  {Howell}, {Infante}, {Ivezic}, {Jacoby}, {Jain}, {R}, {Jedicke}, {Jee},
  {Garrett Jernigan}, {Jha}, {Johnston}, {Jones}, {Juric}, {Kaasalainen},
  {Styliani}, {Kafka}, {Kahn}, {Kaib}, {Kalirai}, {Kantor}, {Kasliwal},
  {Keeton}, {Kessler}, {Knezevic}, {Kowalski}, {Krabbendam}, {Krughoff},
  {Kulkarni}, {Kuhlman}, {Lacy}, {Lepine}, {Liang}, {Lien}, {Lira}, {Long},
  {Lorenz}, {Lotz}, {Lupton}, {Lutz}, {Macri}, {Mahabal}, {Mandelbaum},
  {Marshall}, {May}, {McGehee}, {Meadows}, {Meert}, {Milani}, {Miller},
  {Miller}, {Mills}, {Minniti}, {Monet}, {Mukadam}, {Nakar}, {Neill}, {Newman},
  {Nikolaev}, {Nordby}, {O'Connor}, {Oguri}, {Oliver}, {Olivier}, {Olsen},
  {Olsen}, {Olszewski}, {Oluseyi}, {Padilla}, {Parker}, {Pepper}, {Peterson},
  {Petry}, {Pinto}, {Pizagno}, {Popescu}, {Prsa}, {Radcka}, {Raddick},
  {Rasmussen}, {Rau}, {Rho}, {Rhoads}, {Richards}, {Ridgway}, {Robertson},
  {Roskar}, {Saha}, {Sarajedini}, {Scannapieco}, {Schalk}, {Schindler},
  {Schmidt}, {Schmidt}, {Schneider}, {Schumacher}, {Scranton}, {Sebag},
  {Seppala}, {Shemmer}, {Simon}, {Sivertz}, {Smith}, {Allyn Smith}, {Smith},
  {Spitz}, {Stanford}, {Stassun}, {Strader}, {Strauss}, {Stubbs}, {Sweeney},
  {Szalay}, {Szkody}, {Takada}, {Thorman}, {Trilling}, {Trimble}, {Tyson}, {Van
  Berg}, {Vanden Berk}, {VanderPlas}, {Verde}, {Vrsnak}, {Walkowicz},
  {Wandelt}, {Wang}, {Wang}, {Warner}, {Wechsler}, {West}, {Wiecha},
  {Williams}, {Willman}, {Wittman}, {Wolff}, {Wood-Vasey}, {Wozniak}, {Young},
  {Zentner}, \& {Zhan}}]{lsst09a}
{LSST Science Collaboration}, {Abell}, P.~A., {Allison}, J., {et~al.} 2009,
  arXiv e-prints, arXiv:0912.0201, \dodoi{10.48550/arXiv.0912.0201}

\bibitem[{{McConnachie}(2012)}]{mcconnachie12a}
{McConnachie}, A.~W. 2012, \aj, 144, 4, \dodoi{10.1088/0004-6256/144/1/4}

\bibitem[{{McQuinn} {et~al.}(2015){McQuinn}, {Skillman}, {Dolphin}, {Cannon},
  {Salzer}, {Rhode}, {Adams}, {Berg}, {Giovanelli}, {Girardi}, \&
  {Haynes}}]{mcquinn15b}
{McQuinn}, K. B.~W., {Skillman}, E.~D., {Dolphin}, A., {et~al.} 2015, \apj,
  812, 158, \dodoi{10.1088/0004-637X/812/2/158}

\bibitem[{{Paine} {et~al.}(2020){Paine}, {Bralts-Kelly}, {Adams}, {Cannon},
  {Giovanelli}, {Haynes}, {Janesh}, {Janowiecki}, {Oosterloo}, {Rhode}, \&
  {Salzer}}]{paine20a}
{Paine}, S., {Bralts-Kelly}, L., {Adams}, E., {et~al.} 2020, in American
  Astronomical Society Meeting Abstracts, Vol. 235, American Astronomical
  Society Meeting Abstracts \#235, 168.02

\bibitem[{{Palma} {et~al.}(2002){Palma}, {Majewski}, \& {Johnston}}]{palma02a}
{Palma}, C., {Majewski}, S.~R., \& {Johnston}, K.~V. 2002, \apj, 564, 736,
  \dodoi{10.1086/324137}

\bibitem[{{Peek} {et~al.}(2011){Peek}, {Heiles}, {Douglas}, {Lee}, {Grcevich},
  {Stanimirovi{\'c}}, {Putman}, {Korpela}, {Gibson}, {Begum}, {Saul},
  {Robishaw}, \& {Kr{\v{c}}o}}]{peek11a}
{Peek}, J.~E.~G., {Heiles}, C., {Douglas}, K.~A., {et~al.} 2011, \apjs, 194,
  20, \dodoi{10.1088/0067-0049/194/2/20}

\bibitem[{{Rhode} {et~al.}(2013){Rhode}, {Salzer}, {Haurberg}, {Van Sistine},
  {Young}, {Haynes}, {Giovanelli}, {Cannon}, {Skillman}, {McQuinn}, \&
  {Adams}}]{rhode13a}
{Rhode}, K.~L., {Salzer}, J.~J., {Haurberg}, N.~C., {et~al.} 2013, \aj, 145,
  149, \dodoi{10.1088/0004-6256/145/6/149}

\bibitem[{{Roberts}(1959)}]{roberts59a}
{Roberts}, M.~S. 1959, \nat, 184, 1555, \dodoi{10.1038/1841555a0}

\bibitem[{{Roberts}(1960)}]{roberts60a}
---. 1960, \aj, 65, 457, \dodoi{10.1086/108288}

\bibitem[{{Ryan-Weber} {et~al.}(2008){Ryan-Weber}, {Begum}, {Oosterloo}, {Pal},
  {Irwin}, {Belokurov}, {Evans}, \& {Zucker}}]{ryan-weber08a}
{Ryan-Weber}, E.~V., {Begum}, A., {Oosterloo}, T., {et~al.} 2008, \mnras, 384,
  535, \dodoi{10.1111/j.1365-2966.2007.12734.x}

\bibitem[{{Sand} {et~al.}(2015){Sand}, {Crnojevi{\'c}}, {Bennet}, {Willman},
  {Hargis}, {Strader}, {Olszewski}, {Tollerud}, {Simon}, {Caldwell},
  {Guhathakurta}, {James}, {Koposov}, {McLeod}, {Morrell}, {Peacock},
  {Salinas}, {Seth}, {Stark}, \& {Toloba}}]{sand15a}
{Sand}, D.~J., {Crnojevi{\'c}}, D., {Bennet}, P., {et~al.} 2015, \apj, 806, 95,
  \dodoi{10.1088/0004-637X/806/1/95}

\bibitem[{{Sand} {et~al.}(2017){Sand}, {Seth}, {Crnojevi{\'c}}, {Spekkens},
  {Strader}, {Adams}, {Caldwell}, {Guhathakurta}, {Kenney}, {Randall}, {Simon},
  {Toloba}, \& {Willman}}]{sand17a}
{Sand}, D.~J., {Seth}, A.~C., {Crnojevi{\'c}}, D., {et~al.} 2017, \apj, 843,
  134, \dodoi{10.3847/1538-4357/aa7557}

\bibitem[{{Saul} {et~al.}(2012){Saul}, {Peek}, {Grcevich}, {Putman}, {Douglas},
  {Korpela}, {Stanimirovi{\'c}}, {Heiles}, {Gibson}, {Lee}, {Begum}, {Brown},
  {Burkhart}, {Hamden}, {Pingel}, \& {Tonnesen}}]{saul12a}
{Saul}, D.~R., {Peek}, J.~E.~G., {Grcevich}, J., {et~al.} 2012, \apj, 758, 44,
  \dodoi{10.1088/0004-637X/758/1/44}

\bibitem[{{Schlafly} \& {Finkbeiner}(2011)}]{schlafly11a}
{Schlafly}, E.~F., \& {Finkbeiner}, D.~P. 2011, \apj, 737, 103,
  \dodoi{10.1088/0004-637X/737/2/103}

\bibitem[{{Schlegel} {et~al.}(1998){Schlegel}, {Finkbeiner}, \&
  {Davis}}]{schlegel98a}
{Schlegel}, D.~J., {Finkbeiner}, D.~P., \& {Davis}, M. 1998, \apj, 500, 525,
  \dodoi{10.1086/305772}

\bibitem[{{Sharina} {et~al.}(2018){Sharina}, {Ryabova}, {Maricheva}, \&
  {Gorban}}]{sharina18a}
{Sharina}, M.~E., {Ryabova}, M.~V., {Maricheva}, M.~I., \& {Gorban}, A.~S.
  2018, Astronomy Reports, 62, 733, \dodoi{10.1134/S1063772918110069}

\bibitem[{{Simon}(2019)}]{simon19a}
{Simon}, J.~D. 2019, \araa, 57, 375,
  \dodoi{10.1146/annurev-astro-091918-104453}

\bibitem[{{Skillman} {et~al.}(2013){Skillman}, {Salzer}, {Berg}, {Pogge},
  {Haurberg}, {Cannon}, {Aver}, {Olive}, {Giovanelli}, {Haynes}, {Adams},
  {McQuinn}, \& {Rhode}}]{skillman13a}
{Skillman}, E.~D., {Salzer}, J.~J., {Berg}, D.~A., {et~al.} 2013, \aj, 146, 3,
  \dodoi{10.1088/0004-6256/146/1/3}

\bibitem[{{Smith}(2022)}]{smith22a}
{Smith}, N.~J. 2022, PhD thesis, Indiana University, Bloomington

\bibitem[{{Spergel}(1991)}]{spergel91a}
{Spergel}, D.~N. 1991, \nat, 352, 221, \dodoi{10.1038/352221a0}

\bibitem[{{Stefanon} {et~al.}(2017){Stefanon}, {Yan}, {Mobasher}, {Barro},
  {Donley}, {Fontana}, {Hemmati}, {Koekemoer}, {Lee}, {Lee}, {Nayyeri}, {Peth},
  {Pforr}, {Salvato}, {Wiklind}, {Wuyts}, {Ashby}, {Castellano}, {Conselice},
  {Cooper}, {Cooray}, {Dolch}, {Ferguson}, {Galametz}, {Giavalisco}, {Guo},
  {Willner}, {Dickinson}, {Faber}, {Fazio}, {Gardner}, {Gawiser}, {Grazian},
  {Grogin}, {Kocevski}, {Koo}, {Lee}, {Lucas}, {McGrath}, {Nandra}, {Newman},
  \& {van der Wel}}]{stefanon17a}
{Stefanon}, M., {Yan}, H., {Mobasher}, B., {et~al.} 2017, \apjs, 229, 32,
  \dodoi{10.3847/1538-4365/aa66cb}

\bibitem[{{Tody}(1986)}]{tody86a}
{Tody}, D. 1986, in Society of Photo-Optical Instrumentation Engineers (SPIE)
  Conference Series, Vol. 627, Instrumentation in astronomy VI, ed. D.~L.
  {Crawford}, 733, \dodoi{10.1117/12.968154}

\bibitem[{{Tody}(1993)}]{tody93a}
{Tody}, D. 1993, in Astronomical Society of the Pacific Conference Series,
  Vol.~52, Astronomical Data Analysis Software and Systems II, ed. R.~J.
  {Hanisch}, R.~J.~V. {Brissenden}, \& J.~{Barnes}, 173

\bibitem[{{Tollerud} {et~al.}(2015){Tollerud}, {Geha}, {Grcevich}, {Putman}, \&
  {Stern}}]{tollerud15a}
{Tollerud}, E.~J., {Geha}, M.~C., {Grcevich}, J., {Putman}, M.~E., \& {Stern},
  D. 2015, \apjl, 798, L21, \dodoi{10.1088/2041-8205/798/1/L21}

\bibitem[{{Tollerud} {et~al.}(2016){Tollerud}, {Geha}, {Grcevich}, {Putman},
  {Weisz}, \& {Dolphin}}]{tollerud16a}
{Tollerud}, E.~J., {Geha}, M.~C., {Grcevich}, J., {et~al.} 2016, \apj, 827, 89,
  \dodoi{10.3847/0004-637X/827/2/89}

\bibitem[{{Tollerud} \& {Peek}(2018)}]{tollerud18a}
{Tollerud}, E.~J., \& {Peek}, J.~E.~G. 2018, \apj, 857, 45,
  \dodoi{10.3847/1538-4357/aab3e4}

\bibitem[{{van den Bergh}(2010)}]{vandenbergh10a}
{van den Bergh}, S. 2010, \aj, 140, 1043, \dodoi{10.1088/0004-6256/140/4/1043}

\bibitem[{{van Loon} {et~al.}(2006){van Loon}, {Stanimirovi{\'c}}, {Evans}, \&
  {Muller}}]{vanloon06a}
{van Loon}, J.~T., {Stanimirovi{\'c}}, S., {Evans}, A., \& {Muller}, E. 2006,
  \mnras, 365, 1277, \dodoi{10.1111/j.1365-2966.2005.09815.x}

\bibitem[{{van Loon} {et~al.}(2009){van Loon}, {Stanimirovi{\'c}}, {Putman},
  {Peek}, {Gibson}, {Douglas}, \& {Korpela}}]{vanloon09a}
{van Loon}, J.~T., {Stanimirovi{\'c}}, S., {Putman}, M.~E., {et~al.} 2009,
  \mnras, 396, 1096, \dodoi{10.1111/j.1365-2966.2009.14778.x}

\bibitem[{{Walsh} {et~al.}(2009){Walsh}, {Willman}, \& {Jerjen}}]{walsh09a}
{Walsh}, S.~M., {Willman}, B., \& {Jerjen}, H. 2009, \aj, 137, 450,
  \dodoi{10.1088/0004-6256/137/1/450}

\bibitem[{{York} {et~al.}(2000){York}, {Adelman}, {Anderson}, {Anderson},
  {Annis}, {Bahcall}, {Bakken}, {Barkhouser}, {Bastian}, {Berman}, {Boroski},
  {Bracker}, {Briegel}, {Briggs}, {Brinkmann}, {Brunner}, {Burles}, {Carey},
  {Carr}, {Castander}, {Chen}, {Colestock}, {Connolly}, {Crocker}, {Csabai},
  {Czarapata}, {Davis}, {Doi}, {Dombeck}, {Eisenstein}, {Ellman}, {Elms},
  {Evans}, {Fan}, {Federwitz}, {Fiscelli}, {Friedman}, {Frieman}, {Fukugita},
  {Gillespie}, {Gunn}, {Gurbani}, {de Haas}, {Haldeman}, {Harris}, {Hayes},
  {Heckman}, {Hennessy}, {Hindsley}, {Holm}, {Holmgren}, {Huang}, {Hull},
  {Husby}, {Ichikawa}, {Ichikawa}, {Ivezi{\'c}}, {Kent}, {Kim}, {Kinney},
  {Klaene}, {Kleinman}, {Kleinman}, {Knapp}, {Korienek}, {Kron}, {Kunszt},
  {Lamb}, {Lee}, {Leger}, {Limmongkol}, {Lindenmeyer}, {Long}, {Loomis},
  {Loveday}, {Lucinio}, {Lupton}, {MacKinnon}, {Mannery}, {Mantsch}, {Margon},
  {McGehee}, {McKay}, {Meiksin}, {Merelli}, {Monet}, {Munn}, {Narayanan},
  {Nash}, {Neilsen}, {Neswold}, {Newberg}, {Nichol}, {Nicinski}, {Nonino},
  {Okada}, {Okamura}, {Ostriker}, {Owen}, {Pauls}, {Peoples}, {Peterson},
  {Petravick}, {Pier}, {Pope}, {Pordes}, {Prosapio}, {Rechenmacher}, {Quinn},
  {Richards}, {Richmond}, {Rivetta}, {Rockosi}, {Ruthmansdorfer}, {Sandford},
  {Schlegel}, {Schneider}, {Sekiguchi}, {Sergey}, {Shimasaku}, {Siegmund},
  {Smee}, {Smith}, {Snedden}, {Stone}, {Stoughton}, {Strauss}, {Stubbs},
  {SubbaRao}, {Szalay}, {Szapudi}, {Szokoly}, {Thakar}, {Tremonti}, {Tucker},
  {Uomoto}, {Vanden Berk}, {Vogeley}, {Waddell}, {Wang}, {Watanabe},
  {Weinberg}, {Yanny}, {Yasuda}, \& {SDSS Collaboration}}]{york00a}
{York}, D.~G., {Adelman}, J., {Anderson}, John~E., J., {et~al.} 2000, \aj, 120,
  1579, \dodoi{10.1086/301513}

\bibitem[{{Young} {et~al.}(2014){Young}, {Kotulla}, {Gopu}, \&
  {Liu}}]{young14a}
{Young}, M.~D., {Kotulla}, R., {Gopu}, A., \& {Liu}, W. 2014, in Society of
  Photo-Optical Instrumentation Engineers (SPIE) Conference Series, Vol. 9152,
  Software and Cyberinfrastructure for Astronomy III, ed. G.~{Chiozzi} \& N.~M.
  {Radziwill}, 91522U, \dodoi{10.1117/12.2056763}

\end{thebibliography}
\bibliographystyle{aasjournal}




\begin{deluxetable}{ccccrrrcccc}
\tablecolumns{11}
\tabletypesize{\scriptsize}
\tablecaption{Sample of UCHVCs Analyzed for This Work}
\tablehead{
\colhead{Name} & \colhead{R.A.} & \colhead{Dec.} & \colhead{$S_{21}$} & \colhead{$cz$} 
& \colhead{$w_{50}$} & \colhead{$\bar{a}$} & \colhead{log$\bar{N}_{HI}$} & \colhead{log$M_{HI}$} & \colhead{Obs. Date} & \colhead{Detection} \\
& & & \colhead{[Jy km s$^{-1}$]} & \colhead{[km s$^{-1}$]} & \colhead{[km s$^{-1}$]} & \colhead{[$^\prime$]} & \colhead{[atoms cm$^{-2}$]} & \colhead{[M$_{\odot}$]} &  & \colhead{}
}
\startdata
AGC102994 & 00h54m31.6s & +29d24m02s & 0.58 $\pm$ 0.03 & -290 & 23 $\pm$ 2 & 5.70 & 18.90 & 5.14 & 2016B, 2017B & ND\\
AGC115710 & 01h26m21.5s & +05d23m08s & 4.81 $\pm$ 0.04 & 59 & 21 $\pm$ 1 & 8.65 & 19.45 & 6.05 & 2016B & PD\\
AGC116548 & 01h06m00.7s & +12d26m57s & 0.96 $\pm$ 0.03 & 84 & 11 $\pm$ 1 & 6.15 & 19.05 & 5.35 & 2015B & ND\\
AGC122834 & 02h03m48.8s & +29d13m13s & 3.58 $\pm$ 0.06 & 49 & 16 $\pm$ 5 & 10.70 & 19.14 & 5.93 & 2016B & ND\\
AGC122835 & 02h05m35.5s & +29d13m56s & 1.23 $\pm$ 0.04 & 29 & 23 $\pm$ 5 & 5.60 & 19.24 & 5.46 & 2016B & PD\\
AGC174764 & 07h56m14.8s & +25d09m00s & 0.64 $\pm$ 0.04 & 174 & 21 $\pm$ 1 & 7.00 & 18.76 & 5.18 & 2016B & ND\\
AGC198683 & 09h32m08.0s & +23d37m52s & 0.88 $\pm$ 0.04 & 178 & 19 $\pm$ 1 & 11.10 & 18.50 & 5.32 & 2016B & ND\\
AGC208315 & 10h27m01.1s & +08d47m08s & 4.96 $\pm$ 0.07 & 148 & 20 $\pm$ 2 & 12.60 & 19.14 & 6.07 & 2017A & PD\\
AGC208524 & 10h47m02.5s & +01d46m31s & 0.67 $\pm$ 0.03 & 179 & 14 $\pm$ 1 & 4.85 & 19.10 & 5.20 & 2016A, 2017A & ND \\
AGC208752 & 10h23m09.0s & +20d40m59s & 1.79 $\pm$ 0.03 & -60 & 16 $\pm$ 1 & 7.60 & 19.13 & 5.63 & 2019A & ND\\
AGC219214 & 11h09m29.8s & +05d26m01s & 0.56 $\pm$ 0.03 & 142 & 20 $\pm$ 5 & 5.55 & 18.90 & 5.12 & 2019A & ND\\
AGC227977 & 12h09m20.0s & +04d23m30s & 0.57 $\pm$ 0.05 & -143 & 18 $\pm$ 4 & 10.00 & 18.40 & 5.13 & 2014A & ND\\
AGC227988 & 12h46m22.9s & +04d48m42s & 0.49 $\pm$ 0.05 & 320 & 48 $\pm$ 10 & 4.70 & 18.99 & 5.06 & 2019A & ND\\
AGC233763 & 13h12m42.3s & +13d30m46s & 0.78 $\pm$ 0.04 & 129 & 31 $\pm$ 6 & 5.30 & 19.09 & 5.26 & 2017A & ND\\
AGC233831 & 13h22m41.6s & +11d52m31s & 0.63 $\pm$ 0.02 & 124 & 16 $\pm$ 1 & 4.50 & 19.14 & 5.17 & 2018A & ND\\
AGC249283 & 14h23m57.7s & +05d23m40s & 1.11 $\pm$ 0.07 & 252 & 32 $\pm$ 9 & 13.50 & 18.43 & 5.42 & 2019A & PD\\
AGC257994 & 15h53m54.0s & +14d41m48s & 2.04 $\pm$ 0.04 & 146 & 23 $\pm$ 3 & 9.65 & 18.98 & 5.68 & 2021A & ND\\
AGC268067 & 16h05m29.4s & +16d09m12s & 1.94 $\pm$ 0.05 & 158 & 35 $\pm$ 7 & 8.40 & 19.08 & 5.66 & 2018A & ND\\
AGC268071 & 16h12m36.8s & +14d12m26s & 2.67 $\pm$ 0.08 & 109 & 62 $\pm$ 15 & 9.75 & 19.09 & 5.80 & 2018A & BD\\ 
AGC268213 & 16h22m35.7s & +05d08m48s & 2.90 $\pm$ 0.06 & -139 & 18 $\pm$ 4 & 10.85 & 19.03 & 5.83 & 2017A & ND\\
AGC333604 & 23h11m23.2s & +27d56m45s & 1.74 $\pm$ 0.03 & 66 & 19 $\pm$ 2 & 7.65 & 19.12 & 5.61 & 2018B & ND\\
AGC333651 & 23h55m21.4s & +25d17m26s & 0.87 $\pm$ 0.03 & 45 & 14 $\pm$ 4 & 7.85 & 18.79 & 5.31 & 2015B & ND\\
AGC334257 & 23h02m11.3s & +16d00m48s & 0.68 $\pm$ 0.04 & -452 & 22 $\pm$ 11 & 7.95 & 18.68 & 5.20 & 2018B & ND\\
AGC335755 & 23h14m16.4s & +03d23m07s & 7.95 $\pm$ 0.10 & 42 & 23 $\pm$ 1 & 13.15 & 19.31 & 6.27 & 2018B & ND\\
AGC501816 & 10h05m19.4s & -00d02m26s & 1.70 $\pm$ 0.05 & 107 & 20 $\pm$ 1 & 7.15 & 19.17 & 5.60 & 2019A & ND\\ 
AGC749140 & 00h51m16.4s & +15d11m11s & 1.33 $\pm$ 0.03 & 52 & 17 $\pm$ 4 & 5.95 & 19.22 & 5.50 & 2018B & PD\\
\enddata
\tablecomments{
Following \citet{haynes18a}, the adopted error on the recessional velocities listed in column 5 is half of the w$_{50}$ error. 
The \hi\ mass values in column 9 are calculated at an assumed distance of 1~Mpc. In the "Detection Status" column, objects for which no detection in the field meets our lower-threshold criteria 
are listed as ND; this includes the UCHVC AGC~501816, which is located near the globular cluster Pal~3, and which is discussed in Sec.~\ref{sec: pal3}. Objects categorized as having a possible detection are listed as PD, and objects with our best (most convincing) detections are listed as BD; see Sec.~\ref{sec:assessing} for more information about the categories.}
\label{table: uchvcs sample}
\end{deluxetable}

\begin{deluxetable}{cccccccccc}
\centering 
\tablecolumns{10}
\tabletypesize{\scriptsize}
\tablecaption{UCHVCs With Insufficient Data to Include in the Sample\label{table: insufficient}}
\tablehead{
\colhead{Name} & \colhead{R.A.} & \colhead{Dec.} & \colhead{$S_{21}$} & \colhead{$cz$} & \colhead{$w_{50}$} & \colhead{$\bar{a}$} & \colhead{log$\bar{N}_{HI}$} & \colhead{log$M_{HI}$} & \colhead{Obs. Date} \\
& & & \colhead{[Jy km s$^{-1}$]} & \colhead{[km s$^{-1}$]} & \colhead{[kms$^{-1}$]} & \colhead{[$^\prime$]} & \colhead{[atoms cm$^{-2}$]} & \colhead{[M$_{\odot}$]} &
}
\startdata
AGC219663 & 11h34m29.7s & +20d12m49s & 0.75 $\pm$ 0.03 & 74 & 17$\pm$1 & 7.35 & 18.79 & 5.25 & 2013A \\
AGC229327 & 12h32m31.6s & +17d57m21s & 1.02 $\pm$ 0.05 & 251 & 25$\pm$1 & 12.95 & 18.43 & 5.38 & 2013A \\
AGC232765 & 13h23m09.4s & +15d11m17s & 1.37 $\pm$ 0.03 & 105 & 23$\pm$3 & 5.15 & 19.36 & 5.51 & 2016A \\ 
AGC249326 & 14h31m58.8s & +06d35m20s & 0.70 $\pm$ 0.04 & 136 & 38$\pm$11 & 5.65 & 18.98 & 5.22 & N/A \\
AGC249393 & 14h10m58.1s & +24d12m04s & 1.01 $\pm$ 0.06 & -156 & 36$\pm$1 & 11.90 & 18.50 & 5.38 & 2013A \\
AGC249441 & 14h07m00.8s & +00d13m23s & 0.74 $\pm$ 0.04 & -131 & 16$\pm$2 & 8.40 & 18.66 & 5.24 & g: 2019A \\
 AGC249565 & 14h35m57.6s & +17d10m04s & 1.76 $\pm$ 0.04 & 30 & 18$\pm$1 & 7.65 & 19.12 & 5.62 & 2013A \\
AGC257956 & 15h55m07.5s & +14d29m29s & 1.54 $\pm$ 0.04 & 144 & 25$\pm$6 & 6.70 & 19.18 & 5.56 & i: 2017A\\
AGC258241 & 15h08m24.4s & +11d24m22s & 0.98 $\pm$ 0.07 & 163 & 19$\pm$2 & 12.70 & 18.43 & 5.36 & N/A \\
\enddata

\tablecomments{
The adopted error on recessional velocities listed in column 5 is half of the w$_{50}$ error according to \citet{haynes18a}. The \hi\ mass values in column 9 are calculated at an assumed distance of 1~Mpc.}
\end{deluxetable}

\movetabledown=60mm
\begin{rotatetable}
\begin{deluxetable}{cccccrccccccrc}
\tablecolumns{13}
\tabletypesize{\scriptsize}
\tablecaption{Stellar Overdensities Found in the UCHVC Images Analyzed for This Work\label{table: detections}}
\tablehead{
\colhead{Name} & \colhead{Significance} & \colhead{Distance\tablenotemark{a}} & \colhead{R.A.} & \colhead{Dec.} & \colhead{log M$_{HI}$} & \colhead{M$_V$} & \colhead{M$_V$} & {$g-i$} & {$g-i$} & \colhead{log M$_*$} & \colhead{log M$_*$} & \colhead{M$_{HI}$/M$_*$} & \colhead{M$_{HI}$/M$_*$} \\
& \colhead{(\# stars)} & \colhead{(Range)} & & & & \colhead{(faint)} & \colhead{(bright)} & \colhead{(faint)} & \colhead{(bright)} & \colhead{(faint)} & \colhead{(bright)} & \colhead{(faint)} & \colhead{(bright)}\\
 & \colhead{[\%]}  & \colhead{[Mpc]} & & & \colhead{[M$_\odot$]} & \colhead{[mag]} & \colhead{[mag]} & \colhead{[mag]} & \colhead{[mag]} & \colhead{[M$_\odot$]} & \colhead{[M$_\odot$]} & &
}
\startdata
\multicolumn{12}{c}{Possible Detections} \\
\hline
AGC115710 & 99.55(9) & 0.27 (0.25-0.31) & 01:25:53.7 & +05:25:51.3 & 4.93$^{+0.11}_{-0.08}$ & -2.18$^{+0.19}_{-0.27}$ & -7.01$^{+0.19}_{-0.27}$ & 0.92$\pm$0.02 & 1.19$\pm$0.01 & 3.22$^{+0.11}_{-0.08}$ & 5.35$^{+0.11}_{-0.08}$ & 50.82$\pm0.12$ & 0.38$\pm0.01$\\
AGC122835 & 95.30(10) & 0.87 (0.86-0.92) & 02:05:26.0 & +29:17:57.1 & 5.34$^{+0.05}_{-0.01}$ & -4.47$^{+0.03}_{-0.12}$ & -7.67$^{+0.03}_{-0.12}$ & 1.57$\pm$0.03 & 1.49$\pm$0.07 & 4.62$^{+0.05}_{-0.01}$ & 5.50$^{+0.05}_{-0.01}$ & 5.17$\pm0.01$ & 0.69$\pm0.01$\\
AGC208315 & 86.87(12) & 0.63 (0.59-0.64) & 10:27:08.8 & +08:51:41.2 & 5.67$^{+0.02}_{-0.01}$ & -3.63$^{+0.14}_{-0.04}$ & -6.97$^{+0.14}_{-0.04}$ & 0.26$\pm$0.03 & 0.63$\pm$0.08 & 3.30$^{+0.02}_{-0.05}$ & 4.78$^{+0.02}_{-0.05}$ & 232.23$\pm0.54$ & 7.74$\pm0.02$\\
AGC249283 & 83.13(9) & 0.44 (0.42-0.50) & 14:23:58.7 & +05:22:56.6 & 4.72$^{+0.10}_{-0.04}$ & -3.52$^{+0.11}_{-0.24}$ & -8.15$^{+0.11}_{-0.24}$ & 1.41$\pm$0.02 & 1.59$\pm$0.05 & 4.13$^{+0.10}_{-0.04}$ & 5.74$^{+0.10}_{-0.04}$ & 3.91$\pm0.01$ & 0.09$\pm0.01$\\
AGC749140 & 93.04(9) & 0.49 (0.43-0.49) & 00:51:20.7 & +15:14:46.3 & 4.88$^{+0.01}_{-0.11}$ & -3.18$^{+0.27}_{-0.02}$ & -6.92$^{+0.27}_{-0.02}$ & 0.84$\pm$0.03 & 0.94$\pm$0.05 & 3.56$^{+0.01}_{-0.11}$ & 5.13$^{+0.01}_{-0.11}$ & 20.64$\pm0.10$ & 0.56$\pm0.01$\\
\hline
\multicolumn{12}{c}{Best Detection} \\
\hline
AGC268071 & 97.30(11) & 0.57(0.49-0.59) & 16:12:55.6 & +14:20:45.3 & 5.31$^{+0.04}_{-0.13}$ & -4.29$^{+0.32}_{-0.10}$ & -6.99$^{+0.32}_{-0.10}$ & 1.12$\pm$0.01 & 1.03$\pm$0.01 & 4.21$^{+0.04}_{-0.13}$ & 5.31$^{+0.04}_{-0.13}$ & 12.48$\pm0.03$ & 1.21$\pm0.01$\\
\enddata
\tablenotetext{a}{The distance listed here is the CMD filter distance 
associated with the most significant detection of a given overdensity.  
The numbers within the parentheses represent the range of distances for which a given detection remains above some significance threshold (90\% for overdensities that have a peak significance $\geq$90\%, and 80\% for overdensities with a peak significance between 80$-$90\%.)}

\end{deluxetable}
\end{rotatetable}

\begin{deluxetable}{ccccrrrcccc}
\tablecolumns{11}
\tabletypesize{\scriptsize}
\tablecaption{UCHVCs from J19 That Were Reanalyzed for This Study}
\tablehead{
\colhead{Name} & \colhead{R.A.} & \colhead{Dec.} & \colhead{$S_{21}$} & \colhead{$cz$} & \colhead{$w_{50}$} & \colhead{$\bar{a}$} & \colhead{log$\bar{N}_{HI}$} & \colhead{log$M_{HI}$\tablenotemark{a}} & \colhead{Obs. Date} & \colhead{Detection} \\
& & & \colhead{[Jy km s$^{-1}$]} & \colhead{[km s$^{-1}$]} & \colhead{[km s$^{-1}$]} & \colhead{[$^\prime$]} & \colhead{[atoms cm$^{-2}$]} & \colhead{[M$_{\odot}$]} & 
}
\startdata
AGC198606 & 09h30m05.5s & +16d39m03s & 6.73 $\pm$ 0.67 & 53 & 21$\pm$1 & 9.00 & 19.60 & 6.20 & 2013A & ND\\
AGC215417 & 11h40m08.1s & +15d06m44s & 0.70 $\pm$ 0.07 & 216 & 17$\pm$4 & 9.49 & 18.50 & 5.20 & 2013A & ND\\
AGC219656 & 11h51m24.3s & +20d32m20s & 0.85 $\pm$ 0.08 & 192 & 21$\pm$1 & 8.00 & 18.80 & 5.30 & 2013A & ND\\
AGC249525 & 14h17m50.1s & +17d32m52s & 6.73 $\pm$ 0.67 & 48 & 24$\pm$7 & 9.00 & 19.60 & 6.20 & 2013A & BD\\
AGC268069 & 16h05m32.6s & +14d59m20s & 1.14 $\pm$ 0.11 & 132 & 29$\pm$4 & 7.07 & 19.00 & 5.40 & 2013A & ND\\
\enddata
\tablecomments{
Following \citet{haynes18a}, the adopted error on the recessional velocities listed in column 5 is half of the w$_{50}$ error. 
The \hi\ mass values in column 9 are calculated at an assumed distance of 1~Mpc. In the "Detection Status" column, objects for which no detection in the field meets our lower-threshold criteria 
are listed as ND, and objects with our best (most convincing) detections are listed as BD; see Sec.~\ref{sec:assessing} for more information about the categories.}
\label{table: podi uchvcs sample}
\end{deluxetable}


\movetabledown=60mm
\begin{rotatetable}
\begin{deluxetable}{cccccrccccccrc}
\tablecolumns{13}
\tabletypesize{\scriptsize}
\tablecaption{Stellar Overdensities Found in the Reanalyzed UCHVC Images from J19}
\tablehead{
\colhead{Name} & \colhead{Significance} & \colhead{Distance\tablenotemark{a}} & \colhead{R.A.} & \colhead{Dec.} & \colhead{log M$_{HI}$} & \colhead{M$_V$} & \colhead{M$_V$} & {$g-i$} & {$g-i$} & \colhead{log M$_*$} & \colhead{log M$_*$} & \colhead{M$_{HI}$/M$_*$} & \colhead{M$_{HI}$/M$_*$} \\
& \colhead{(\# stars)} & \colhead{(Range)} & & & & \colhead{(faint)} & \colhead{(bright)} & \colhead{(faint)} & \colhead{(bright)} & \colhead{(faint)} & \colhead{(bright)} & \colhead{(faint)} & \colhead{(bright)}\\
 & \colhead{[\%]}  & \colhead{[Mpc]} & & & \colhead{[M$_\odot$]} & \colhead{[mag]} & \colhead{[mag]} & \colhead{[mag]} & \colhead{[mag]} & \colhead{[M$_\odot$]} & \colhead{[M$_\odot$]} & &
}
\startdata
AGC249525 & 99.26(25) & 2.09 (1.93-2.12) & 14:17:53.9 & +17:32:46.0 & 6.80$^{+0.01}_{-0.03}$ & -5.99$^{+0.07}_{-0.03}$ & -6.87$^{+0.07}_{-0.03}$ & 0.96$\pm$0.03 & 0.92$\pm$0.02 & 4.77$^{+0.34}_{-0.38}$ & 5.09$^{+0.01}_{-0.05}$ & 106.40$\pm2.15$ & 50.84$\pm0.54$\\
\enddata
\tablenotetext{a}{The distance listed here is the CMD filter distance 
associated with the most significant detection of the overdensity.  
The numbers within the parentheses represent the range of distances for which the detection remains above 90\% significance. 
}
\label{table: podi detection}
\end{deluxetable}
\end{rotatetable}

\end{document}